# Lattice-Spring Analogy for Isotropic Elasticity

D. M. LI [a, b, *], Meng-Cheng HE [a]

a. *School of Civil Engineering and Architecture, Wuhan University of Technology, Wuhan, 430070, China*

b. *Sanya Science and Education Innovation Park of Wuhan University of Technology, Sanya, 572000, China*

**Abstract:** This study introduces an innovative Isotropic Elastic Lattice Spring Model (IELSM) that addresses the fundamental limitation of classical lattice spring models: the constraint of fixed Poisson's ratio. By amending the total strain energy within the Lattice Spring Model (LSM) framework, IELSM provides a theoretically self-consistent formulation for simulating isotropic elastic materials with arbitrary Poisson's ratios. The model's core innovation lies in augmenting classical axial spring frameworks with explicit additional volumetric constraints, establishing a direct and exact mapping between IELSM's parameters and macroscopic elastic constants. This enables simulation across the entire physically admissible Poisson's ratio spectrum: $-1 < \nu < 1$ under plane stress conditions and $-1 < \nu < 0.5$ under plane strain conditions. Eigenvalue analysis indicates that the IELSM has better numerical stability compared to the standard bilinear quadrilateral element and the constant strain triangular element. The characteristic of the numerical implementation lies in directly decomposing the additional volumetric constraints into an equivalent combination of standard mechanical components (axial, shear and rotational springs), laying the foundation for the realization of fracture simulation based on discrete methods. Comprehensive validation through uniaxial tension, pure shear, stress concentration around a circular hole, and stress singularity analyses for central and crucifix-shaped cracks demonstrates IELSM's exceptional accuracy, convergence analysis, and computational robustness. The model exhibits excellent performance in stress intensity factor calculations at crack tips, validating its effectiveness for singular stress field analysis. This work bridges the gap between LSM and continuum mechanics, establishing an analog framework that maintains theoretical consistency while offering computational accuracy for the solution of elastic boundary value problems.

**Keywords:** Lattice spring model; Isotropic elasticity; Poisson's ratio; Additional volumetric constraint; Discrete-continuum equivalence

[*]Corresponding to D. M. LI (dongmli2-c@my.cityu.edu.hk, domili@whut.edu.cn).

## 1.0 Introduction

In recent years, the lattice spring model (LSM) has regained the attention of researchers for its natural adaptability to fracture simulation in the context of crack propagation prediction and simulation. Unlike continuum-based numerical simulation methods such as the finite element method (FEM), LSM employs a discontinuous grid system and simulates the elastic behavior of a continuum through a structure composed of springs (trusses), effectively avoiding issues related to singularity. The idea of this method can be traced back to the mid-20th century: Hrennikoff (1941) initiated the frame method, laying the foundation for using simple truss networks to simulate continuous elastic bodies; Mchenry (1943) further systematized the relaxation method solution process and refined the post-processing relationship for calculating stress from lattice displacements, enhancing the method's engineering practicality; McCormick (1963) extended this method to digital computer platforms, achieving independent control of Poisson's ratio by introducing the bending stiffness of side bars and establishing an efficient solution algorithm based on narrowband storage and triangular decomposition, laying the computational foundation for the development of LSM into a general numerical method. Since then, the application scope of LSM has continuously expanded, extending from traditional small deformation analysis (Attar et al., 2014) to complex responses such as large deformation, nonlinear behavior, and dynamic fracture (Braun and Ariza, 2019; Fournier et al., 2007; Zhang, 2019). In the field of material modeling, its applicability far exceeds the linear elastic domain, and it can be applied to hyperelastic solids (Hartquist et al., 2025; Zhang, 2019), porous elastoplastic geotechnical materials (Hadzalić et al., 2019; Li et al., 2024), quasi-brittle composites (Braun and Ariza, 2019; Gaetani et al., 2019), and stochastic heterogeneous materials (Fascetti et al., 2022). In engineering practice, LSM has been proven to be effectively applied in areas such as acoustic emission simulation of concrete structures (Zhou et al., 2025), structural topology optimization (Isikli et al., 2026), and shell form-finding (Sá Marques et al., 2019). These applications fully demonstrate the outstanding multi-functionality of LSM in addressing various mechanical challenges.

However, LSM faces a fundamental theoretical limitation: when attempting to simulate isotropic macroscopic media, the simulated Poisson's ratio is constrained to a fixed value (as shown by Hrennikoff (1941), for square grids with diagonal components, $v$ = 1/3 under plane stress conditions and $v$ = 1/4 under plane strain conditions). The theoretical root of this limitation can be traced back to 1827, when Poisson, based on the assumption of central interaction between molecules, derived that the ratio of lateral contraction to longitudinal extension of an isotropic elastic body under uniaxial tension is a constant (i.e., Poisson's ratio), and thus wrongly concluded that the Poisson's ratio of all materials is constant (Greaves, 2013). Since then, the controversy over whether Poisson's ratio is an inherent constant of materials has sparked extensive experimental measurements and in-depth research on elastic theory modeling, gradually revealing the fact that Poisson's ratio can vary within a certain range. However, the Poisson's ratio limitation of pure central interaction systems has persisted to this day. Although Hrennikoff (1941) proposed a corresponding solution, it requires introducing an additional degree of freedom and only achieves limited range control of Poisson's ratio, and the resulting computational overhead has severely hindered LSM's ability to simulate general engineering materials. Due to its inherent limitations, as the finite element method made transformative progress in variational principles, triangular/quadrilateral elements, geometric adaptability, material multifunctionality, and generality of solutions during the same period (Courant, 1943; Feng, 1965; Turner et al., 1956), LSM gradually became marginalized. However, attempts to bridge continuous media and discrete systems have always existed and thrived in various numerical methods. Discrete element method (DEM) (Cundall, 1971), peridynamics (PD) (Silling, 2000), and the modern revival of LSM are all based on the idea of discretization. These methods based on discrete interactions all retain the Poisson's ratio limitation of their theoretical sources. To advance LSM from being focused on fracture applications to a general method for describing continuous medium problems, overcoming this limitation is crucial, which has also become the motivation of this study.

Due to the single constitutive relationship of the axial force spring model, LSM essentially cannot independently control the two elastic parameters in isotropic constitutive relations, that is, it cannot capture the independent shear response mode in the continuous medium (Braun and Ariza, 2019). This has prompted extensive research to expand the LSM's ability to characterize elastic parameters. Meanwhile, parallel research on related discrete methods has developed alternative strategies to address similar Poisson's ratio limitations. PD tackles this challenge by using non-local integral formulas instead of local gradient descriptions, employing techniques such as state-based models (which have collective bond states) (Zhou and Tian, 2021) and micropolar forms with rotational degrees of freedom (Diana and Casolo, 2019; Gerstle et al., 2007). DEM, on the other hand, adopts structural elements with additional stiffness components, including beam elements (Griffiths and Mustoe, 2001), angular springs (Wang et al., 2020), and uses non-local grid particle models (Chen et al., 2016) or introduces normal-tangential coupling springs to simulate anisotropic material behavior (Sun et al., 2026). Within the LSM paradigm specifically, recent extensions commonly discretize the continuum into interacting particles and enable either pairwise interparticle forces or finite particle rotations. Strictly speaking, therefore, although such models are frequently labeled "LSM" in the literature, they represent hybrid formulations that integrate classical LSM concepts with principles drawn from other discrete numerical frameworks.

While maintaining the basic discrete form and physical principles, LSM has stimulated many systematic strategies for achieving arbitrary Poisson's ratios, among which the most straightforward theoretical approach is to introduce independent shear springs (Griffiths and Mustoe, 2001; Braun and Fernández-Sáez, 2014). Although this method can achieve arbitrary Poisson's ratio mapping under macroscopic energy equivalence conditions, it faces a critical rotational invariance issue (Keating, 1966; Jagota and Scherer, 1993). Research has pointed out (Pan et al., 2018) that simple shear springs cannot distinguish between rigid body rotation and pure shear deformation, and rigid body rotation can erroneously generate strain energy within the shear springs, which goes against physical intuition. To address this problem, two methods can be employed: Firstly, beam-type interactions can be introduced (McCormick, 1963; Gerstle et al., 2007; Ostoja-Starzewski, 2007), which add rotational degrees of freedom at the nodes and apply auxiliary shear spring balancing moments to solve the rotational invariance problem; secondly, the calculation of shear displacements can be corrected, such as in the Distinct Lattice Spring Model (DLSM)(Zhao et al., 2011), which directly eliminates the coupled rigid body rotation-related terms in the shear displacement between two particles and instead uses the least squares method to approximate the local strain of each particle, and through hyperelastic derivation, allows the shear stiffness to become negative, thereby covering the entire range of Poisson's ratios. Due to the additional node degrees of freedom, the ordinary springs are effectively replaced by beam elements in the above beam-type interactions, so this method can also be called the Lattice Beam Model (LBM) (Schlangen and Garboczi, 1997; Lilliu and van Mier, 2003; Karihaloo et al., 2003).

Drawing from the beam lattice method and the effective use of angular interactions in previous studies (Gerstle et al., 2007; Zhang et al., 2014), a more straightforward approach is to impose constraints on the rotational degrees of freedom by introducing angular springs. Li et al. (2020) and Wang et al. (2020) both adopted this strategy, mapping the rotational degrees of freedom to the stiffness matrix of the finite element to determine the spring constants in the discrete model. The rotational invariance of the angular spring itself ensures that the model can distinguish between rigid body rotation and shear deformation, thereby introducing parameters independent of the axial spring stiffness to achieve precise matching of the isotropic elastic tensor. At the same time, some researchers have found that introducing angular springs does not require rotational degrees of freedom, but rather establish a geometric relationship between angular change and displacement, thereby greatly reducing the number of degrees of freedom of the LSM with angular springs: Wang et al. (2009) derived the axial spring - angular spring interaction (α-β) model for various grid geometries in 2D; Nguyen et al. (2025) provided an exact analytical solution for the free vibration of 3D rectangular grid based on the Gazis model (Gazis et al., 1960); Reck (2017) removed the previous unit shape restrictions by arranging a sufficient number of independent axial and angular springs within irregular triangular or tetrahedral elements and solving the energy equivalence equation; Desmoulins and Kochmann (2017) developed a nonlocal LSM based on the extended Cauchy-Born rule (Bardenhagen and Triantafyllidis, 1994), in which the deformation of the representative volume element contains second gradient information, capable of accurately simulating inelastic behavior and localization phenomena in truss networks, while naturally adapting to complex constitutive laws.

Based on the axial springs, shear springs and angular springs in the aforementioned LSM, new methods can also be derived from the perspective of the number of particles involved in the interaction. The introduced axial and shear springs are equivalent to the central or tangential interaction between two particles, while the angular spring represents the interaction among three particles. Following this line of thought, mechanisms involving more particle interactions have also been developed: one is the surface interaction among four nodes considered by Challamel et al. (2024) in a square grid; the other is the nonlocal idea borrowed from PD, where Chen et al. (2016) considered a nonlocal many-body potential to account for a similar horizon as in PD, with the stiffness of the spring depending on the deformation of its adjacent springs; Zhou and Fu (2023) further proposed a force-vector state-based nonlocal lattice model, which no longer defines the interaction between particles as a simple spring force but introduces the force-vector state from PD, and the interaction between each particle and all its neighbors within the horizon is uniformly calculated through the deformation gradient and stress tensor.

In addition to the above-mentioned mainstream methods, several innovative ideas have emerged in the academic community, aiming to address the issue of fixed Poisson’s ratio. Li and Zhu (2024) achieved adjustable Poisson’s ratio while maintaining the locality of LSM by enriching the constitutive relationship of bonds to incorporate the influence of lateral deformation, distributing a certain number of virtual internal bonds within triangular elements, and balancing the bond forces to the nodes. Asahina et al. (2015) proposed an iterative fictitious stress method, which first solves the model with a Poisson’s ratio of 0 and then achieves the target Poisson’s ratio by iteratively applying fictitious forces derived from principal stresses. Zhao (2017) introduced a fourth dimension to the three-dimensional lattice model and proposed the four-dimensional lattice spring model (4D-LSM). This model only uses axial springs, preserving the pure central interaction between particles. To overcome the limitation of fixed Poisson’s ratio in the classical LSM, the original model and its parallel mapping version in the fourth dimension are connected by additional 4D springs, and the mapping relationship between spring parameters and isotropic elastic parameters is established through hyperelastic analysis. The innovation of 4D-LSM lies in introducing the concept of extra dimensions from modern physics into classical solid mechanics. The additional 4D springs not only enhance the robustness and flexibility of the model but also eliminate the limitation of fixed Poisson's ratio in the classical LSM.

Despite differing implementation pathways, these LSM improvement strategies share the core objective of correctly mapping macroscopic continuum strain and constitutive relationships onto deformation and energy frameworks borne by discontinuously connected discrete spring networks. While they have significantly enriched LSM's elastic simulation capabilities, these approaches typically incur substantial costs. They suffer from increased model complexity and computational overhead. This occurs whether through the addition of degrees of freedom in beam elements, or through the introduction of nonlocal dependencies associated with angular springs and multi-body potentials. Ultimately, these complexities erode the simplicity and computational efficiency advantages that are characteristic of classical LSM. Many methods (e.g., those requiring negative shear stiffness for high Poisson's ratios) lack intuitive physical meaning, complicating parameter calibration and physical interpretation (Li and Zhu, 2024). Most critically, the majority of existing improvement schemes are self-contained systems that lack native compatibility with the widely adopted standard FEM theoretical framework and computational workflows. Therefore, this study aims to develop a novel LSM formulation that fundamentally addresses these limitations by providing exact equivalence with continuum isotropic elasticity across the full parameter range, maintaining clear physical interpretations for all model components, and enabling seamless integration with conventional finite element frameworks. These objectives together address the pressing challenges in advancing LSM toward practical engineering applications.

To overcome the intrinsic limitations of traditional LSM in representing isotropic elasticity with arbitrary Poisson's ratios, this paper introduces an Isotropic Elastic Lattice Spring Model (IELSM). This model introduces an additional volumetric constraint based on the classical LSM to regulate the volume deformation capacity of the original system, where the additional volumetric constraint is characterized by the introduced additional bulk modulus parameter. This approach establishes a direct and physically transparent connection between discrete spring networks and continuum elasticity. The principal contributions and their outcomes are as follows: First, the model enables independent tuning of Poisson's ratio across its entire physically admissible range while rigorously maintaining rotational invariance of the strain energy. Second, the formulation retains clear physical interpretations for all components, including the additional volumetric constraint and the auxiliary springs that emerge in the numerical implementation. Third, the additional volumetric constraints is elegantly decomposed into equivalent combinations of standard mechanical elements—axial, shear, and rotational springs—providing a straightforward pathway for embedding IELSM within established finite element workflows. This decomposition not only simplifies computational implementation but also lays a foundation for future extensions, such as incorporating damage or fracture criteria.

The paper is structured as follows: Section 2 first re-examines the fundamental reasons for classical LSM's fixed Poisson's ratio constraint from a strain energy density perspective within elasticity theory, then systematically analyzes rotation invariance conditions for shear-spring-enhanced LSM models within the same framework. Section 3 presents the fundamental formulation of the Isotropic Elastic Lattice Spring Model (IELSM). This section details the model's physical construction and rigorously derives the analytical relationships between its core parameters, specifically the axial stiffness and the additional bulk modulus, and the macroscopic continuum elastic constants including Young's modulus and Poisson's ratio. Section 4 focuses on model implementation: through complete derivation of IELSM's element stiffness matrix, we propose an equivalent assembly scheme decomposing it into standard spring elements (axial, shear, and rotational springs), significantly simplifying integration into computational procedures. For model validation and robustness assessment, Section 5 designs and executes a series of benchmark examples covering continuum deformation and crack-tip stress singularity problems, systematically evaluating IELSM's accuracy, convergence, and application potential. Finally, Section 6 summarizes key findings and discusses potential future research directions.

## 2.0 Revisiting Poisson's Ratio Constraints in Lattice Spring Models

Consider a classical two-dimensional LSM based on the computationally tractable and geometrically simple square lattice configuration presented in Fig. 1(a), which has been widely adopted since its original formulation (Hrennikoff, 1941; Mchenry, 1943). We adopt the unit cell delimited by the red region in Fig. 1(*a*) as the representative volume element (RVE), as depicted in Fig. 1(*b*). Based on this configuration, we herein re-derive the Poisson's ratio constraint inherent to classical LSMs employing exclusively axial stiffness springs (detailed derivation provided in Appendix A).

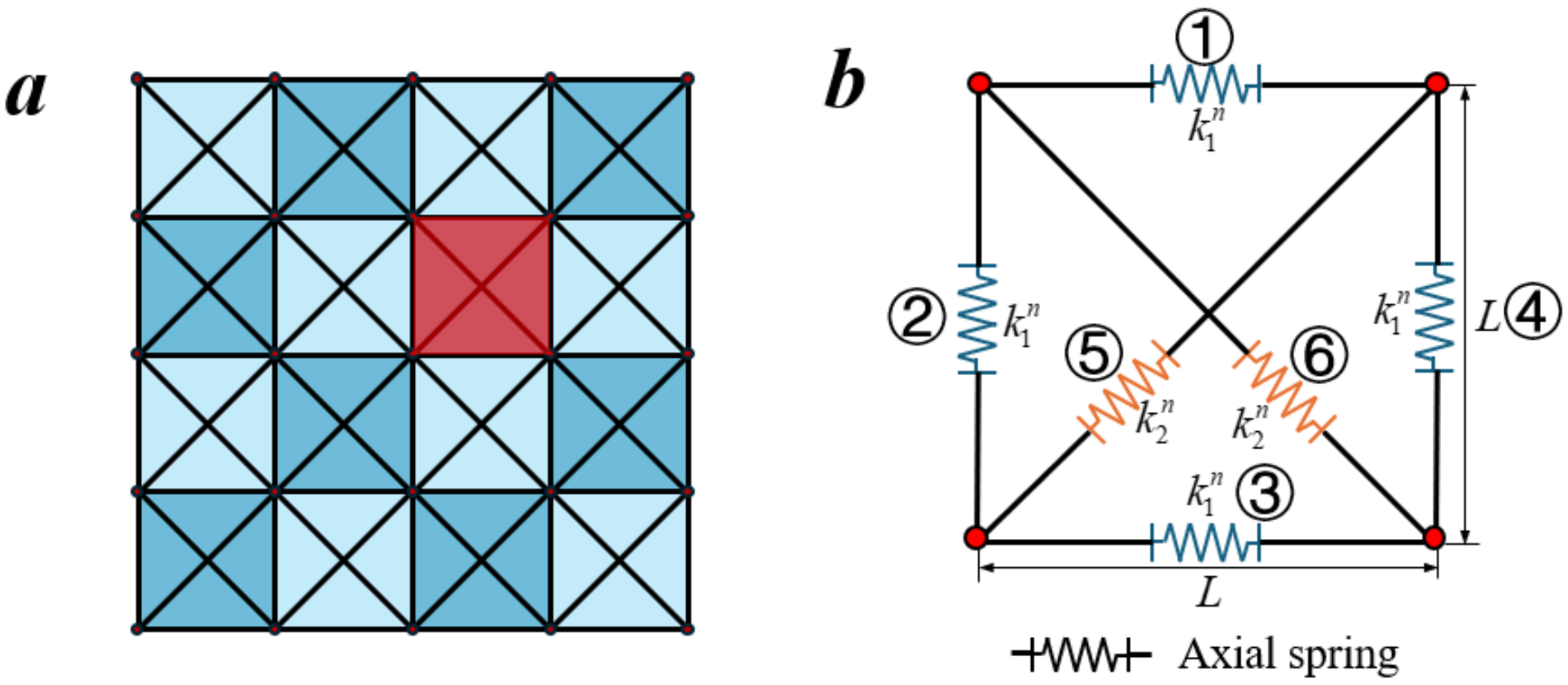

Fig. 1 (*a*) Schematic illustration of LSM; (*b*) Arrangement of axial springs within the RVE.

Assuming axial springs respond solely to axial loading, with material obeying linear elasticity, homogeneity, isotropy, and small-deformation conditions, the RVE encompasses perimeter members (①~④) and diagonal members (⑤, ⑥), as shown in Fig. 1(*b*). Each member is equivalent to an axial spring with stiffnesses $k_1^n$ and $k_2^n$, and lengths $L$ and $\sqrt{2}L$, respectively.

The relationship between axial relative displacement of two nodes and strain components is established following Griffiths (2001):

$$\varepsilon_\theta = \frac{\Delta u^n}{L} = \cos^2\theta\ \varepsilon_{xx} + \sin\theta\cos\theta(\varepsilon_{xy} + \varepsilon_{yx}) + \sin^2\theta\ \varepsilon_{yy} \tag{1}$$

where $\varepsilon_\theta$ denotes the axial strain of the member, $\varepsilon_{ij}$ represent global strain components, $\Delta u^n$ is the projection of the relative displacement $\Delta \boldsymbol{u}$ between two nodes along the axis of the member. The strain energy of a single member is thus $W(\theta) = k^n {\varepsilon_\theta}^2 L^2 / 2$. Superimposing strain energies from members at different orientations yields the total strain energy within the RVE. For an RVE oriented at angle $\theta$ relative to the *x*-axis, the total strain energy density is expressed as:

$$W_{\text{LSM}} = \frac{L^2}{4}\left( \underbrace{\begin{array}{l} (k_1^n - k_2^n)(\varepsilon_{xx} + \varepsilon_{xy} + \varepsilon_{yx} - \varepsilon_{yy})(\varepsilon_{xx} - \varepsilon_{xy} - \varepsilon_{yx} - \varepsilon_{yy})\cos(4\theta) \\ +2(k_1^n - k_2^n)(\varepsilon_{xx} - \varepsilon_{yy})(\varepsilon_{xy} + \varepsilon_{yx})\sin(4\theta) \end{array}}_{\textit{angle dependent items}} + \underbrace{(k_1^n + k_2^n)(3\varepsilon_{xx}^2 + 3\varepsilon_{yy}^2 + \varepsilon_{xy}^2 + \varepsilon_{yx}^2 + 2\varepsilon_{xx}\varepsilon_{yy} + 2\varepsilon_{xy}\varepsilon_{yx})}_{\textit{angle independent items}} \right) \tag{2}$$

This formulation reveals that the RVE's strain energy comprises both angle-independent and angle-dependent terms. To satisfy isotropic behavior, we enforce $k_1^n = k_2^n$ to eliminate angle-dependent contributions. Deriving the stress-strain relationship from total strain energy and comparing with two-dimensional Hooke's law yields the relation between spring stiffness and elastic parameters:

$$\textit{Plane strain}\begin{cases} E = \dfrac{5k_1^n}{2} \\ v = \dfrac{1}{4} \end{cases}, \ \textit{Plane stress}\begin{cases} E = \dfrac{8k_1^n}{3} \\ v = \dfrac{1}{3} \end{cases} (k_1^n = k_2^n) \tag{3}$$

These equations demonstrate that isotropic Hooke's law imposes the constraint $k_1^n = k_2^n$, restricting Poisson's ratio to $v = 1/3$ under plane stress conditions and $v = 1/4$ under plane strain conditions. While this square RVE formulation inherently fails to represent arbitrary isotropic behavior, it naturally captures anisotropic constitutive relations. Braun and Ariza (2019) demonstrated that rotating lattice orientations accommodates arbitrary material directions for generally anisotropic materials

## 3.0 Isotropic Elastic Lattice Spring Model (IELSM)

Several approaches have been developed to address the Poisson's ratio limitation inherent in LSM. The earliest systematic attempt was made by Hrennikoff (1941), who introduced four additional internal nodes within each square unit, connected by auxiliary horizontal and vertical rods to modify the system's elastic response. Later, McCormick (1963) proposed a plane stress model using one-dimensional elements with both axial and bending stiffnesses. However, to account for bending deformations, each node in McCormick's model carries an additional rotational degree of freedom. Subsequent extensions include beam elements developed by Griffiths and Mustoe (2001), micropolar rods introduced by Gerstle et al. (2007), and angular springs proposed by Wang et al. (2020). However, these methods inevitably increase the number of degrees of freedom, substantially elevating computational costs. An alternative approach was proposed by Zhao et al., (2011) in their DLSM, which employs local strain reconstruction to circumvent rotational invariance issues. Nevertheless, its reliance on least-squares approximation of displacement-strain transformation matrix introduces additional computational overhead while compromising simulation accuracy.

In this work, we propose the Isotropic Elastic Lattice Spring Model (IELSM) by enhancing the classical LSM with an additional volumetric constraint to more accurately analog isotropic elastic continua. Fig. 2 presents the physical decomposition underlying IELSM: isotropic elastic continua are equivalently represented through two complementary components: (1) classical LSM with axial springs, and (2) additional volumetric constraints associated with square RVEs. Classical LSM suffers from a fixed Poisson's ratio, which inevitably constrains its volumetric deformation behavior. This can be intuitively seen in the strain energy expression, taking the plane strain condition as an example:

$$\begin{aligned} W_{\mathrm{LSM}} &= \frac{L^2}{4}(k_1^n + k_2^n)(3\varepsilon_{xx}^2 + 3\varepsilon_{yy}^2 + 2\varepsilon_{xx}\varepsilon_{yy} + 4\varepsilon_{xy}^2) \\ &= \frac{L^2}{2}K^0(\varepsilon_{xx}+\varepsilon_{yy})^2 + G\left(\varepsilon_{xx}^2 + \varepsilon_{yy}^2 - \frac{1}{3}(\varepsilon_{xx}+\varepsilon_{yy})^2 + 2\varepsilon_{xy}^2\right)L^2 \end{aligned} \tag{4}$$

where $K^0 = \dfrac{E}{3(1-2v)} = \dfrac{5}{3}k_1^n$ is the initial bulk modulus, $G = \dfrac{E}{2(1+v)} = k_1^n$ is the shear modulus, and under plane stress conditions $K^0$ can be directly replaced by $K^0 + \dfrac{4G}{3}$. In order for the strain energy of the LSM in Eq. (4) to be equivalent to that of a continuum, its parameters $K^0$ and $G$ are actually coupled, which restricts the Poisson's ratio to a fixed value. To address this, the proposed IELSM introduces an additional bulk modulus $k^v$ to directly modify the initial bulk modulus $K^0$, thereby adjusting the composition of the strain energy and removing the limitation on Poisson's ratio, i.e.,

$$\begin{aligned} W_{\mathrm{IELSM}} &= \frac{L^2}{2}(K^0 + k^v)(\varepsilon_{xx}+\varepsilon_{yy})^2 + G\left(\varepsilon_{xx}^2 + \varepsilon_{yy}^2 - \frac{1}{3}(\varepsilon_{xx}+\varepsilon_{yy})^2 + 2\varepsilon_{xy}^2\right)L^2 \\ &= W_{\mathrm{LSM}} + \frac{L^2}{2}k^v(\varepsilon_{xx}+\varepsilon_{yy})^2 \end{aligned} \tag{5}$$

The bulk modulus of the IELSM is given by $K = K^0 + k^v$. Taking plane strain conditions as an example, the mechanism of the additional bulk modulus in IELSM is illustrated in Fig. 3. For an RVE of IELSM subjected to a uniform volumetric pressure $\Delta P$, when $k^v = 0$, the strain energy of the IELSM equals that of the LSM, and the Poisson's ratio is 1/4. When $k^v > 0$, $K > K^0$, which enhances the resistance to volumetric deformation compared to the LSM, leading to a Poisson's ratio greater than 1/4 and approaching 1/2. When $k^v < 0$, $K < K^0$, which weakens the resistance to volumetric deformation, resulting in a Poisson's ratio less than 1/4 and capable of taking negative values. The same principle applies to the additional bulk modulus under plane stress conditions.

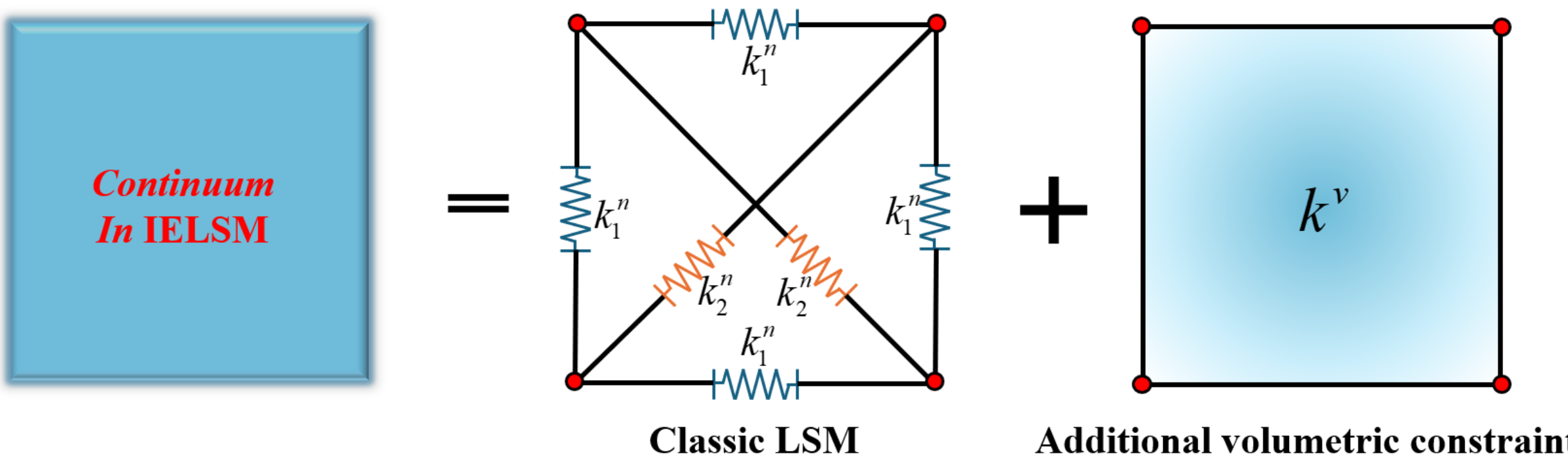

Fig. 2 Schematic representation of IELSM RVE.

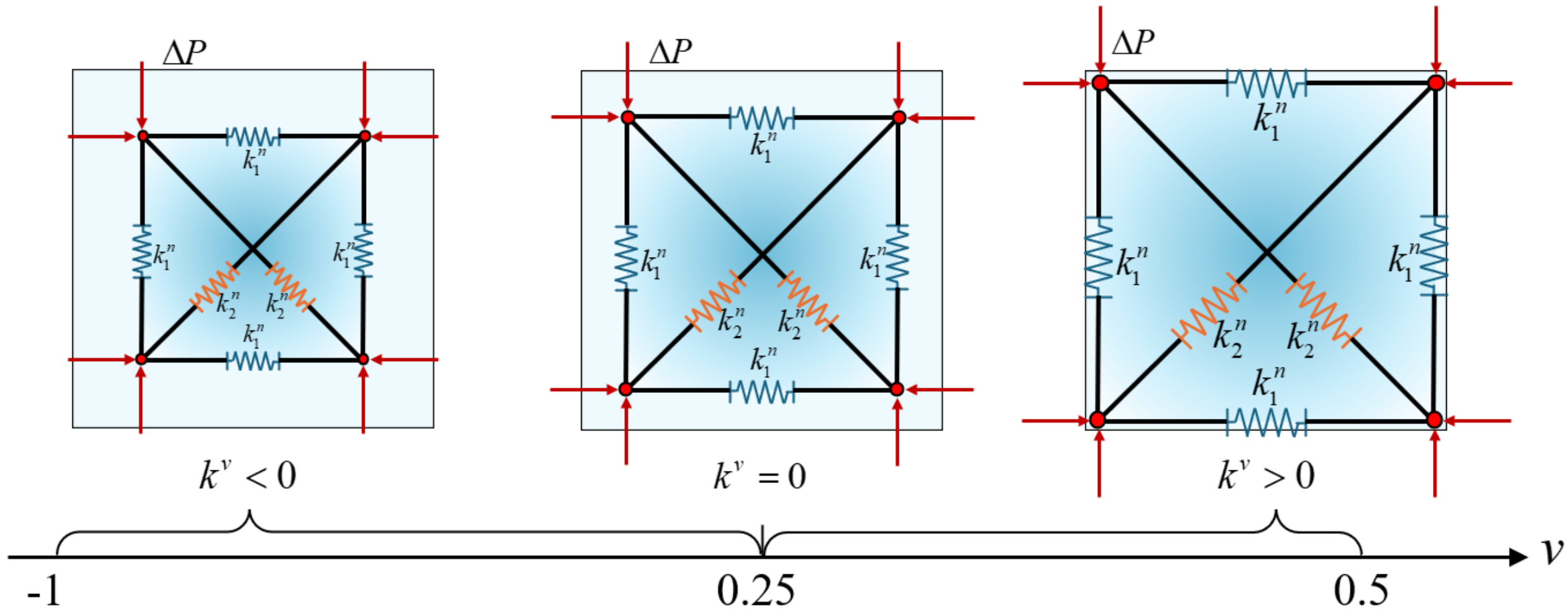

Fig. 3 Schematic representation of additional bulk modulus

According to Eq. (5), the strain energy with additional volumetric constraints can be defined as:

$$W_{\text{volume}} = \frac{1}{2}k^v{\varepsilon_v}^2L^2 = \begin{cases} \frac{1}{2}k^v_{plain\ strain}(\varepsilon_{xx}+\varepsilon_{yy})^2L^2 & (plain\ strain) \\ \frac{1}{2}k^v_{plain\ stress}\left(\frac{1-2v}{1-v}\right)^2(\varepsilon_{xx}+\varepsilon_{yy})^2L^2 & (plain\ stress) \end{cases} \tag{6}$$

where $\varepsilon_v$ denotes volumetric strain. Under small-deformation assumptions: $\varepsilon_v = \varepsilon_{xx}+\varepsilon_{yy}+\varepsilon_{zz}$, $\varepsilon_{zz}=0$ for plane strain conditions, thus $\varepsilon_v = \varepsilon_{xx}+\varepsilon_{yy}$; and $\varepsilon_{zz} = -\frac{v}{1-v}(\varepsilon_{xx}+\varepsilon_{yy})$ for plane stress conditions, thus $\varepsilon_v = \frac{1-2v}{1-v}(\varepsilon_{xx}+\varepsilon_{yy})$. To unify these formulations, we define $k^v = k^v_{plain\ strain} = k^v_{plain\ stress}\left(\frac{1-2v}{1-v}\right)^2$ as the normalized additional bulk modulus.

Superimposing $W_{\text{volume}}$ onto classical LSM strain energy $W_{\text{LSM}}$ yields the total strain energy:

$$
\begin{aligned}
W_{\text{IELSM}} &= W_{\text{LSM}} + W_{\text{volume}} \\
&= \frac{L^2}{4}\left((3k_1^n + 3k_2^n + 2k^v)(\varepsilon_{xx}^2 + \varepsilon_{yy}^2) + 2(k_1^n + k_2^n + 2k^v)\varepsilon_{xx}\varepsilon_{yy} + (k_1^n + k_2^n)\gamma_{xy}^2\right)
\end{aligned} \tag{7}
$$

Differentiating with respect to strain components provides the stress-strain relationship:

$$
\begin{cases}
\sigma_{xx} = \dfrac{(3k_1^n + 3k_2^n + 2k^v)}{2}\varepsilon_{xx} + \dfrac{(k_1^n + k_2^n + 2k^v)}{2}\varepsilon_{yy} \\
\sigma_{yy} = \dfrac{(k_1^n + k_2^n + 2k^v)}{2}\varepsilon_{xx} + \dfrac{(3k_1^n + 3k_2^n + 2k^v)}{2}\varepsilon_{yy} \\
\tau_{xy} = \dfrac{(k_1^n + k_2^n)}{2}\gamma_{xy}
\end{cases} \tag{8}
$$

Comparison with isotropic Hooke's law yields the following equivalences:

$$
Plane\ strain \begin{cases}
\dfrac{E(1-v)}{(1+v)(1-2v)} \equiv \dfrac{(3k_1^n + 3k_2^n + 2k^v)}{2} \\
\dfrac{Ev}{(1+v)(1-2v)} \equiv \dfrac{(k_1^n + k_2^n + 2k^v)}{2} \\
\dfrac{E}{2(1+v)} \equiv \dfrac{(k_1^n + k_2^n)}{2}
\end{cases}
,\ Plane\ stress \begin{cases}
\dfrac{E}{(1-v^2)} \equiv \dfrac{(3k_1^n + 3k_2^n + 2k^v)}{2} \\
\dfrac{Ev}{(1-v^2)} \equiv \dfrac{(k_1^n + k_2^n + 2k^v)}{2} \\
\dfrac{E}{2(1+v)} \equiv \dfrac{(k_1^n + k_2^n)}{2}
\end{cases} \tag{9}
$$

Solving these equations provides the explicit parameter mapping:

$$
Plane\ strain \begin{cases}
k_1^n = \dfrac{E}{2(1+v)} \\
k_2^n = \dfrac{E}{2(1+v)} \\
k^v = \dfrac{E(4v-1)}{2(1-2v)(1+v)}
\end{cases}
,\ Plane\ stress \begin{cases}
k_1^n = \dfrac{E}{2(1+v)} \\
k_2^n = \dfrac{E}{2(1+v)} \\
k^v = \dfrac{E(3v-1)}{2(1-v^2)}
\end{cases} \tag{10}
$$

or

$$
\begin{aligned}
&Plane\ strain \begin{cases}
v = \dfrac{k_1^n + k^v}{4k_1^n + 2k^v} \\
E = k_1^n\left(\dfrac{5k_1^n + 3k^v}{2k_1^n + k^v}\right)
\end{cases} \quad (k_1^n = k_2^n) \\
&Plane\ stress \begin{cases}
v = \dfrac{2k_1^n + 2k^v}{6k_1^n + 2k^v} \\
E = 4k_1^n\left(\dfrac{2k_1^n + k^v}{k_1^n + k^v}\right)
\end{cases} \quad (k_1^n = k_2^n)
\end{aligned} \tag{11}
$$

For non-negative axial stiffness (both $k_1^n$ and $k_2^n$ $\geq 0$), the lower bound of Poisson's ratio is $v > -1$. As $k^v \gg k_1^n$ (dominant additional volumetric constraint), Poisson's ratio asymptotically approaches $v \to 1$ under plane stress and $v \to 1/2$ under plane strain. When $k^v = 0$ (negligible additional volumetric constraint), IELSM reverts to classical LSM with $v = 1/3$ (plane stress) and $v = 1/4$ (plane strain). When $k^v < 0$ (promote compression/expansion), the Poisson's ratio can range from −1 to 1/3 under plane stress and range from −1 to 1/4 under plane strain. Consequently, IELSM spans the physically admissible ranges: $-1 < v < 1$ for plane stress and $-1 < v < 1/2$ for plane strain, while maintaining exactly two parameters matching isotropic elasticity theory.

## 4.0 Implementation within Standard Finite Element Analysis

This section delineates the numerical discretization implementation of IELSM. We first derive element stiffness matrices for axial spring rods and additional volumetric constraints, then detail the assembly procedures for global stiffness formulation.

### 4.1 Derivation of the Element Stiffness Matrix

Following the classical direct stiffness method established in pioneering finite element research (Turner et al., 1956) for deriving element stiffness matrices from fundamental beam and rod elements, we first consider conventional LSM rod elements with pure axial springs. The local coordinate element stiffness matrix $\mathbf{K}_n^e$ is given by:

$$\mathbf{K}_n^e = \begin{bmatrix} k^n & 0 & -k^n & 0 \\ 0 & 0 & 0 & 0 \\ -k^n & 0 & k^n & 0 \\ 0 & 0 & 0 & 0 \end{bmatrix} \tag{12}$$

For rods incorporating series-connected axial and shear springs (as depicted in Fig. 2), the local stiffness matrix aggregates contributions from both spring types:

$$\mathbf{K}_{ns}^e = \begin{bmatrix} k^n & 0 & -k^n & 0 \\ 0 & k^s & 0 & -k^s \\ -k^n & 0 & k^n & 0 \\ 0 & -k^s & 0 & k^s \end{bmatrix} \tag{13}$$

where $k^s$ is the stiffness of the shear spring. Transformation to global coordinates via rotation matrix $\mathbf{T}$ yields:

$$\mathbf{K}_n(k^n, \theta) = \mathbf{T}^{\mathrm{T}} \mathbf{K}_n^e \mathbf{T}, \quad \mathbf{K}_{ns}(k^n, k^s, \theta) = \mathbf{T}^{\mathrm{T}} \mathbf{K}_{ns}^e \mathbf{T} \tag{14}$$

where the 2D coordinate transformation matrix is:

$$
\mathbf{T}=\begin{bmatrix} \cos\theta & \sin\theta & 0 & 0 \\ -\sin\theta & \cos\theta & 0 & 0 \\ 0 & 0 & \cos\theta & \sin\theta \\ 0 & 0 & -\sin\theta & \cos\theta \end{bmatrix} \tag{15}
$$

Here, we exploit the equivalence between the volumetric strain and the displacement field to construct the strain energy associated with the additional volumetric constraint, yielding a quadratic form in terms of the displacements and thereby enabling a direct derivation of the corresponding element stiffness matrix. As shown in Fig. 4, consider a square RVE whose nodal coordinates are $P_1(L,\ L)$, $P_2(0,\ L)$, $P_3(0,\ 0)$ and $P_4(L,\ 0)$. Let the nodal displacements after deformation be $(u_i,\ v_i)$. The deformed area of the RVE is then given by:

$$
\begin{aligned}
S' &= \frac{1}{2}\overrightarrow{P_1'P_3'} \otimes \overrightarrow{P_2'P_4'} \\
&= \frac{1}{2}(u_3-u_1-L, v_3-v_1-L) \otimes (u_4-u_2+L, v_4-v_2-L) \\
&= \frac{(u_3-u_1-L)(v_4-v_2-L)-(v_3-v_1-L)(u_4-u_2+L)}{2}
\end{aligned} \tag{16}
$$

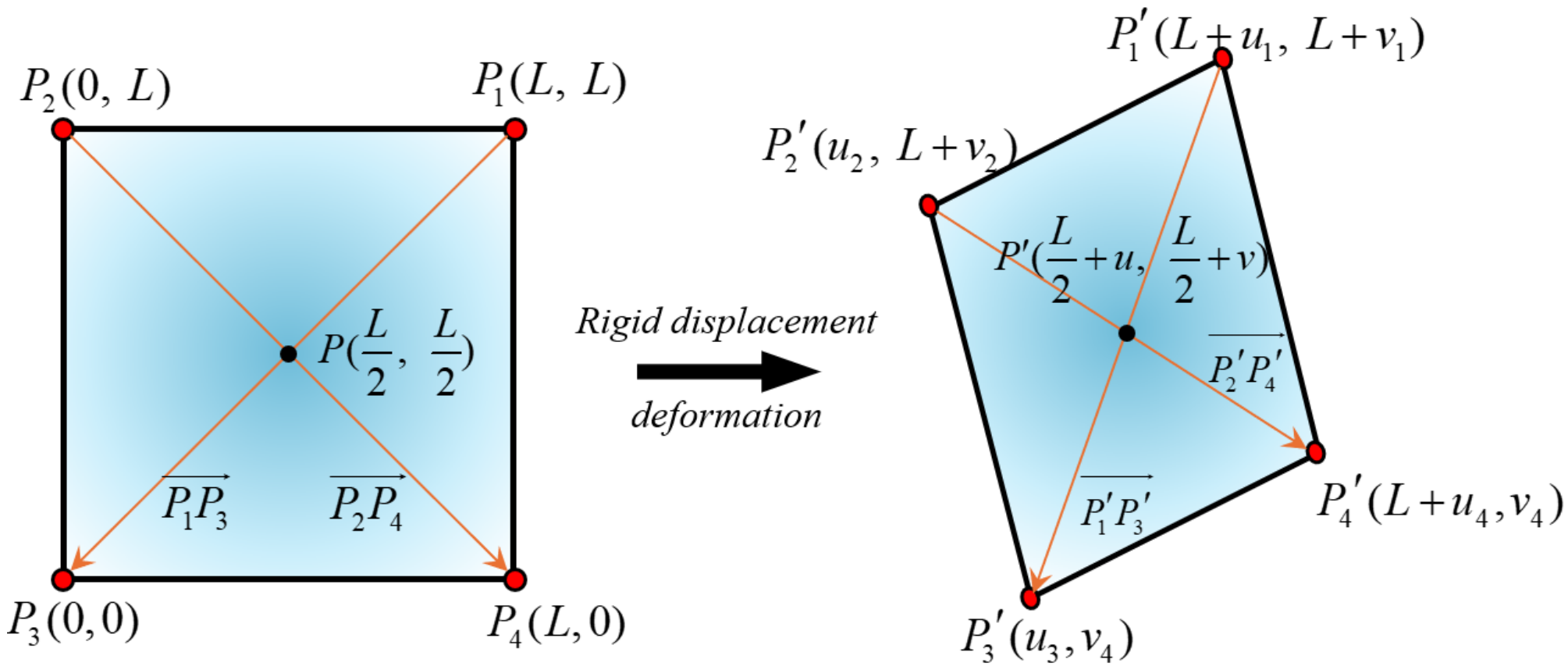

Fig. 4 Illustration of rigid body displacement and deformation in square RVE.

Volumetric strain is defined as:

$$
\begin{aligned}
\varepsilon_v &= \frac{S'-S}{S} \\
&= \underbrace{\frac{1}{2L}(u_1-u_2-u_3+u_4+v_1+v_2-v_3-v_4)}_{linear\ terms} \\
&\quad + \underbrace{\frac{1}{2L^2}(u_1v_2-u_1v_4-v_1u_2+v_3u_2-u_3v_2+u_3v_4+v_1u_4-v_3u_4)}_{nonlinear\ terms}
\end{aligned} \tag{17}
$$

Nonlinear displacement terms in volumetric strain correspond to second-order displacement gradients (detailed in Appendix B). Under small-deformation assumptions, these higher-order terms are neglected, yielding volumetric strain energy:

$$
\begin{aligned}
W_{volume} &\simeq \frac{k^v {\varepsilon_v}^2 L^2}{2} = \frac{k_v (u_1 - u_2 - u_3 + u_4 + v_1 + v_2 - v_3 - v_4)^2}{8} \\
&= \frac{k^v}{8} (\mathbf{U}^{\mathrm{T}}\mathbf{c})(\mathbf{c}^{\mathrm{T}}\mathbf{U}) = \frac{1}{2}\mathbf{U}^{\mathrm{T}}(\mathbf{c}^{\mathrm{T}} \frac{k^v}{4}\mathbf{c})\mathbf{U}
\end{aligned}
\tag{18}
$$

where $\mathbf{U}= \begin{bmatrix} u_1 & v_1 & u_2 & v_2 & u_3 & v_3 & u_4 & v_4 \end{bmatrix}^{\mathrm{T}}$ , $\mathbf{c}= \begin{bmatrix} 1 & 1 & -1 & 1 & -1 & -1 & 1 & -1 \end{bmatrix}^{\mathrm{T}}$ . This expression leads to the additional volumetric constraint stiffness matrix:

$$
\begin{aligned}
\mathbf{K}_{volume} &= \frac{\partial^2 W_{volume}}{\partial \mathbf{U}^2} \\
&= \mathbf{c}\frac{k^v}{4}\mathbf{c}^{\mathrm{T}} \\
&= \frac{k^v}{4}\begin{bmatrix}
1 & 1 & -1 & 1 & -1 & -1 & 1 & -1 \\
1 & 1 & -1 & 1 & -1 & -1 & 1 & -1 \\
-1 & -1 & 1 & -1 & 1 & 1 & -1 & 1 \\
1 & 1 & -1 & 1 & -1 & -1 & 1 & -1 \\
-1 & -1 & 1 & -1 & 1 & 1 & -1 & 1 \\
-1 & -1 & 1 & -1 & 1 & 1 & -1 & 1 \\
1 & 1 & -1 & 1 & -1 & -1 & 1 & -1 \\
-1 & -1 & 1 & -1 & 1 & 1 & -1 & 1
\end{bmatrix}
\end{aligned}
\tag{19}
$$

To facilitate assembly, we decompose the 8×8 additional volumetric constraint matrix $\mathbf{K}^e_{volume}$ into ten elemental contributions:

$$
\begin{aligned}
\mathbf{K}_{volume} &= \mathbf{K}_{v,1}+\mathbf{K}_{v,2}+\mathbf{K}_{v,3}+\mathbf{K}_{v,4}+\mathbf{K}_{v,5}+\mathbf{K}_{v,6}+\mathbf{K}_{v,7}+\mathbf{K}_{v,8}+\mathbf{K}_{v,9}+\mathbf{K}_{v,10} \\
&= \underbrace{\frac{k^v}{4}\begin{bmatrix} 1 & 0 & -1 & 0 \\ 0 & -1 & 0 & 1 \\ -1 & 0 & 1 & 0 \\ 0 & 1 & 0 & -1 \end{bmatrix}\begin{matrix} u_1 \\ v_1 \\ u_2 \\ v_2 \end{matrix}}_{\substack{u_1\ v_1\ u_2\ v_2 \\ \textcircled{1}}}
+\underbrace{\frac{k^v}{4}\begin{bmatrix} -1 & 0 & 1 & 0 \\ 0 & 1 & 0 & -1 \\ 1 & 0 & -1 & 0 \\ 0 & -1 & 0 & 1 \end{bmatrix}\begin{matrix} u_2 \\ v_2 \\ u_3 \\ v_3 \end{matrix}}_{\substack{u_2\ v_2\ u_3\ v_3 \\ \textcircled{2}}}
+\underbrace{\frac{k^v}{4}\begin{bmatrix} 1 & 0 & -1 & 0 \\ 0 & -1 & 0 & 1 \\ -1 & 0 & 1 & 0 \\ 0 & 1 & 0 & -1 \end{bmatrix}\begin{matrix} u_3 \\ v_3 \\ u_4 \\ v_4 \end{matrix}}_{\substack{u_3\ v_3\ u_4\ v_4 \\ \textcircled{3}}} \\
&+\underbrace{\frac{k^v}{4}\begin{bmatrix} -1 & 0 & 1 & 0 \\ 0 & 1 & 0 & -1 \\ 1 & 0 & -1 & 0 \\ 0 & -1 & 0 & 1 \end{bmatrix}\begin{matrix} u_1 \\ v_1 \\ u_4 \\ v_4 \end{matrix}}_{\substack{u_1\ v_1\ u_4\ v_4 \\ \textcircled{4}}}
+\underbrace{\frac{k^v}{4}\begin{bmatrix} 1 & 1 & -1 & -1 \\ 1 & 1 & -1 & -1 \\ -1 & -1 & 1 & 1 \\ -1 & -1 & 1 & 1 \end{bmatrix}\begin{matrix} u_1 \\ v_1 \\ u_3 \\ v_3 \end{matrix}}_{\substack{u_1\ v_1\ u_3\ v_3 \\ \textcircled{5}}}
+\underbrace{\frac{k^v}{4}\begin{bmatrix} 1 & -1 & -1 & 1 \\ -1 & 1 & 1 & -1 \\ -1 & 1 & 1 & -1 \\ 1 & -1 & -1 & 1 \end{bmatrix}\begin{matrix} u_2 \\ v_2 \\ u_4 \\ v_4 \end{matrix}}_{\substack{u_2\ v_2\ u_4\ v_4 \\ \textcircled{6}}} \\
&+\underbrace{\frac{k^v}{4}\begin{bmatrix} 0 & 0 & 0 & 1 \\ 0 & 0 & -1 & 0 \\ 0 & -1 & 0 & 0 \\ 1 & 0 & 0 & 0 \end{bmatrix}\begin{matrix} u_1 \\ v_1 \\ u_2 \\ v_2 \end{matrix}}_{\substack{u_1\ v_1\ u_2\ v_2 \\ \textcircled{7}}}
+\underbrace{\frac{k^v}{4}\begin{bmatrix} 0 & 0 & 0 & 1 \\ 0 & 0 & -1 & 0 \\ 0 & -1 & 0 & 0 \\ 1 & 0 & 0 & 0 \end{bmatrix}\begin{matrix} u_2 \\ v_2 \\ u_3 \\ v_3 \end{matrix}}_{\substack{u_2\ v_2\ u_3\ v_3 \\ \textcircled{8}}}
+\underbrace{\frac{k^v}{4}\begin{bmatrix} 0 & 0 & 0 & 1 \\ 0 & 0 & -1 & 0 \\ 0 & -1 & 0 & 0 \\ 1 & 0 & 0 & 0 \end{bmatrix}\begin{matrix} u_3 \\ v_3 \\ u_4 \\ v_4 \end{matrix}}_{\substack{u_3\ v_3\ u_4\ v_4 \\ \textcircled{9}}} \\
&+\underbrace{\frac{k^v}{4}\begin{bmatrix} 0 & 0 & 0 & -1 \\ 0 & 0 & 1 & 0 \\ 0 & 1 & 0 & 0 \\ -1 & 0 & 0 & 0 \end{bmatrix}\begin{matrix} u_1 \\ v_1 \\ u_4 \\ v_4 \end{matrix}}_{\substack{u_1\ v_1\ u_4\ v_4 \\ \textcircled{10}}}
\end{aligned}
\tag{20}
$$

The segmentation in eq.(20) corresponds to the ten units shown in Fig. 5, where Part 1 includes rods ❶~❹ with series axial and shear springs (axial stiffness $k^v/4$, shear stiffness $-k^v/4$) and rods ❺ and ❻ with only axial springs (axial stiffness $k^v/2$); Part 2 consists of rods ❼~❿ equipped with rotational springs that resist rigid body rotation (stiffness $k^v/4$). According to Eq. (14), the element stiffness matrices of components ❶~❻ satisfy the following relationship:

$$
\begin{gathered}
\mathbf{K}_{v,1}=\mathbf{K}_{ns}(\frac{k^v}{4},-\frac{k^v}{4},0),\ \mathbf{K}_{v,2}=\mathbf{K}_{ns}(\frac{k^v}{4},-\frac{k^v}{4},\frac{\pi}{2}),\ \mathbf{K}_{v,3}=\mathbf{K}_{ns}(\frac{k^v}{4},-\frac{k^v}{4},\pi) \\
\mathbf{K}_{v,4}=\mathbf{K}_{ns}(\frac{k^v}{4},-\frac{k^v}{4},\frac{3\pi}{2}),\ \ \mathbf{K}_{v,5}=\mathbf{K}_{n}(\frac{k^v}{2},\frac{\pi}{4}),\ \ \mathbf{K}_{v,6}=\mathbf{K}_{n}(\frac{k^v}{2},\frac{3\pi}{4}).
\end{gathered}
\tag{21}
$$

For components ❼~❿ (rotational springs), we can define their local element stiffness matrix as:

$$
\mathbf{K}_r^e(k^r) = \begin{bmatrix} 0 & 0 & 0 & -k^r \\ 0 & 0 & k^r & 0 \\ 0 & k^r & 0 & 0 \\ -k^r & 0 & 0 & 0 \end{bmatrix} \tag{22}
$$

where $k^r$ is the stiffness of the rotational spring. It should be noted that $\mathbf{K}_r^e$ is independent of the angle at which it is placed in the global coordinate system, that is, it satisfies:

$$
\mathbf{K}_r(k^r) = \mathbf{T}^{\mathrm{T}}\mathbf{K}_r^e\mathbf{T} = \mathbf{K}_r^e \tag{23}
$$

therefore, the element stiffness matrix of components ❼~❿ is:

$$
\begin{gathered}
\mathbf{K}_{v,7} = \mathbf{K}_{v,8} = \mathbf{K}_{v,9} = \mathbf{K}_r(\frac{k^v}{4}), \\
\mathbf{K}_{v,10} = -\mathbf{K}_r(\frac{k^v}{4})
\end{gathered} \tag{24}
$$

The sign reversal observed for rod ❿ compared to rods ❼~❾ originates from differences in local-to-global node sequencing, a detail that will be clarified in the subsequent assembly discussion (see Fig. 6).

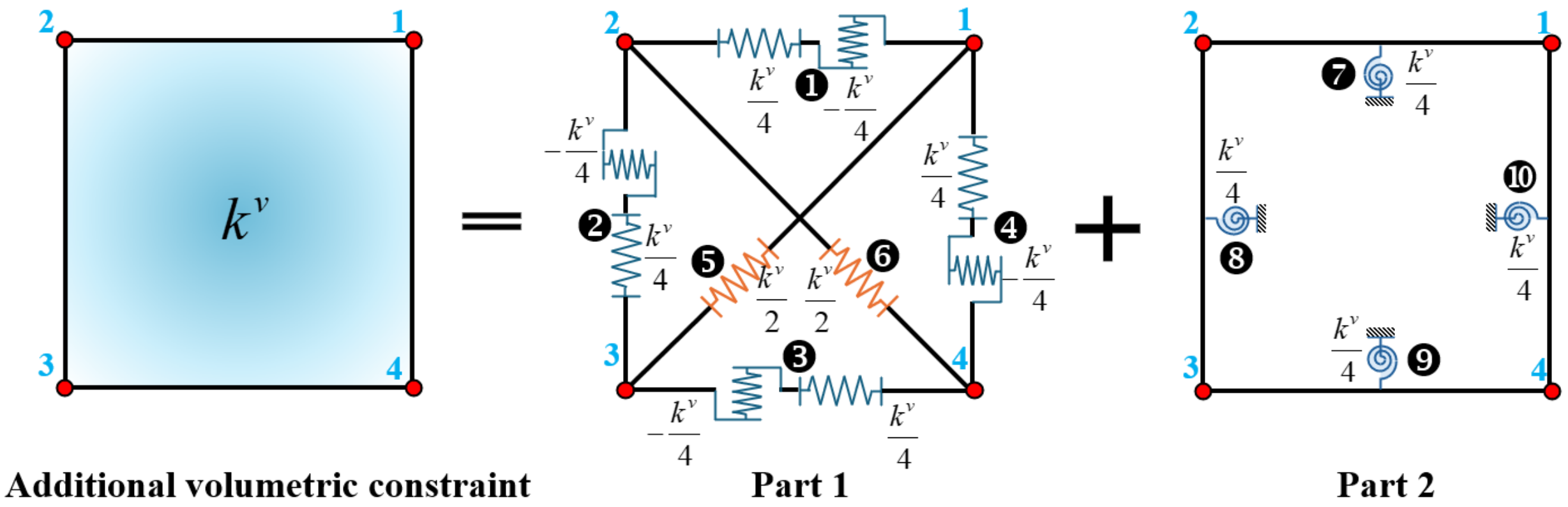

Fig. 5 Numerical discretization of additional volumetric constraint.

### 4.2 Element Stiffness Assembly

Employing strain energy-based construction of elemental stiffness matrices, we assemble the global system **KU**=**F** via the direct stiffness method. This approach initially positions springs from each RVE category, subsequently superimposing contributions at coincident locations to minimize assembly operations.

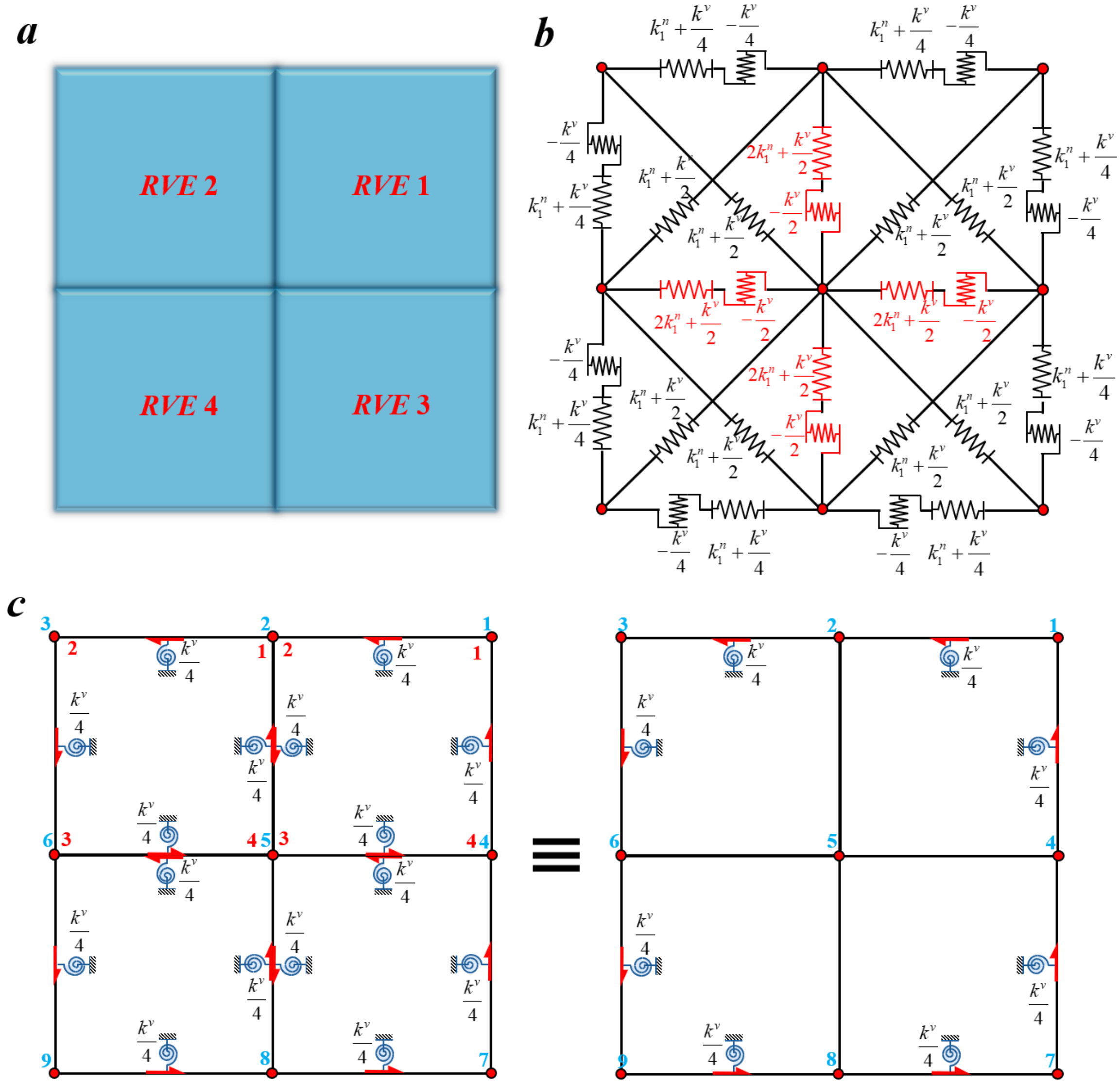

Fig. 6 Element stiffness matrix assembly schematic.

Fig. 6 illustrates assembly procedures for a 2×2 RVE configuration: (*a*) overall arrangement, (*b*) axial and shear spring assembly, (*c*) rotational spring assembly. Axial and shear springs in IELSM originate partially from classical LSM and partially from discretized additional volumetric constraints. Their assembly (Fig. 6(*b*)) exhibits non-directionality, permitting direct superposition of stiffness coefficients for springs occupying identical positions within individual RVEs. Along shared boundaries (red springs), contributions from adjacent RVEs double the stiffness relative to peripheral edges. The assembly pattern for axial and shear springs follows:

(1) Peripheral boundary rods: axial stiffness $k_1^n + \frac{k^v}{4}$, shear stiffness $-\frac{k^v}{4}$;

(2) Internal orthogonal rods: axial stiffness $2k_1^n + \frac{k^v}{2}$, shear stiffness $-\frac{k^v}{2}$;

(3) Internal diagonal rods: axial stiffness $k_1^n + \frac{k^v}{2}$.

As shown in Fig. 6($c$), the rotational spring assembly in Part 2 uses red numerals to denote local node indices and blue numerals to denote global node indices. Within each RVE, the local node ordering adopted in the stiffness matrix of the rotational spring follows a counterclockwise convention. For example, in RVE-1, the member connecting local node 2 to local node 3 is associated with a local element stiffness matrix $\mathbf{K}^{e}_{r,RVE-1}$ whose nodal sequence is 2→3. The corresponding nodal sequence in the assembled global stiffness matrix $\mathbf{K}_{r,RVE-1}$ is 2→5, which is consistent with the local ordering; hence $\mathbf{K}_{r,RVE-1} = \mathbf{K}^{e}_{r,RVE-1}$. In contrast, in RVE-2, the member connecting local node 1 and local node 4 has a nodal sequence 4→1 in its local element stiffness matrix $\mathbf{K}^{e}_{r,RVE-2}$, whereas the corresponding nodal sequence in the global stiffness matrix $\mathbf{K}_{r,RVE-2}$ is 2→5. Because the global ordering is opposite to the local convention in this case, $\mathbf{K}_{r,RVE-2} = -\mathbf{K}^{e}_{r,RVE-2}$.The explicit expressions are given below:

$$
\begin{aligned}
\mathbf{K}^{r}_{RVE-1} = k^{r}\begin{bmatrix} & & & 1 \\ & & -1 & \\ & -1 & & \\ 1 & & & \end{bmatrix}\begin{matrix} Local\ u_2 \\ Local\ v_2 \\ Local\ u_3 \\ Local\ v_3 \end{matrix} \quad &\Rightarrow \quad \overline{\mathbf{K}}^{r}_{RVE-1} = k^{r}\begin{bmatrix} & & & 1 \\ & & -1 & \\ & -1 & & \\ 1 & & & \end{bmatrix}\begin{matrix} Global\ u_2 \\ Global\ v_2 \\ Global\ u_5 \\ Global\ v_5 \end{matrix} \\
\mathbf{K}^{r}_{RVE-2} = k^{r}\begin{bmatrix} & & & 1 \\ & & -1 & \\ & -1 & & \\ 1 & & & \end{bmatrix}\begin{matrix} Local\ u_4 \\ Local\ v_4 \\ Local\ u_1 \\ Local\ v_1 \end{matrix} \quad &\Rightarrow \quad \overline{\mathbf{K}}^{r}_{RVE-2} = -k^{r}\begin{bmatrix} & & & 1 \\ & & -1 & \\ & -1 & & \\ 1 & & & \end{bmatrix}\begin{matrix} Global\ u_2 \\ Global\ v_2 \\ Global\ u_5 \\ Global\ v_5 \end{matrix}
\end{aligned}
\tag{25}
$$

Consequently, $\mathbf{K}_{r,RVE-1}$ and $\mathbf{K}_{r,RVE-2}$ cancel each other in the assembly. Likewise, as illustrated in Fig. 6($c$), the rotational springs placed on the overlapping boundary ultimately cancel out. This is equivalent to applying rotational springs only along the boundary of the model.

It is noteworthy that for the decomposition with additional volumetric constraints, axial springs and shear springs with the same sign as the additional bulk modulus are introduced, as well as shear springs with the opposite sign. After assembling the components decomposed from the additional volumetric constraints with the axial springs of the LSM, the total stiffness of the axial springs will remain positive. According to Eq. (10), we have:

$$
plain\ strain \begin{cases} k_1^n + \dfrac{k^v}{4} = \dfrac{E}{2(1+v)} + \dfrac{1}{4}\dfrac{E(4v-1)}{2(1-2v)(1+v)} = \dfrac{E(3-4v)}{8(1-2v)(1+v)} > 0 \\ k_1^n + \dfrac{k^v}{2} = \dfrac{E}{2(1+v)} + \dfrac{1}{2}\dfrac{E(4v-1)}{2(1-2v)(1+v)} = \dfrac{E}{4(1-2v)(1+v)} > 0 \end{cases}, \quad (-1 < v < \frac{1}{2})
$$

$$
plain\ stress \begin{cases} k_1^n + \dfrac{k^v}{4} = \dfrac{E}{2(1+v)} + \dfrac{1}{4}\dfrac{E(3v-1)}{2(1-v^2)} = \dfrac{E(3-v)}{8(1-v^2)} > 0 \\ k_1^n + \dfrac{k^v}{2} = \dfrac{E}{2(1+v)} + \dfrac{1}{2}\dfrac{E(3v-1)}{2(1-v^2)} = \dfrac{E}{4(1-v)} > 0 \end{cases}, \quad (-1 < v < 1) \tag{26}
$$

In the proposed IELSM, a positive additional bulk modulus leads to a negative shear spring stiffness. This can be understood from the fact that enhancing the volumetric stiffness (i.e., $k^v > 0$) under plane strain conditions correspondingly reduces the shear resistance in the discrete spring network, manifesting as a negative shear stiffness. This situation is consistent with the case in the DLSM reported by Zhao and Zhao (2012). They took silver with a face-centered cubic structure as an example, whose experimentally measured Poisson's ratio is 0.37, corresponding to a negative shear spring stiffness in the DLSM. Through curvature analysis of the potential function in molecular dynamics, Zhao and Zhao (2012) demonstrated that the negative shear spring physically corresponds to the "concave downward" behavior of the interatomic shear potential energy. It is a remarkable fact that the negative shear spring appearing in this work exists only in the numerical discretization process and does not involve any approximation to the theoretical model. Subsequent calculation results also show that this negative stiffness is stable and feasible in the overall numerical computation.

### 4.3 Displacement and Stress/Strain Fields for IELSM

Since theoretical derivations presented in preceding sections remain independent of specific displacement fields, IELSM accommodates various displacement field assumptions. Numerical computation of stress-strain fields requires assumed displacement fields to construct strain-displacement transformation matrices. To assess IELSM's performance under different approximations, we examine two displacement field formulations:

$$
linear\ displacement\ field \quad \begin{cases} u = a_0 + a_1 x + a_2 y \\ v = b_0 + b_1 x + b_2 y \end{cases} \tag{27}
$$

$$
bilinear\ displacement\ field \quad \begin{cases} u = a_0 + a_1 x + a_2 y + a_3 xy \\ v = b_0 + b_1 x + b_2 y + b_3 xy \end{cases} \tag{28}
$$

For linear displacement fields, with six parameters $a_0$, $a_1$, $a_2$, $b_0$, $b_1$, $b_2$ but eight nodal displacement components (two per node), the system is overdetermined. As shown in Fig. 7, we partition the square element into two triangular elements: *Tri*-1 (nodes $P_1, P_2, P_3$) and *Tri*-2 (nodes $P_3$, $P_4$, $P_1$). The strain-displacement transformation matrices for the two triangular elements are:

$$
\begin{aligned}
\mathbf{B}_{Tri-1} &= \frac{1}{L}\begin{bmatrix} 1 & & -1 & & & \\ & & & 1 & & -1 \\ & 1 & 1 & -1 & -1 & \end{bmatrix} \\
\mathbf{B}_{Tri-2} &= \frac{1}{L}\begin{bmatrix} -1 & & 1 & & & \\ & & & -1 & & 1 \\ & -1 & -1 & 1 & 1 & \end{bmatrix}
\end{aligned}
\tag{29}
$$

where $\mathbf{B}_{Tri-1}$ and $\mathbf{B}_{Tri-2}$ is the strain-displacement transformation matrices for the *Tri*-1 and *Tri*-2, respectively. Strains within the two triangular elements are:

$$
\begin{aligned}
\boldsymbol{\varepsilon}_{Tri-1} = \mathbf{B}_{Tri-1}\mathbf{U}_{Tri-1} &= \frac{1}{L}\begin{bmatrix} u_1 - u_2 \\ v_2 - v_3 \\ v_1 + u_2 - v_2 - u_3 \end{bmatrix} \\
\boldsymbol{\varepsilon}_{Tri-2} = \mathbf{B}_{Tri-2}\mathbf{U}_{Tri-2} &= \frac{1}{L}\begin{bmatrix} u_4 - u_3 \\ v_1 - v_4 \\ u_1 - v_3 - u_4 + v_4 \end{bmatrix}
\end{aligned}
\tag{30}
$$

where $\mathbf{U}_{Tri-1} = \begin{bmatrix} u_1 & v_1 & u_2 & v_2 & u_3 & v_3 \end{bmatrix}^{\mathrm{T}}$, $\mathbf{U}_{Tri-2} = \begin{bmatrix} u_3 & v_3 & u_4 & v_4 & u_1 & v_1 \end{bmatrix}^{\mathrm{T}}$. Averaging strains from both triangles yields the mean strain for the entire square element:

$$
\begin{aligned}
\boldsymbol{\varepsilon}_{linear} &= \frac{\boldsymbol{\varepsilon}_{Tri-1} + \boldsymbol{\varepsilon}_{Tri-2}}{2} = \frac{1}{2L}\begin{bmatrix} u_1 - u_2 - u_3 + u_4 \\ v_1 + v_2 - v_3 - v_4 \\ u_1 + v_1 + u_2 - v_2 - u_3 - v_3 - u_4 + v_4 \end{bmatrix} \\
&= \frac{1}{2L}\begin{bmatrix} 1 & 0 & -1 & 0 & -1 & 0 & 1 & 0 \\ 0 & 1 & 0 & 1 & 0 & -1 & 0 & -1 \\ 1 & 1 & 1 & -1 & -1 & -1 & -1 & 1 \end{bmatrix}\begin{bmatrix} u_1 \\ v_1 \\ u_2 \\ v_2 \\ u_3 \\ v_3 \\ u_4 \\ v_4 \end{bmatrix} \\
&= \mathbf{B}_{linear}\mathbf{U}
\end{aligned}
\tag{31}
$$

where $\boldsymbol{\varepsilon}_{linear} = \begin{bmatrix} \varepsilon_{xx} & \varepsilon_{yy} & \gamma_{xy} \end{bmatrix}^{\mathrm{T}}$ is the strain field obtained based on the linear displacement field, $\mathbf{B}_{linear}$ is the strain-displacement transformation matrix for linear displacement fields.

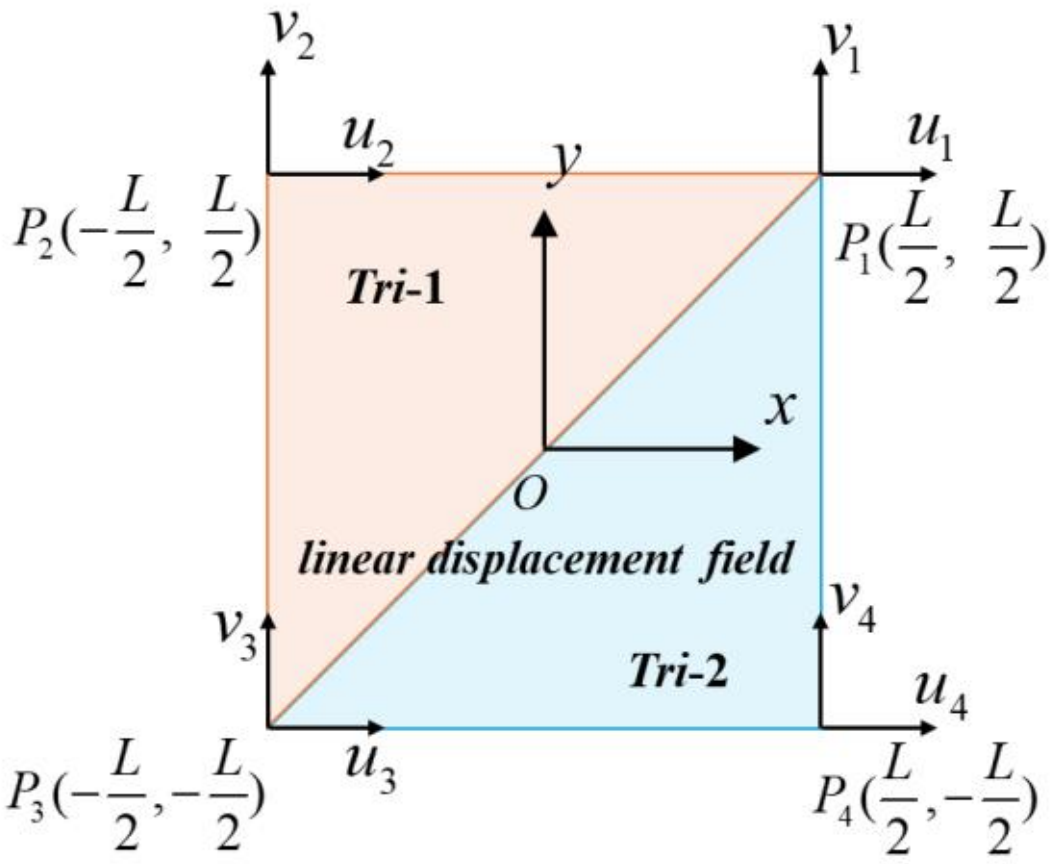

Fig. 7 Discretization of square representative unit into two triangles under linear displacement field assumption.

For bilinear displacement fields:

$$\begin{cases} u = a_0 + a_1 x + a_2 y + a_3 xy \\ v = b_0 + b_1 x + b_2 y + b_3 xy \end{cases} \tag{32}$$

where the number of parameters equals the number of nodal displacement components, enabling direct solution. The node coordinates and numbers in Figure C1 are still adopted, the strain-displacement transformation matrix is derived as:

$$\mathbf{B}_{bilinear} = \frac{1}{2L}\begin{bmatrix} 1+\frac{2y}{L} & 0 & -(1+\frac{2y}{L}) & 0 & -(1-\frac{2y}{L}) & 0 & 1-\frac{2y}{L} & 0 \\ 0 & 1+\frac{2x}{L} & 0 & 1-\frac{2x}{L} & 0 & -(1-\frac{2x}{L}) & 0 & -(1+\frac{2x}{L}) \\ 1+\frac{2x}{L} & 1+\frac{2y}{L} & 1-\frac{2x}{L} & -(1+\frac{2y}{L}) & -(1-\frac{2x}{L}) & -(1-\frac{2y}{L}) & -(1+\frac{2x}{L}) & 1-\frac{2y}{L} \end{bmatrix} \tag{33}$$

where $\mathbf{B}_{bilinear}$ is the strain-displacement transformation matrix for bilinear displacement fields. Element strains are obtained via:

$$\begin{aligned} \boldsymbol{\varepsilon}_{bilinear} &= \mathbf{B}_{bilinear}\mathbf{U} \\ &= \frac{1}{2L}\begin{bmatrix} u_1 - u_2 - u_3 + u_4 + \frac{2y}{L}(u_1 - u_2 + u_3 - u_4) \\ v_1 + v_2 - v_3 - v_4 + \frac{2x}{L}(v_1 - v_2 + v_3 - v_4) \\ u_1 + v_1 + u_2 - v_2 - u_3 - v_3 - u_4 + v_4 + \frac{2x}{L}(u_1 - u_2 + u_3 - u_4) + \frac{2x}{L}(v_1 - v_2 + v_3 - v_4) \end{bmatrix} \end{aligned} \tag{34}$$

where $\boldsymbol{\varepsilon}_{bilinear} = \begin{bmatrix} \varepsilon_{xx} & \varepsilon_{yy} & \gamma_{xy} \end{bmatrix}^{\mathrm{T}}$ is the strain field obtained based on the bilinear displacement field. Evaluating Eq. (33) and (34) at element center coordinates (0, 0) yields:

$$\mathbf{B}_{bilinear}\Big|_{\substack{x=0\\y=0}} = \mathbf{B}_{linear}$$
$$\boldsymbol{\varepsilon}_{bilinear}\Big|_{\substack{x=0\\y=0}} = \boldsymbol{\varepsilon}_{linear} \tag{35}$$

indicating that constant strain components under linear displacement field assumption equal strain components at the element center under bilinear displacement field assumption.

Based on the physical equations of plane problems in elasticity, the stress field of a representative element can be obtained as:

$$\boldsymbol{\sigma} = \begin{bmatrix} \sigma_{xx} \\ \sigma_{yy} \\ \tau_{xy} \end{bmatrix} = \mathbf{D}\,\boldsymbol{\varepsilon} \tag{36}$$

where **D** is the elastic coefficient matrix. For plane stress problems, **D** is given by:

$$\mathbf{D} = \frac{E}{1-v^2}\begin{bmatrix} 1 & v & 0 \\ v & 1 & 0 \\ 0 & 0 & \dfrac{1-v}{2} \end{bmatrix} \tag{37}$$

for plane strain problems, it is sufficient to replace $E$ in the above expression with $E/(1\text{-}v^2)$ and $v$ with $v/(1-v)$.

It is noteworthy that the square IELSM RVE has exactly four nodes, which inherently restricts the highest-order displacement approximation to bilinear terms. To achieve higher-order displacement approximations, the basic RVE would require additional nodes and establish interconnections through axial rods, followed by the application of additional volume constraints that affect all nodes within the RVE. Although the IELSM proposed in this paper does not involve the calculation of shape functions during its derivation via the direct stiffness method, nor does it assume a displacement distribution within the void regions between the springs, the direct stiffness method itself inherently implies a linear displacement field along the springs. To interpolate the displacement field within the void regions of the RVE, an additional displacement field assumption must be introduced. However, if the assumed displacement field is of an order higher than linear, the interpolation based on this higher-order field may become incompatible with the actual nodal displacement solution obtained from the IELSM along the spring positions, thereby yielding interpolation results that violate physical reality. In other words, when reconstructing the full-field displacement within the RVE voids, only a linear displacement field assumption consistent with the axial displacement mode of the springs is permissible. Furthermore, in the post-processing stage of this work, both linear and bilinear displacement field assumptions are employed to compute the stress and strain distributions. This operation is a typical stress recovery technique in finite element theory, aimed at improving the continuity and accuracy of the local stress and strain fields.

### 4.4 Comparative Analysis of Stiffness Matrices and Numerical Accuracy

In the process of establishing the aforementioned stress and strain fields, linear and bilinear displacement fields were introduced. Under this discretization strategy, it is necessary to further discuss the relationship between the stiffness matrix $\mathbf{K}_{\mathrm{IELSM}}$ of the IELSM RVE and the stiffness matrices of the following two finite element types: (1) the stiffness matrix $\mathbf{K}_{\mathrm{CST}}$ assembled from two Constant Strain Triangle (CST) elements based on a linear displacement field; (2) the stiffness matrix $\mathbf{K}_{\mathrm{Q4}}$ of a single four-node rectangular element (Q4) based on a bilinear displacement field. Taking plane stress conditions as an example and based on the node numbering and coordinates shown in Fig. 8, $\mathbf{K}_{\mathrm{CST}}$, $\mathbf{K}_{\mathrm{Q4}}$, and $\mathbf{K}_{\mathrm{IELSM}}$ are derived below.

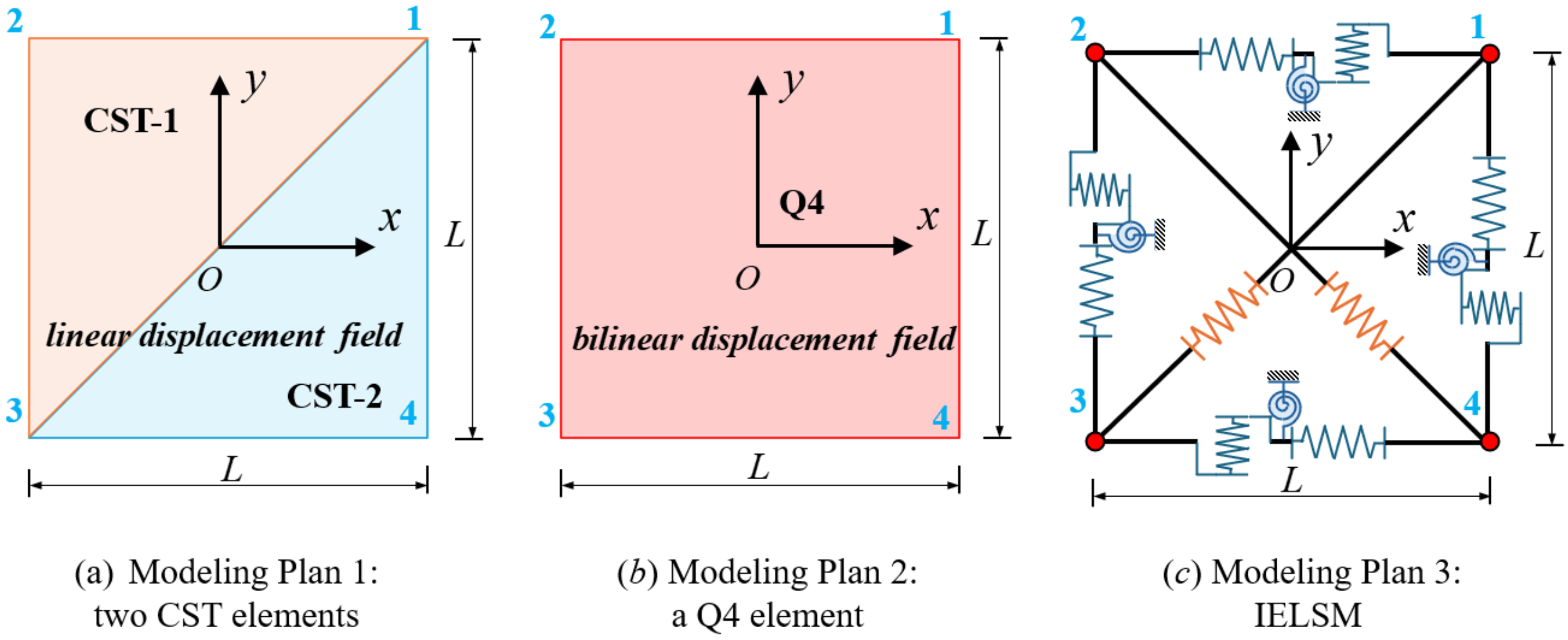

(a) Modeling Plan 1: two CST elements

(b) Modeling Plan 2: a Q4 element

(c) Modeling Plan 3: IELSM

Fig. 8 Element division and node numbering of planar rectangular structures.

According to Eq. (29) and Eq. (37), we have:

$$\mathbf{K}_{\mathrm{CST}} = \int_{\Omega_1} \mathbf{B}_{Tri-1}^{\mathrm{T}} \mathbf{D}\, \mathbf{B}_{Tri-1} \mathrm{d}\Omega_1 + \int_{\Omega_2} \mathbf{B}_{Tri-2}^{\mathrm{T}} \mathbf{D}\, \mathbf{B}_{Tri-2} \mathrm{d}\Omega_2$$

$$= \frac{E}{4(1-v^2)} \begin{bmatrix} 3-v & & & & & & & \\ 0 & 3-v & & & & & & \\ -2 & 1-v & 3-v & & & sym & & \\ 2v & v-1 & -v-1 & 3-v & & & & \\ 0 & -1-v & v-1 & 1-v & 3-v & & & \\ -1-v & 0 & 2v & -2 & 0 & 3-v & & \\ v-1 & 2v & 0 & 0 & -2 & 1-v & 3-v & \\ 1-v & -2 & 0 & 0 & 2v & v-1 & -1-v & 3-v \end{bmatrix} \tag{38}$$

where $\Omega_1$ is the domain of CST-1, $\Omega_2$ is the domain of CST-2, and $\Omega = \Omega_1 \bigcup \Omega_2$, where $\Omega$ is the domain of the square RVE.

According to Eq.(33) and (37), we have:

$$\mathbf{K}_{\mathrm{Q4}} = \int_\Omega \mathbf{B}_{bilinear}^{\mathrm{T}} \mathbf{D}\, \mathbf{B}_{bilinear} \mathrm{d}\Omega$$
$$= \frac{E}{24(1-v^2)} \begin{bmatrix} 12-4v & & & & & & & \\ 3+3v & 12-4v & & & & & & \\ -6-2v & 3-9v & 12-4v & & & sym & & \\ -3+9v & 4v & -3v-3 & 12-4v & & & & \\ -6+2v & -3-3v & 4v & 3-9v & 12-4v & & & \\ -3-3v & -6+2v & -3+9v & -6-2v & 3+3v & 12-4v & & \\ 4v & -3+9v & -6+2v & 3+3v & -6-2v & 3-9v & 12-4v & \\ 3-9v & -6-2v & 3+3v & -6+2v & -3+9v & 4v & -3-3v & 12-4v \end{bmatrix} \tag{39}$$

Within the square RVE of IELSM, there are 6 axial springs (with stiffness matrix $\mathbf{K}_{\mathrm{LSM},i}$) and 10 components derived from the split additional volumetric constraints; therefore:

$$\mathbf{K}_{\mathrm{IELSM}} = \mathbf{K}_{\mathrm{LSM}} + \mathbf{K}_{volume} = \sum_{i=1}^{6} \mathbf{K}_{\mathrm{LSM},i} + \sum_{j=1}^{10} \mathbf{K}_{v,j}$$
$$= \frac{1}{4} \begin{bmatrix} 4k_1^n+2k_2^n+k^v & & & & & & & \\ 2k_2^n+k^v & 4k_1^n+2k_2^n+k^v & & & & & & \\ -4k_1^n-k^v & -k^v & 4k_1^n+2k_2^n+k^v & & & sym & & \\ k^v & k^v & -2k_2^n-k^v & 4k_1^n+2k_2^n+k^v & & & & \\ -2k_2^n-k^v & -2k_2^n-k^v & k^v & -k^v & 4k_1^n+2k_2^n+k^v & & & \\ -2k_2^n-k^v & -2k_2^n-k^v & k^v & -4k_1^n-k^v & -k^v & 4k_1^n+2k_2^n+k^v & & \\ k^v & k^v & -2k_2^n-k^v & 2k_2^n+k^v & -4k_1^n-k^v & -k^v & 4k_1^n+2k_2^n+k^v & \\ -k^v & -4k_1^n-k^v & 2k_2^n+k^v & -2k_2^n-k^v & k^v & k^v & -2k_2^n-k^v & 4k_1^n+2k_2^n+k^v \end{bmatrix} \tag{40}$$

where

$$\begin{aligned} &\mathbf{K}_{\mathrm{LSM},1} = \mathbf{K}_n(k_1^n, 0),\ \mathbf{K}_{\mathrm{LSM},2} = \mathbf{K}_n(k_1^n, \frac{\pi}{2}),\ \mathbf{K}_{\mathrm{LSM},3} = \mathbf{K}_n(k_1^n, \pi), \\ &\mathbf{K}_{\mathrm{LSM},4} = \mathbf{K}_n(k_1^n, \frac{3\pi}{2}),\ \mathbf{K}_{\mathrm{LSM},5} = \mathbf{K}_n(k_2^n, \frac{\pi}{4}),\ \mathbf{K}_{\mathrm{LSM},6} = \mathbf{K}_n(k_2^n, \frac{3\pi}{4}), \end{aligned} \tag{41}$$

Substituting Eq. (10) into Eq. (40) yields:

$$\mathbf{K}_{\text{IELSM}} = \frac{E}{8(1-v^2)} \begin{bmatrix} 5-3v & & & & & & & \\ 1+v & 5-3v & & & & & & \\ v-3 & 1-3v & 5-3v & & & sym & & \\ 3v-1 & 3v-1 & -1-v & 5-3v & & & & \\ -1-v & -1-v & 3v-1 & 1-3v & 5-3v & & & \\ -1-v & -1-v & 3v-1 & v-3 & 1+v & 5-3v & & \\ 3v-1 & 3v-1 & -1-v & 1+v & v-3 & 1-3v & 5-3v & \\ 1-3v & v-3 & 1+v & -1-v & 3v-1 & 3v-1 & -1-v & 5-3v \end{bmatrix} \tag{42}$$

Comparing $\mathbf{K}_{\text{CST}}$, $\mathbf{K}_{\text{Q4}}$, and $\mathbf{K}_{\text{IELSM}}$, it can be observed that in the total stiffness matrix, $\mathbf{K}_{\text{CST}}$ assembled from the two CST elements, zero blocks appear at the degrees of freedom corresponding to the original quadrilateral diagonal nodes, whereas $\mathbf{K}_{\text{Q4}}$, and $\mathbf{K}_{\text{IELSM}}$ are full matrices. To evaluate the numerical stability and deformation characteristics of the three types of element stiffness matrices, it is necessary to solve the corresponding standard eigenvalue problem. The specific solution equation is:

$$\mathbf{K}\phi_i = \lambda_i \phi_i \tag{43}$$

where $\mathbf{K}$ represents the stiffness matrix to be analyzed, $\lambda_i$ indicates the $i$-th eigenvalue, and $\phi_i$ represents the corresponding eigenvector. The specific calculation results are shown in Table 1. The results show that all three discretization schemes have pure volumetric deformation modes, with their eigenvalues all being $E/(1-v)$, which is in line with the theoretical expectations of continuum mechanics, indicating that each element has consistency in describing isotropic expansion/contraction. Except for the unstable modes, the majority of the eigenvalues and corresponding mode shapes of the CST element and IELSM are relatively close, reflecting a certain inheritance in their constant strain representation capabilities. Both the CST element and the Q4 element have unstable modes: the $\lambda_5$ mode of the CST element shows severe geometric distortion of a single CST sub-element, that is, the aspect ratio of the triangle increases extremely, presenting a flattened shape. This mode shows an abnormal increase in the eigenvalue under high Poisson's ratio conditions, which is a typical numerical manifestation of directional parasitic shear locking; the $\lambda_4$ and $\lambda_5$ modes of the Q4 element present typical hourglass modes, corresponding to zero-energy modes. However, the $\lambda_2$ to $\lambda_5$ modes of IELSM remain stable throughout the Poisson's ratio range ($-1 < v < 1$), without showing abnormal elevation or zeroing, indicating that this model effectively suppresses the two types of numerical defects of parasitic shear and hourglass.

Table 1 Comparison of stiffness matrix eigenvalues and modal shape (plain stress).

| | CST | | Q4 | | IELSM | |
|---|---|---|---|---|---|---|
| $\lambda_i$ | Value | Mode shape | Value | Mode shape | Value | Mode shape |
| $\lambda_1$ | $\frac{E}{1-v}$ | Volumetric deformation | $\frac{E}{1-v}$ | Volumetric deformation | $\frac{E}{1-v}$ | Volumetric deformation |
| $\lambda_2$ | $\frac{E}{1+v}$ | Shear Deformation | $\frac{E}{1+v}$ | Shear Deformation | $\frac{E}{1+v}$ | Mixed deformation mode |
| $\lambda_3$ | $\frac{E}{1+v}$ | Mixed deformation mode | $\frac{E}{1+v}$ | Shear Deformation | $\frac{E}{1+v}$ | Mixed deformation mode |
| $\lambda_4$ | $\frac{E}{1+v}$ | Mixed deformation mode | $\frac{E(3-v)}{6(1-v^2)}$ | Hourglass | $\frac{E}{1+v}$ | Mixed deformation mode |
| $\lambda_5$ | $\frac{2E}{1-v^2}$ | Parasitic shear | $\frac{E(3-v)}{6(1-v^2)}$ | Hourglass | $\frac{E}{1+v}$ | Mixed deformation mode |

To further compare the numerical accuracy of the different discretization schemes, the total potential energy under the boundary conditions shown in Fig. 9 (with unit thickness) is calculated for each scheme. Based on the boundary conditions in Fig. 9, the global nodal displacement and force vectors are:

$$\begin{aligned} \mathbf{q} &= \begin{bmatrix} u_1 & v_1 & 0 & v_2 & 0 & 0 & u_4 & v_4 \end{bmatrix}^{\mathrm{T}}, \\ \mathbf{P} &= \begin{bmatrix} 1 & 0 & R_{2x} & 0 & R_{3x} & R_{3y} & -1 & 0 \end{bmatrix}^{\mathrm{T}} \end{aligned} \tag{44}$$

where $u_1$, $v_1$, $v_2$, $u_4$, $v_4$, $R_{2x}$, $R_{3x}$, $R_{3y}$ are unknowns.

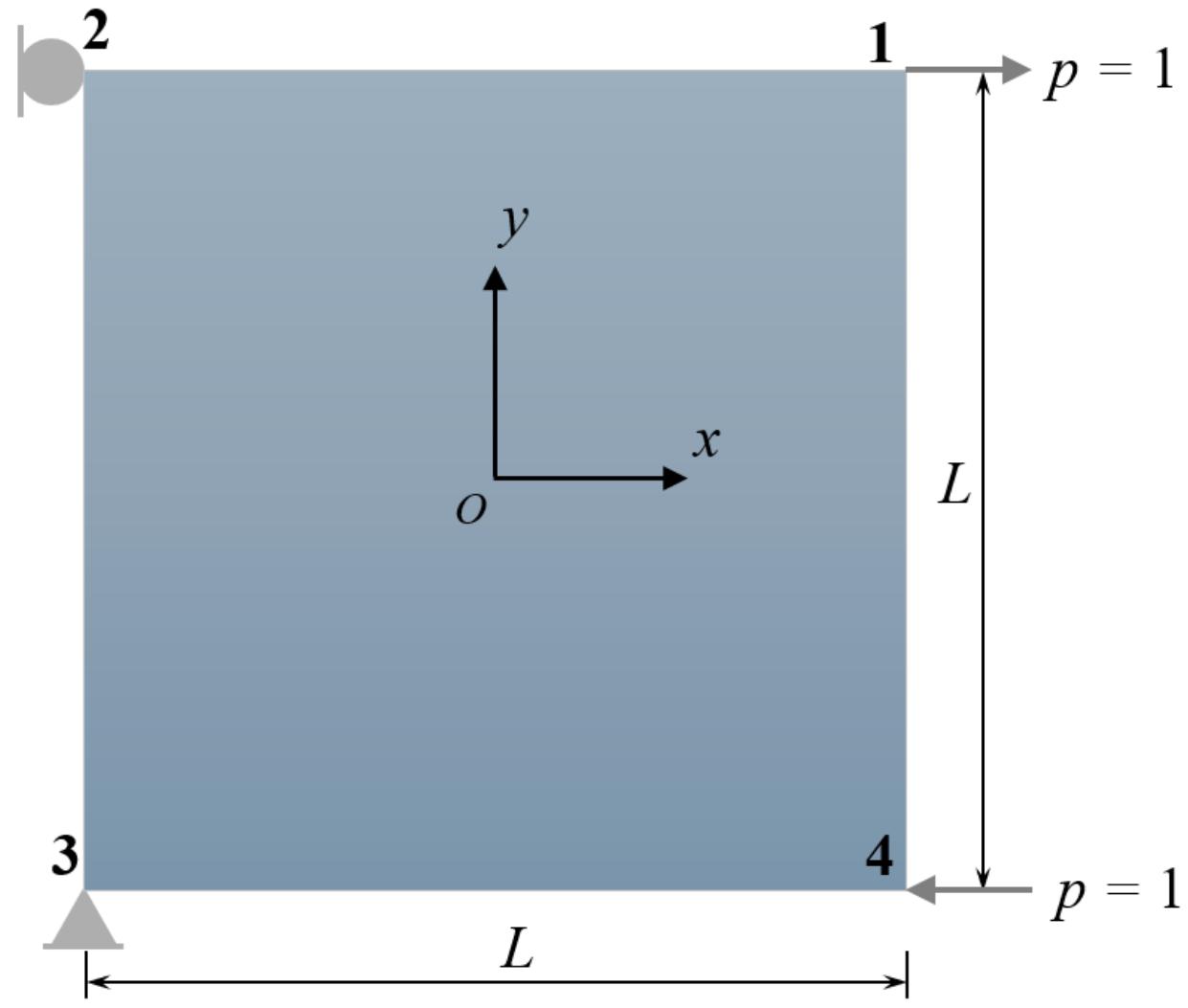

Fig. 9 Boundary conditions of planar rectangular structures.

The stiffness equations for the three discretization schemes are constructed as:

$$
\begin{aligned}
\mathbf{K}_{\mathrm{CST}} \cdot \mathbf{q}_1 &= \mathbf{P}_1, \\
\mathbf{K}_{\mathrm{Q4}} \cdot \mathbf{q}_2 &= \mathbf{P}_2, \\
\mathbf{K}_{\mathrm{IELSM,RVE}} \cdot \mathbf{q}_3 &= \mathbf{P}_3
\end{aligned} \tag{45}
$$

Solving these equations respectively yields:

$$
\begin{aligned}
&\mathbf{q}_1 = \frac{(v+1)}{2E}\begin{bmatrix}3-v & v-3 & 0 & -1-v & 0 & 0 & v-3 & -1\end{bmatrix}^{\mathrm{T}},\ \mathbf{P}_1 = \begin{bmatrix}1 & 0 & -1 & 0 & 1 & 0 & -1 & 0\end{bmatrix}^{\mathrm{T}}; \\
&\mathbf{q}_2 = \frac{12(v+1)}{E(3-v)}\begin{bmatrix}1-v & v-1 & 0 & 0 & 0 & 0 & v-1 & v-1\end{bmatrix}^{\mathrm{T}},\ \mathbf{P}_2 = \begin{bmatrix}1 & 0 & -1 & 0 & 1 & 0 & -1 & 0\end{bmatrix}^{\mathrm{T}}; \\
&\mathbf{q}_3 = \frac{2(v+1)}{E}\begin{bmatrix}1 & -1 & 0 & 0 & 0 & 0 & -1 & -1\end{bmatrix}^{\mathrm{T}},\ \mathbf{P}_3 = \begin{bmatrix}1 & 0 & -1 & 0 & 1 & 0 & -1 & 0\end{bmatrix}^{\mathrm{T}}
\end{aligned} \tag{46}
$$

The total potential energy of the system for the three discretization schemes is:

$$
\begin{aligned}
\Pi_{\mathrm{CST}} &= \frac{1}{2}\mathbf{q}_1^{\mathrm{T}}\mathbf{K}_{\mathrm{CST}}\mathbf{q}_1 - \mathbf{P}_1^{\mathrm{T}}\mathbf{q}_1 = -\frac{(1+v)(3-v)}{2E}, \\
\Pi_{\mathrm{Q4}} &= \frac{1}{2}\mathbf{q}_2^{\mathrm{T}}\mathbf{K}_{\mathrm{Q4}}\mathbf{q}_2 - \mathbf{P}_2^{\mathrm{T}}\mathbf{q}_2 = -\frac{12(1-v^2)}{(3-v)E}, \\
\Pi_{\mathrm{IELSM}} &= \frac{1}{2}\mathbf{q}_3^{\mathrm{T}}\mathbf{K}_{\mathrm{IELSM}}\mathbf{q}_3 - \mathbf{P}_3^{\mathrm{T}}\mathbf{q}_3 = -\frac{2(1+v)}{E}
\end{aligned} \tag{47}
$$

Fig. 10 plots the relationship between the total potential energy of the system (taking $E$=1) and Poisson's ratio for the three discretization schemes under both plane stress and plane strain assumptions. The total potential energy for the plane strain problem is obtained by replacing ($E$, $v$) in Eq. (46) with ($E/(1-v^2)$, $v/(1-v)$). According to the principle of minimum potential energy, for a given external force, a lower potential energy indicates higher overall computational accuracy. A comparison of the potential energies obtained from the three schemes in Fig. 10 shows that, for the same number of nodal degrees of freedom, IELSM exhibits superior computational accuracy compared to the CST element over the entire range of Poisson's ratio. Regarding the Q4 element and IELSM, their relative accuracy depends on the value of Poisson's ratio: under plane stress conditions, the accuracy of the Q4 element is higher than that of IELSM when $v < 0.6$; under plane strain conditions, the Q4 element is more accurate when $v < 0.375$. Conversely, when Poisson's ratio exceeds these critical values, the accuracy of the IELSM element surpasses that of the Q4 element. This phenomenon can be attributed to the fact that IELSM effectively avoids the zero-energy mode that seriously plagues the Q4 element. As shown in Table 1, the stiffness matrix of Q4 suffers from rank deficiency and presents hourglass modes ($\lambda_4$, $\lambda_5$), which are singular in eigenvalue under high Poisson's ratio conditions, resulting in the element having deformation but the calculated strain energy being zero. In contrast, IELSM shows a stable eigenvalue spectrum on all higher-order modes ($\lambda_2$ to $\lambda_5$), indicating that the additional volumetric constraint introduced in this work successfully eliminates the aforementioned hourglass instability and avoids the parasitic shear locking in the CST combination, thereby achieving more robust and accurate solution capabilities in the simulation of near-incompressible material deformation behavior.

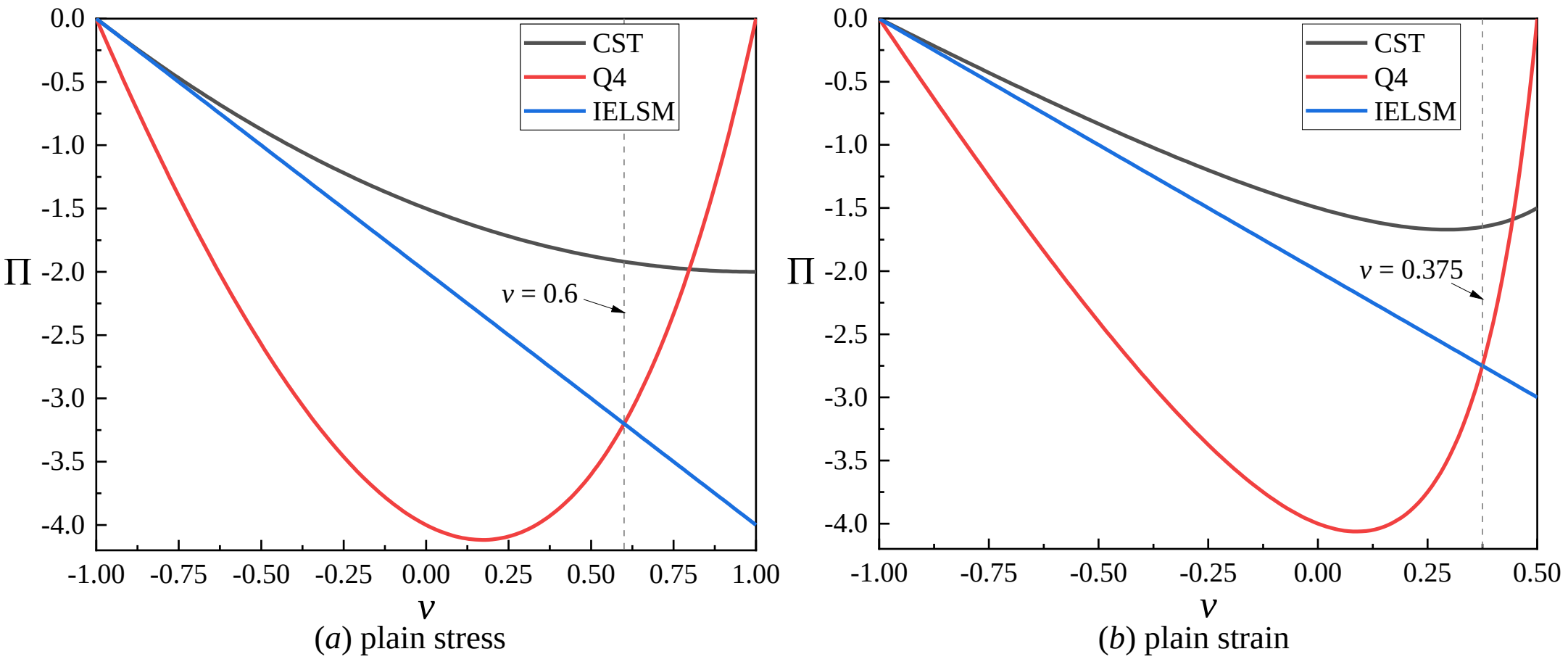

Fig. 10 Relationship between total potential energy and Poisson’s ratio for the three discretization schemes ($E$=1).

## 5.0 Numerical Validation

The validation suite comprises five benchmark problems: (1) uniaxial tension, (2) pure shear, (3) infinite plate with central circular hole, (4) infinite center-cracked plate, and (5) finite plate with crucifix-shaped cracks. For the stress concentration around a circular hole, comparisons include high-resolution FEM solutions; for crack problems, IELSM predictions are compared with those from FEM quadrilateral (Q4) elements, with equivalent degrees of freedom and element counts (IELSM RVEs and Q4 elements). Selected material properties span Poisson’s ratios from nearly 0 to 0.5 to test IELSM’s adaptability. In these benchmark problems, the width and height of the rectangular plate (or quarter-plate) are consistently denoted as $H$, while the side length of the square RVE is denoted as $L$. The ratio $H/L$ serves as a direct indicator of IELSM mesh density across different configurations. All cases undergo convergence analysis. Elastic parameters and their corresponding IELSM model parameters under plane stress conditions are summarized in Table 2, while those for plane strain conditions are presented in Table 3. Notably, when the additional bulk modulus equals zero, IELSM reverts to the classical LSM model with its inherent Poisson’s ratio limitation. This is observed in Table 2 where the Poisson’s ratio is 1/3 (recorded as 0.333333 with six-decimal precision) at $k^v = 0$ under plane stress conditions, and in Table 3 where $v$ = 1/4 under plane strain conditions. Furthermore, as the magnitude of $k^v$ increases, particularly when $k^v$ >0, the macroscopic Poisson’s ratio predicted by IELSM progressively increases, effectively broadening the accessible range beyond the classical LSM constraints.

Table 2 Elastic parameters and corresponding IELSM parameter values (plane stress).

| $E$(GPa) | $\nu$ | $k_1^n$ (GPa) | $k_2^n$ (GPa) | $k^v$ (GPa) |
|---|---|---|---|---|
| 1 | 0.01 | 0.495050 | 0.495050 | −0.485049 |
| 1 | 0.1 | 0.454545 | 0.454545 | −0.353535 |
| 1 | 0.2 | 0.416667 | 0.416667 | −0.208333 |
| 1 | 0.3 | 0.384615 | 0.384615 | −0.054945 |
| 1 | 0.333333 | 0.375000 | 0.375000 | 0 |
| 1 | 0.4 | 0.357143 | 0.357143 | 0.119048 |
| 1 | 0.49 | 0.335570 | 0.335570 | 0.309251 |
| 200 | 0.286 | 77.760498 | 77.760498 | −15.464973 |

Table 3 Elastic parameters and corresponding IELSM parameter values (plane strain).

| $E$(GPa) | $\nu$ | $k_1^n$ (GPa) | $k_2^n$ (GPa) | $k^v$ (GPa) |
|---|---|---|---|---|
| 1 | 0.01 | 0.495050 | 0.495050 | −0.484946 |
| 1 | 0.1 | 0.454545 | 0.454545 | −0.340909 |
| 1 | 0.2 | 0.416667 | 0.416667 | −0.138889 |
| 1 | 0.25 | 0.384615 | 0.384615 | 0 |
| 1 | 0.3 | 0.384615 | 0.384615 | 0.192308 |
| 1 | 0.4 | 0.357143 | 0.357143 | 1.071429 |
| 1 | 0.49 | 0.335570 | 0.335570 | 16.107383 |

### 5.1 Uniaxial tension tests

The geometric configuration and boundary conditions for uniaxial tension are illustrated in Fig. 11. A rectangular plate of unit thickness undergoes uniform tensile traction $\sigma$=1MPa along the top edge, with the bottom left corner fixed in the $x$-direction and the entire bottom edge fixed in the $y$-direction. For plane stress analysis, distributed traction is discretized into nodal forces: outer nodes receive $F_y = \sigma L/2$ while interior nodes receive $F_y = \sigma L$, where $L$ is the RVE size.

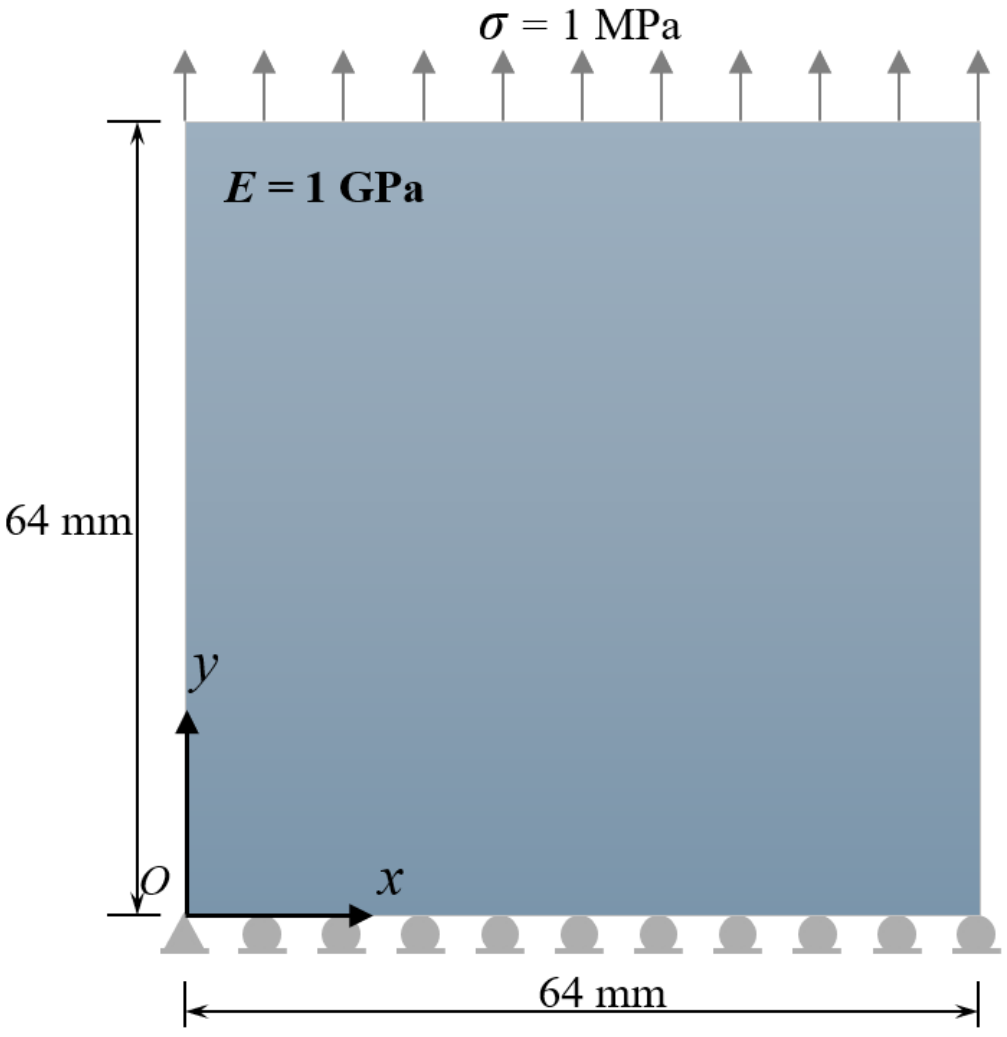

Fig. 11 Uniaxial tension tests: geometry and boundary conditions.

The analytical displacement field is given by:

$$u_y(y) = \frac{\sigma}{E} y \ , \quad u_x(x) = -v \frac{\sigma}{E} x \tag{48}$$

Convergence analysis for $v$ = 0.333333 with varying mesh sizes ($L$=1, 2, 4, 8mm) demonstrates IELSM's mesh-convergent behavior, as shown in Fig. 12($a$). IELSM solutions coincide exactly with analytical predictions across all mesh resolutions (mesh topology depicted in Fig. 13). Fig. 12($b$) further validates IELSM's capability across the full Poisson's ratio spectrum ($v$ = 0.01, 0.1, 0.2, 0.3, 0.333333, 0.4, 0.49) using $L$ = 1mm mesh resolution.

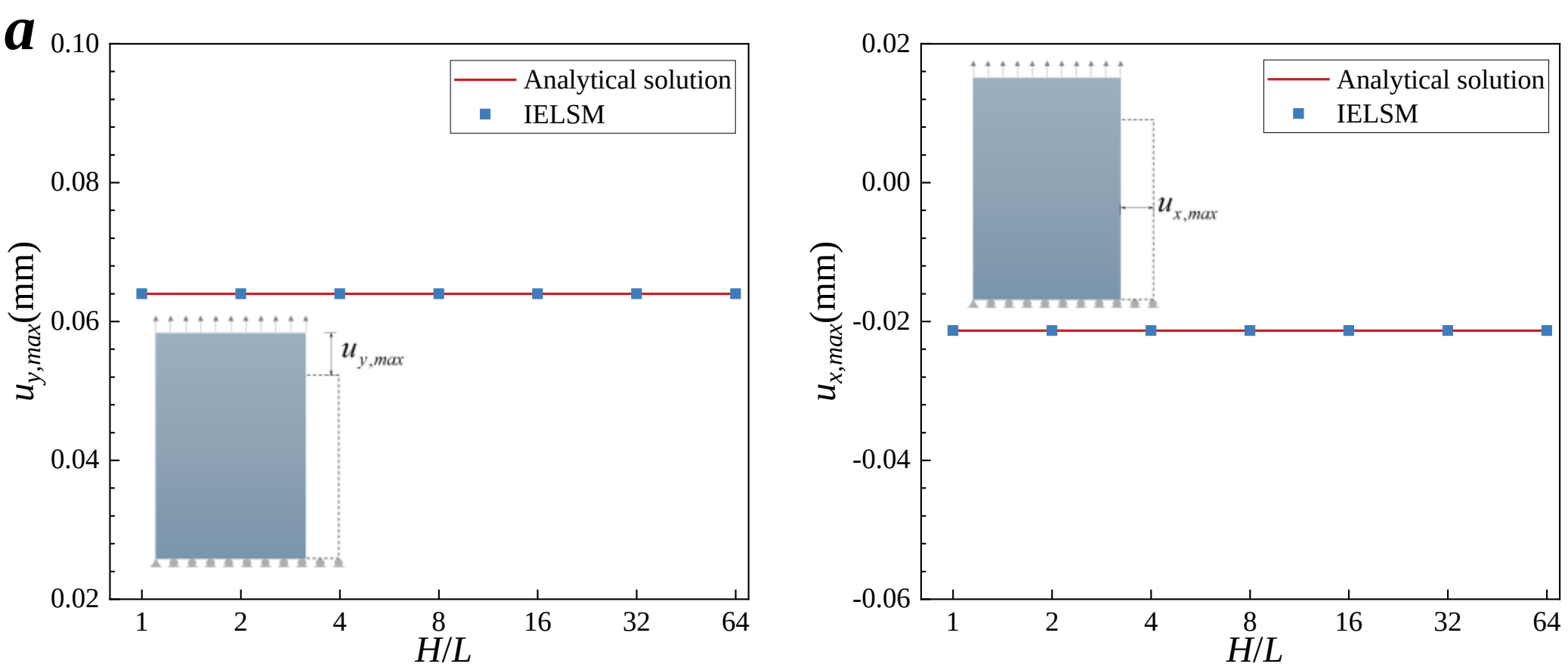

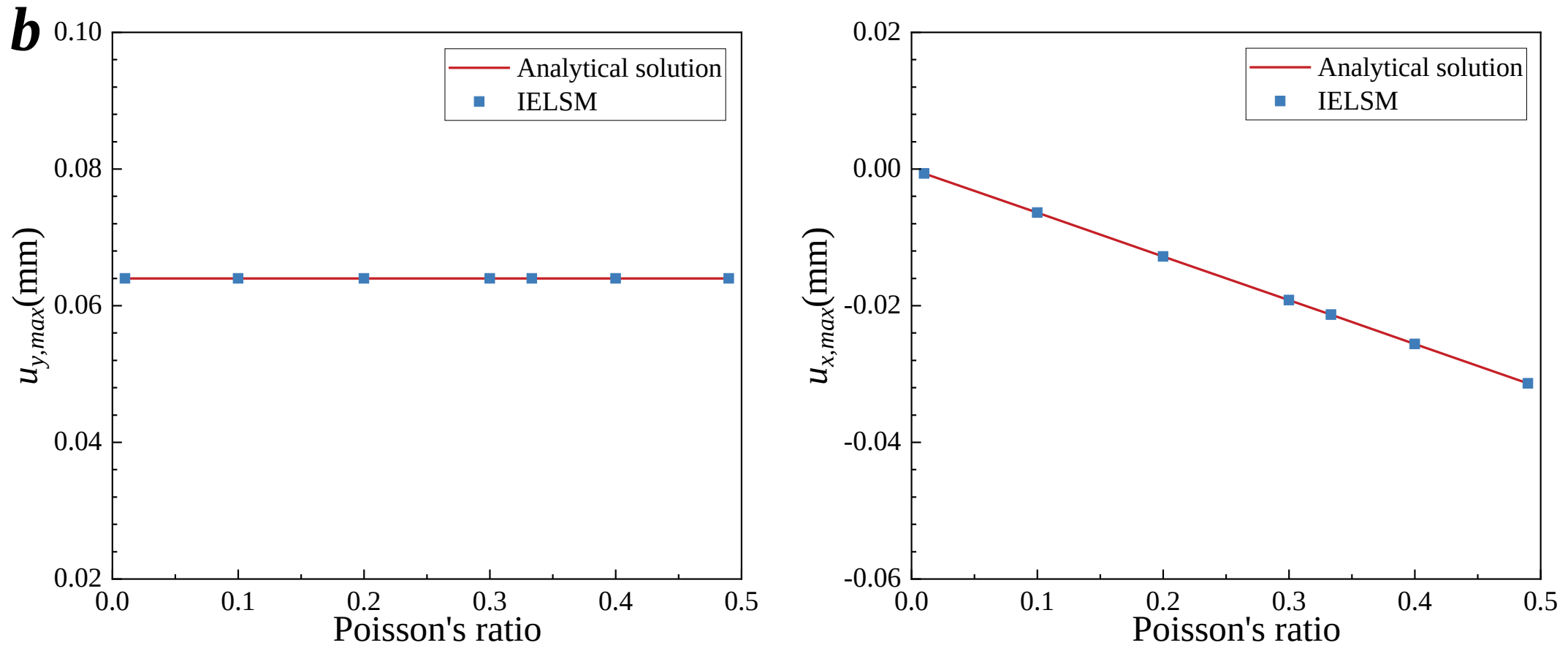

Fig. 12 Uniaxial tension test: (*a*) convergence analysis ($v$ = 0.333333); (*b*) maximum *x*- and *y*-direction displacements for varying Poisson's ratios.

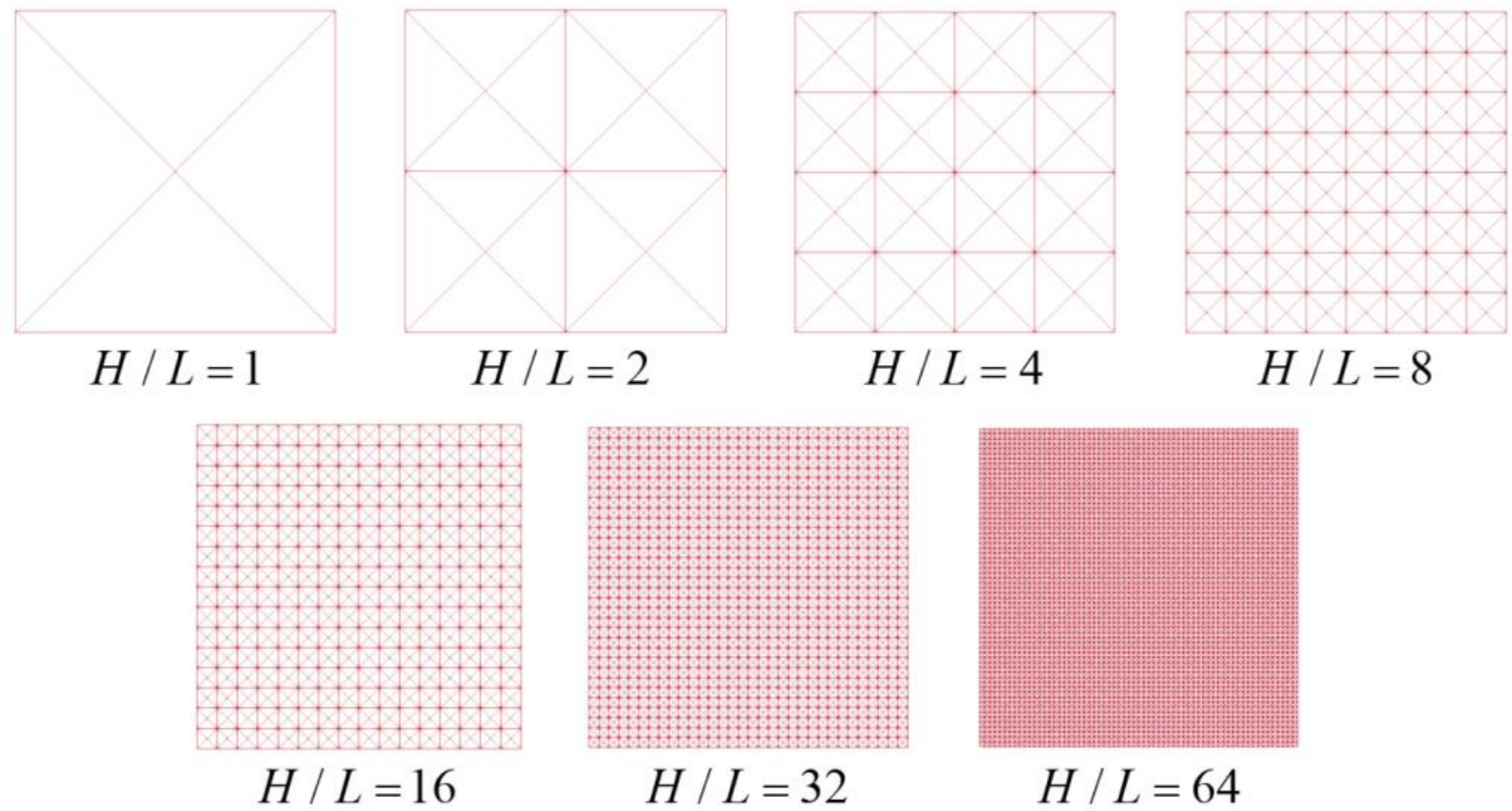

Fig. 13 IELSM mesh arrangement for rectangular plate.

### 5.2 Pure-shear tests

As shown in Fig. 14, a rectangular plate with free upper, left, and right edges undergoes uniform shear traction $\tau$ = 1MPa, while the bottom edge remains fully fixed. This analysis assumes plane strain conditions with a uniform plate thickness of 64mm. Nodal force discretization follows the same principle as the tension case. The analytical displacement field for pure shear is:

$$u_x(y) = \frac{2\tau(1+v)}{E} y \tag{49}$$

Convergence analysis for $v$=0.25 with identical mesh resolutions confirms IELSM's consistent agreement with analytical solutions, as evidenced in Fig. 15(*a*). Across varied Poisson's ratios ($v$ = 0.01, 0.1, 0.2, 0.25, 0.3, 0.4, 0.49), IELSM maintains precise correspondence with analytical predictions (Fig. 15(*b*)), validating its shear response modeling capability.

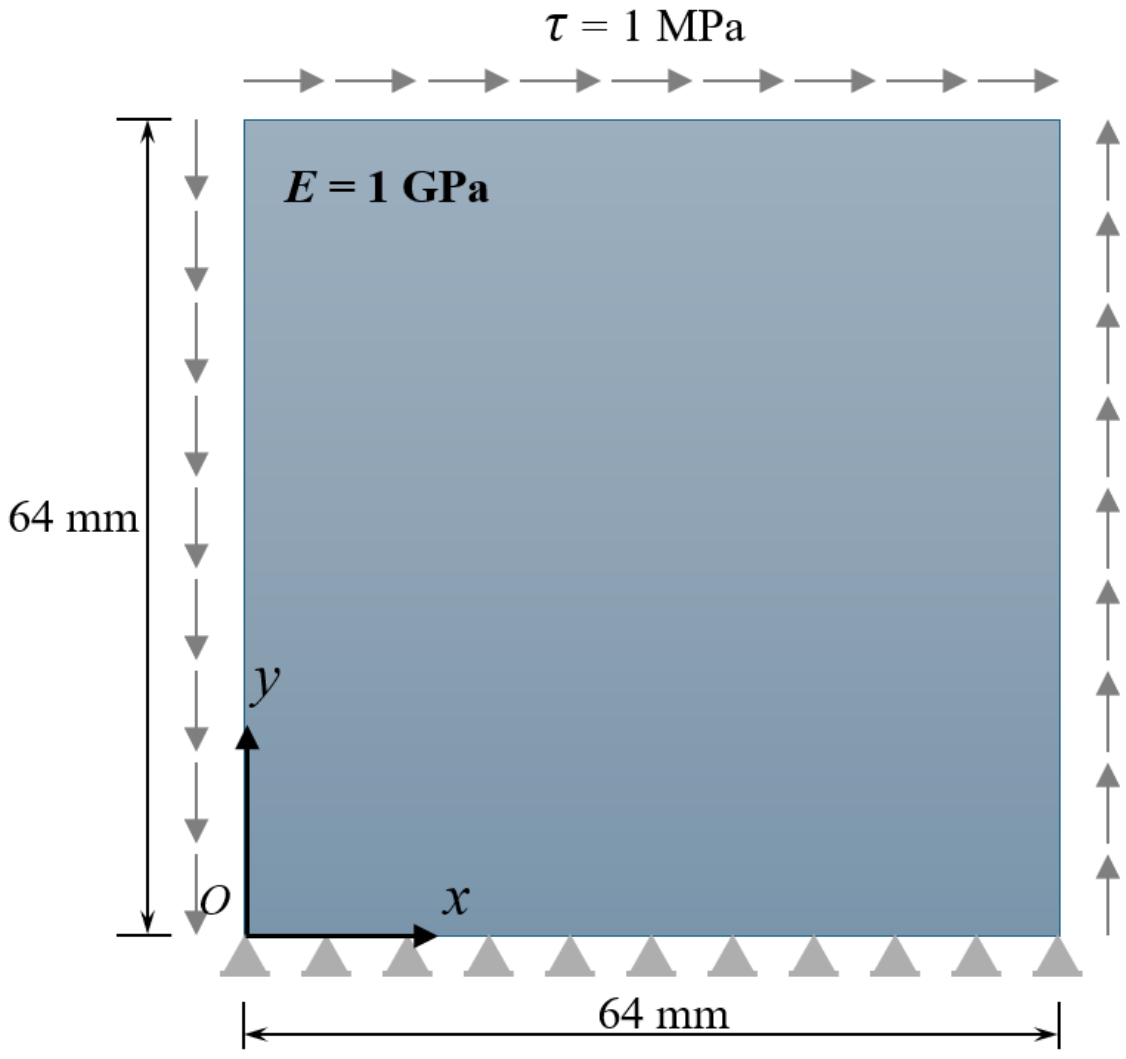

Fig. 14 Pure-shear tests: geometry and boundary conditions.

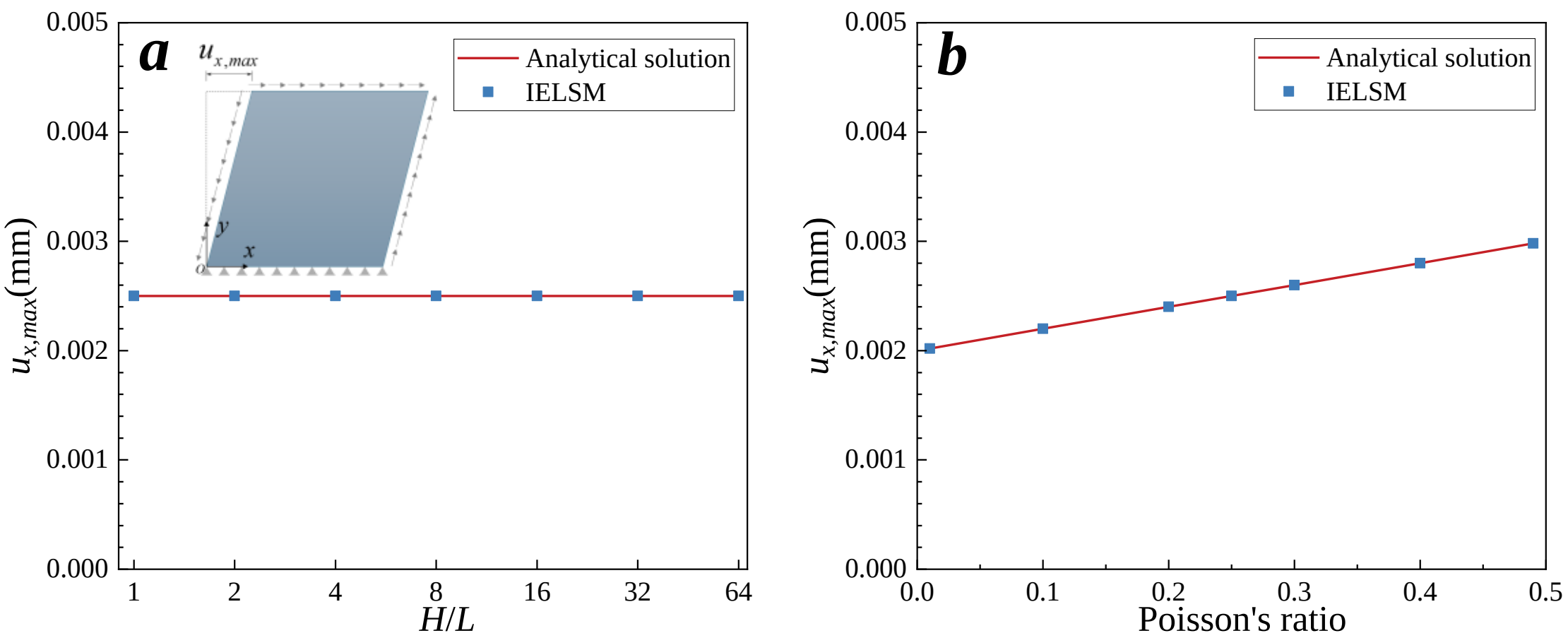

Fig. 15 Pure-shear tests: ($a$) convergence analysis ($v$ = 0.25); ($b$) maximum $x$-direction displacement for varying Poisson's ratios.

### 5.3 Circular hole plate

Consider an infinite thin plate with central circular hole of radius $R$=1m, subjected to uniform tensile stress $\sigma$ = 1MPa. To simulate infinite domain conditions, a plate width $H$=10$R$ is adopted, with quarter-symmetry exploited (Fig. 16). Boundary conditions enforce symmetry: $u_x = 0$ along the left edge, $u_y = 0$ along the bottom edge, and $\sigma_x = \sigma$ along the right edge. The layout of the IELSM mesh arrangement in this example is shown in Figure 17.

Cartesian stress components for the infinite plate with central hole under uniaxial tension are given by (Li and Zhu, 2024; Ren et al., 2020):

$$
\begin{cases}
\sigma_{xx}(\rho,\theta)=\sigma\left(1-\dfrac{3R^2}{2\rho^2}\cos 2\theta-\dfrac{R^2}{\rho^2}\cos 4\theta+\dfrac{3R^4}{2\rho^4}\cos 4\theta\right)\\
\sigma_{yy}(\rho,\theta)=\sigma\left(-\dfrac{R^2}{2\rho^2}\cos 2\theta+\dfrac{R^2}{\rho^2}\cos 4\theta-\dfrac{3R^4}{2\rho^4}\cos 4\theta\right)\\
\tau_{xy}(\rho,\theta)=\sigma\left(-\dfrac{R^2}{2\rho^2}\sin 2\theta+\dfrac{R^2}{\rho^2}\sin 4\theta+\dfrac{3R^4}{2\rho^4}\sin 4\theta\right)
\end{cases}
\tag{50}
$$

where $(\rho,\theta)$ are polar coordinates.

Displacement field solutions are available for plane stress conditions:

$$
\begin{cases}
u_x(\rho,\theta)=\dfrac{\sigma R}{8G}\left((\dfrac{\rho}{R}+\dfrac{2R}{\rho})(\kappa+1)\cos\theta+(1-\dfrac{R^2}{\rho^2})\dfrac{2R}{\rho}\cos 3\theta\right)\\
u_y(\rho,\theta)=\dfrac{\sigma R}{8G}\left(\left(\dfrac{\rho}{R}(\kappa-3)+\dfrac{2R}{\rho}(1-\kappa)\right)\sin\theta+(1-\dfrac{R^2}{\rho^2})\dfrac{2R}{\rho}\sin 3\theta\right)
\end{cases}
\tag{51}
$$

where $\kappa=\dfrac{3-v}{1+v}$, $G=\dfrac{E}{2(1+v)}$.

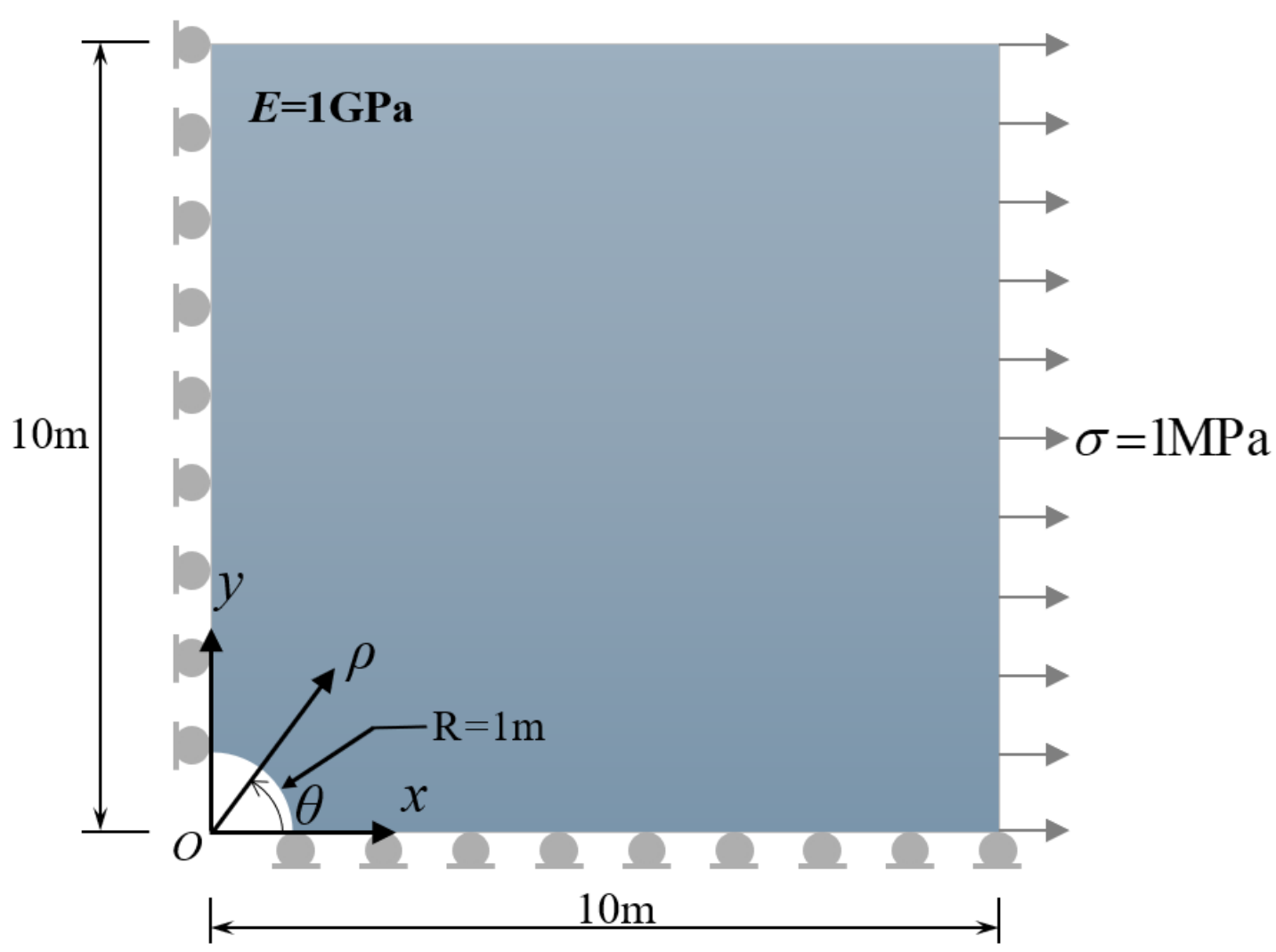

Fig. 16 Circular hole plate: geometry and boundary conditions.

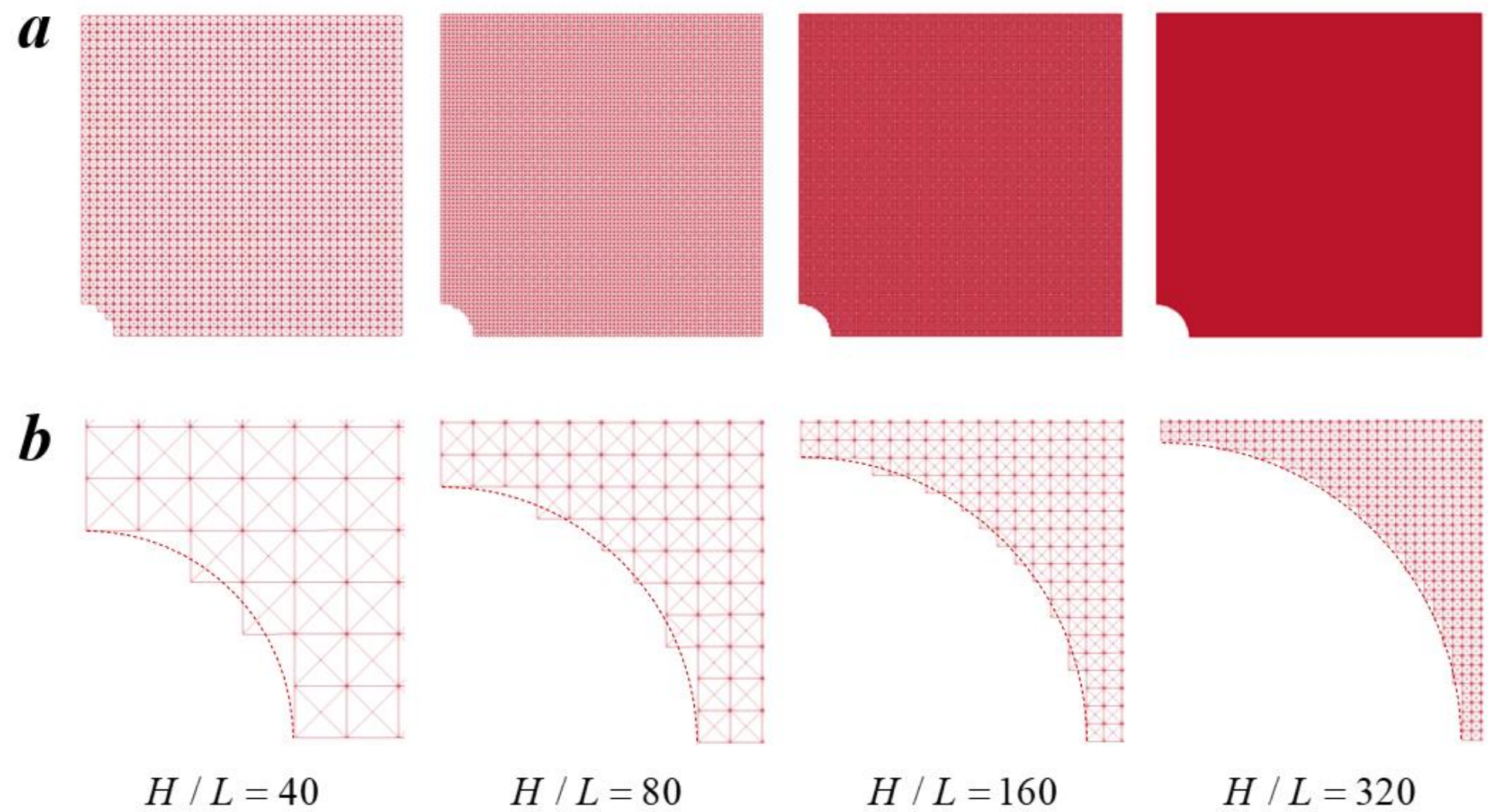

Fig. 17 Circular hole plate: (*a*) quarter-plate mesh arrangement; (*b*) mesh refinement around the hole.

Convergence analysis for $v = 0.1$ compares $\sigma_{xx}$ and $\sigma_{yy}$ distributions along the left boundary ($x = 0$), as shown in Fig. 18. The strain field therein is calculated based on the assumption of a linear displacement field. The reference high-resolution FEM solution employs Q4 elements with a total of 13580 nodes and 13344 elements. Near the hole boundary, high-resolution FEM closely approximates analytical $\sigma_{xx}$ values, while IELSM slightly underestimates them, potentially due to stair-step approximations of curved boundaries. Conversely, IELSM provides superior $\sigma_{yy}$ predictions matching analytical solutions more closely than FEM near the hole. Both methods demonstrate satisfactory consistency away from the hole, with minor deviations attributable to finite-domain approximations of infinite plates. Overall, IELSM exhibits excellent convergence characteristics.

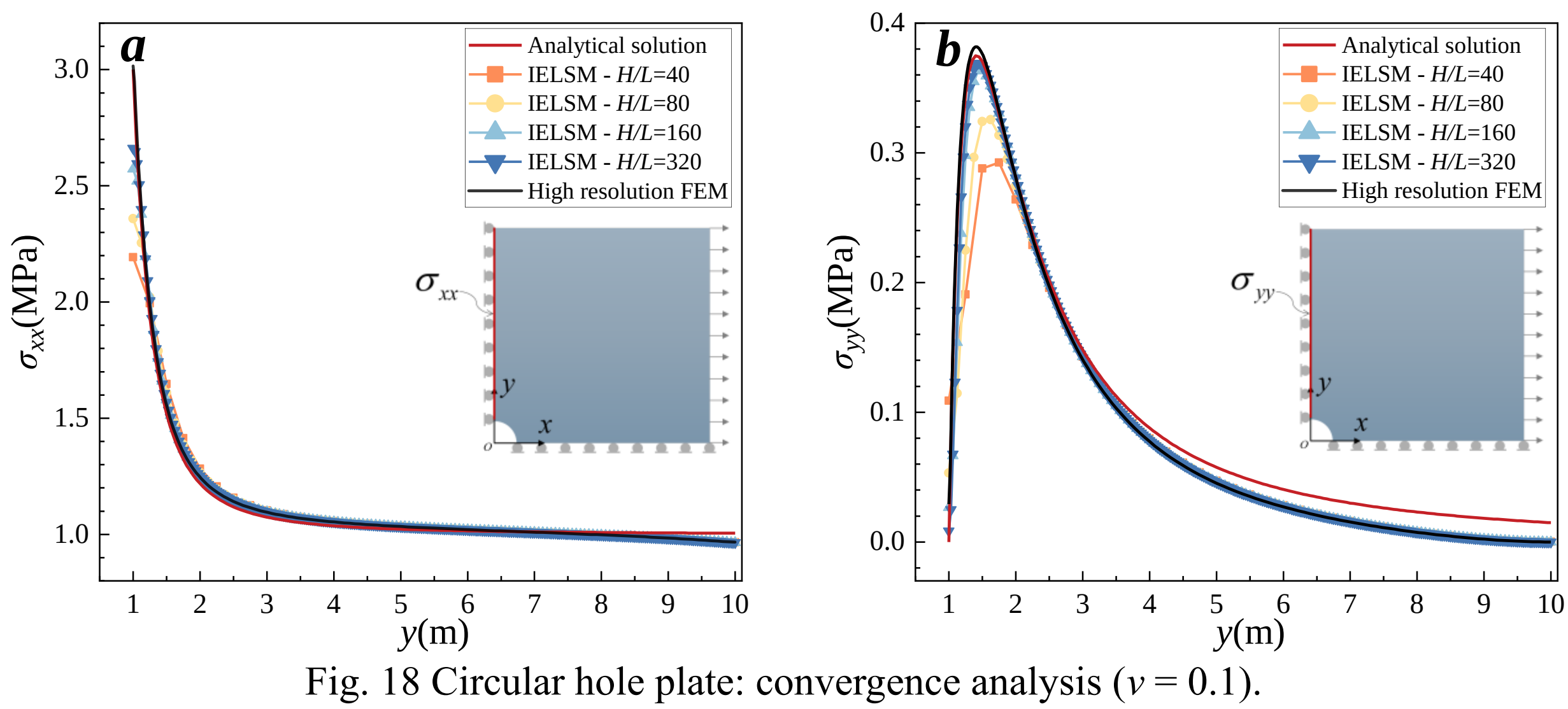

Fig. 18 Circular hole plate: convergence analysis ($v = 0.1$).

Fig. 19 presents displacement field contour comparisons for $v = 0.1$. The IELSM solution (Fig. 19*b*) with mesh refinement *H/L* = 320 demonstrates close agreement with high-resolution

FEM predictions (Fig. 19*c*). Both numerical approaches exhibit comparable deviations from the analytical solution, particularly near the hole boundary where geometric approximations induce local discretization errors. Further away from the hole, both high-resolution FEM and IELSM predictions show systematic deviations from the analytical displacement field. This behavior is consistent with the $\sigma_{xx}$ and $\sigma_{yy}$ convergence results presented in Fig. 18, where both methods similarly underestimate the stress components away from the hole boundary. As noted in the convergence analysis, these discrepancies arise from the finite-domain approximation of an infinite plate, an inherent limitation of the computational setup rather than a deficiency of either method. These minor discrepancies diminish with increasing mesh refinement, confirming IELSM's convergence characteristics.

Since strain represents a more sensitive metric than displacement, strain field comparisons are presented in Fig. 20. High-resolution FEM strain contours (Fig. 20(*b*)) align closely with analytical solutions (Fig. 20(*a*)). Linear displacement field IELSM (Fig. 20(*c*)) exhibits minor deviations, while bilinear displacement field IELSM (Fig. 20(*d*)) overestimates strain magnitudes near stress concentration regions. This indicates that the bilinear form allows higher strain variability, which in this configuration appears to artificially amplify the stress concentration effect. The discrete, stair-step approximation of the circular hole boundary inherent to lattice models likely contributes to this behavior by creating localized geometric irregularities that interact with the assumed displacement field. Overall, while both IELSM variants capture the essential strain patterns, the linear formulation provides smoother and more conservative estimates of peak strains near geometric discontinuities.

To verify IELSM's coverage of the complete Poisson's ratio spectrum, Fig. 21 compares *y*-direction displacement $u_y$ along the left boundary ($x = 0$) for various Poisson's ratios against high-resolution FEM solutions. Remarkably, IELSM maintains consistent predictive accuracy across the entire physically admissible range of Poisson's ratios ($\nu$ = 0.01 to 0.49 for plane stress), demonstrating its robust representation of elastic parameters and theoretical completeness.

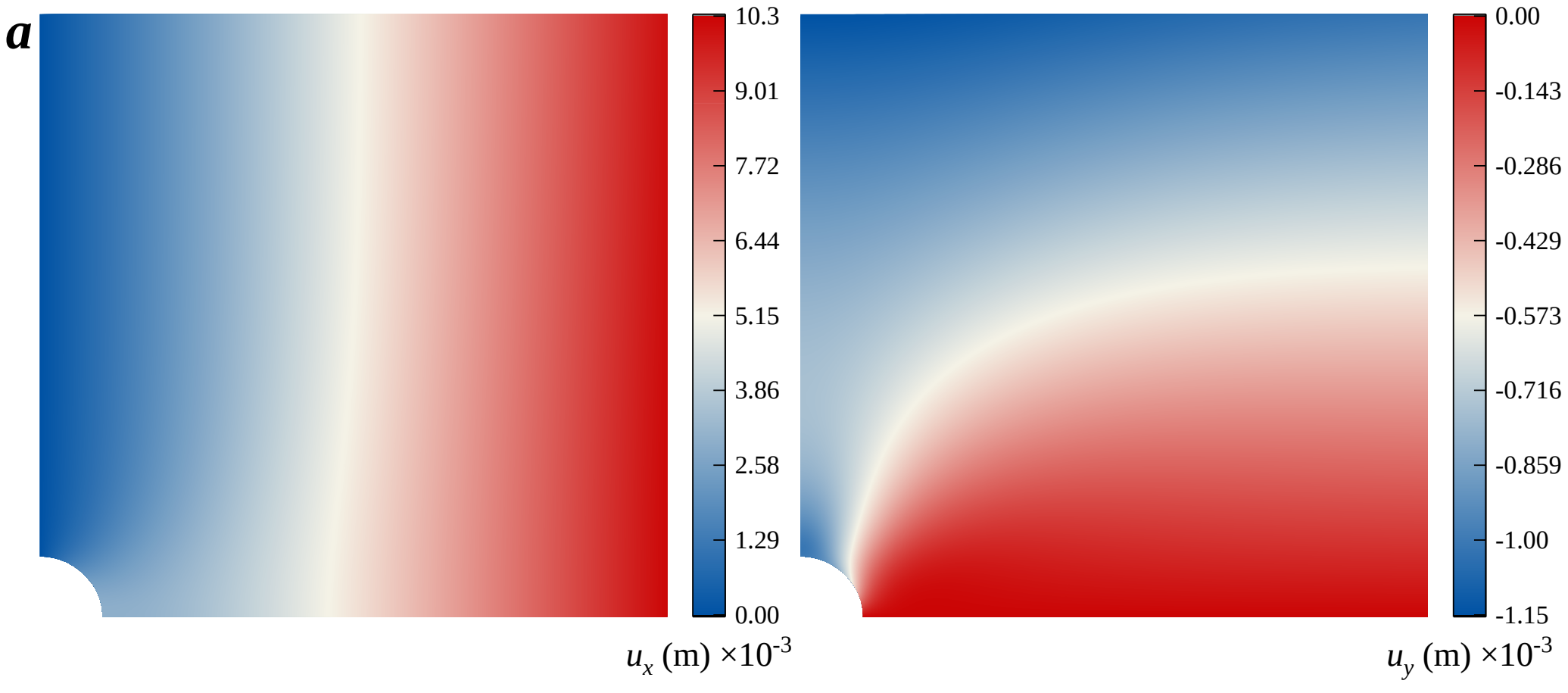

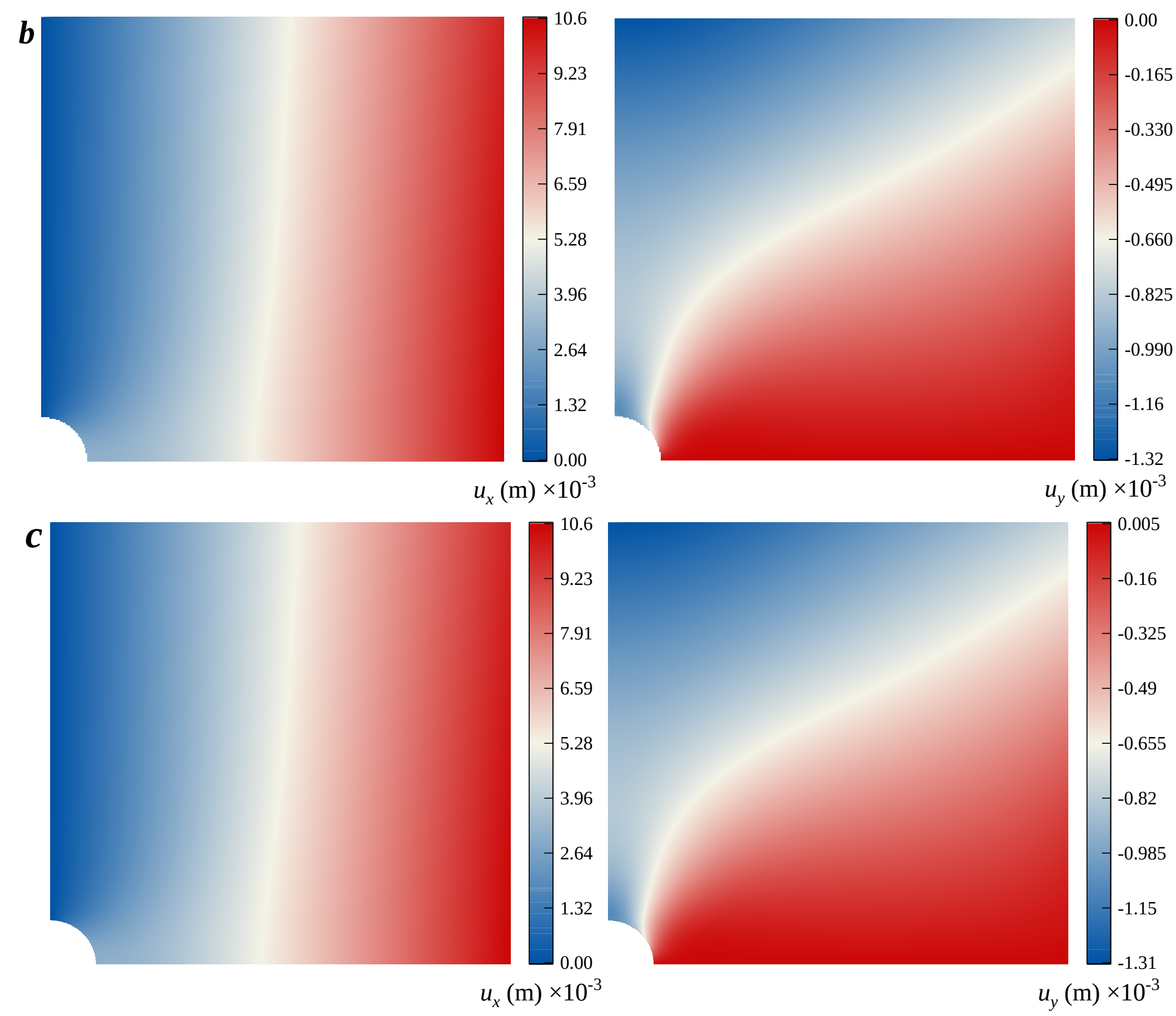

Fig. 19 Circular hole plate: (*a*) exact displacement field contour maps; (*b*) IELSM-predicted displacement field; (*c*) high-resolution FEM-predicted displacement field ($v$ = 0.1).

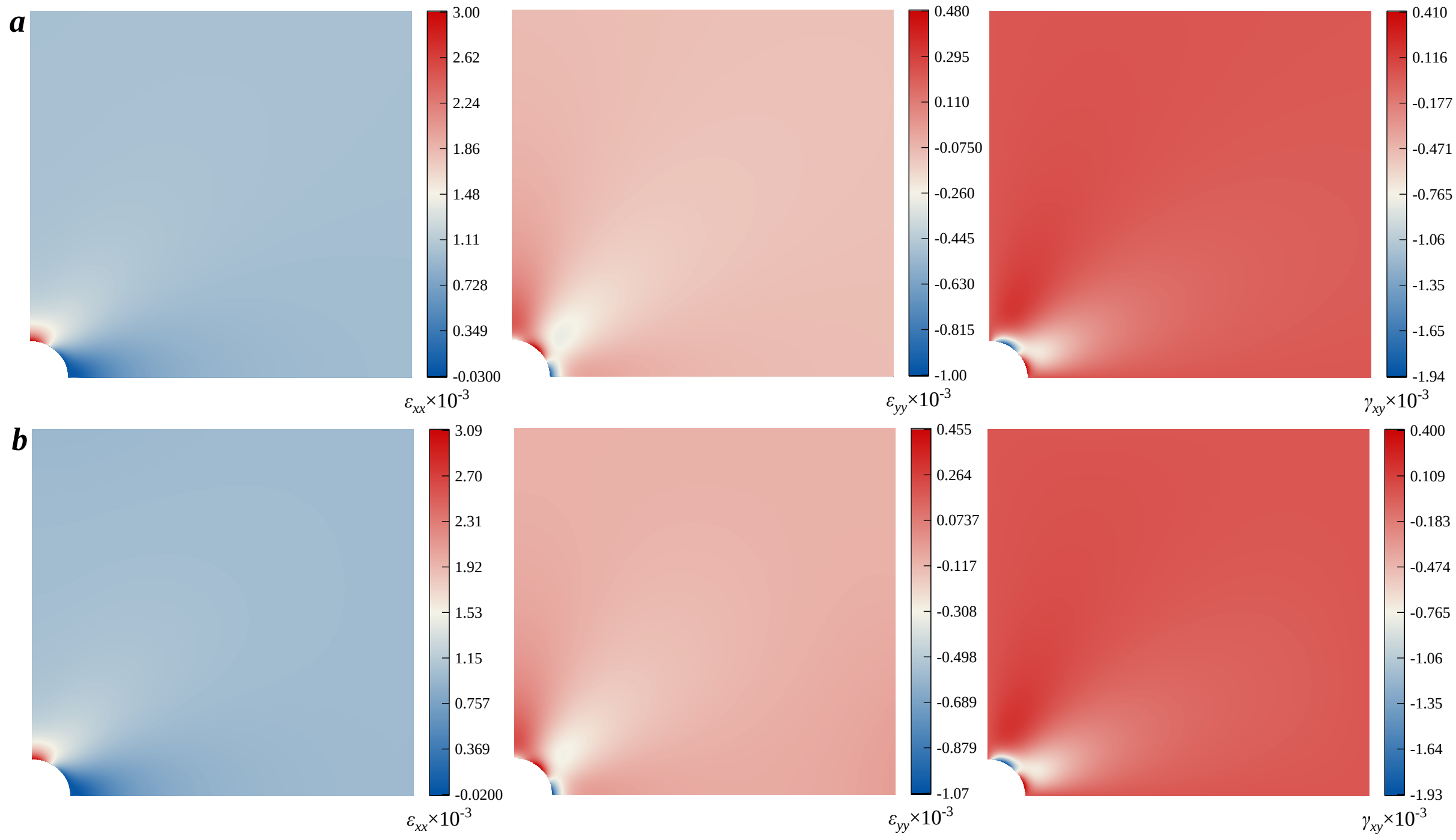

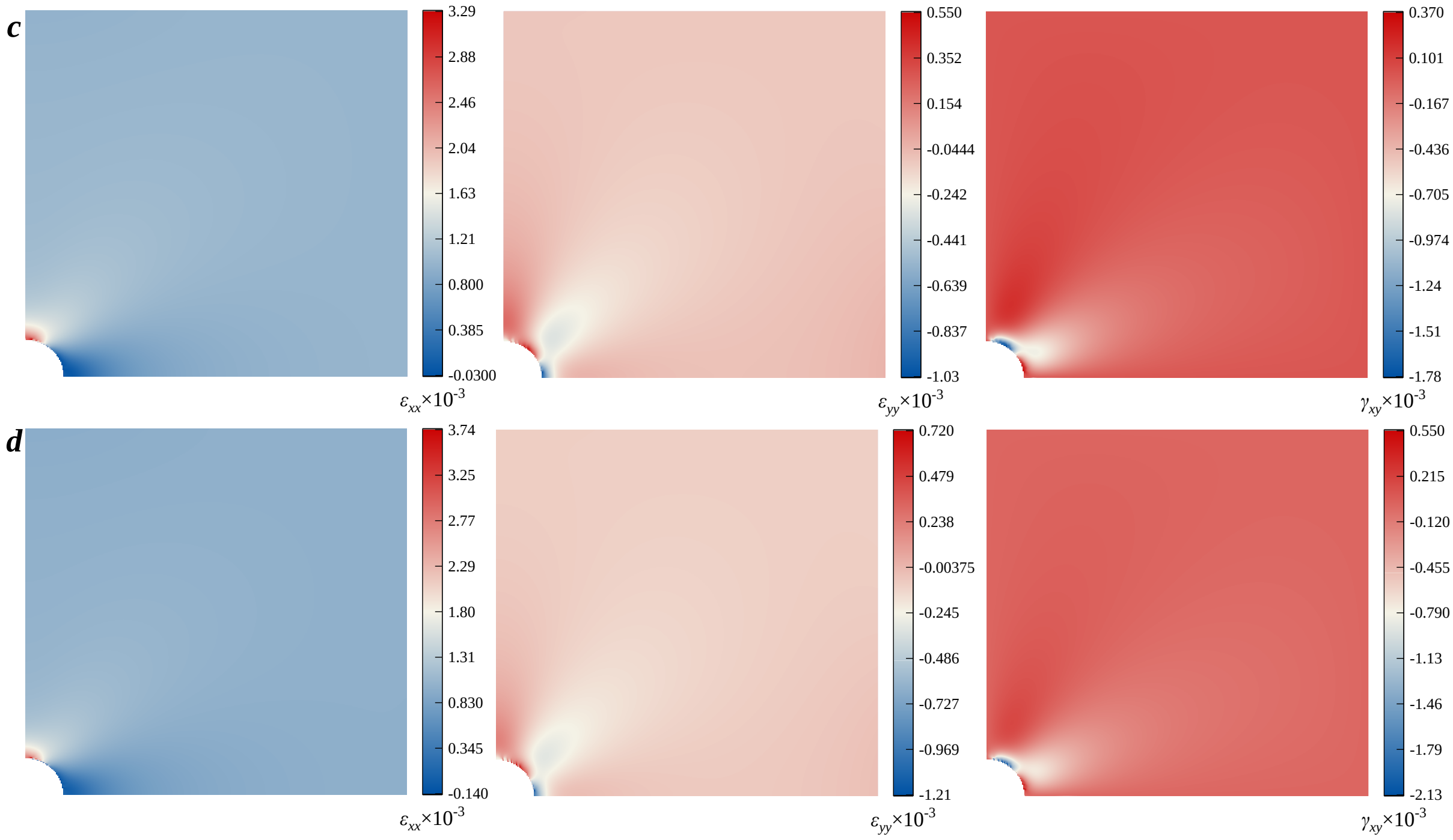

Fig. 20 Circular hole plate: (*a*) exact strain field contour maps; (*b*) high-resolution FEM-predicted strain field; (*c*) linear displacement field IELSM-predicted strain field; (*d*) bilinear displacement field IELSM-predicted strain field ($v$ = 0.1).

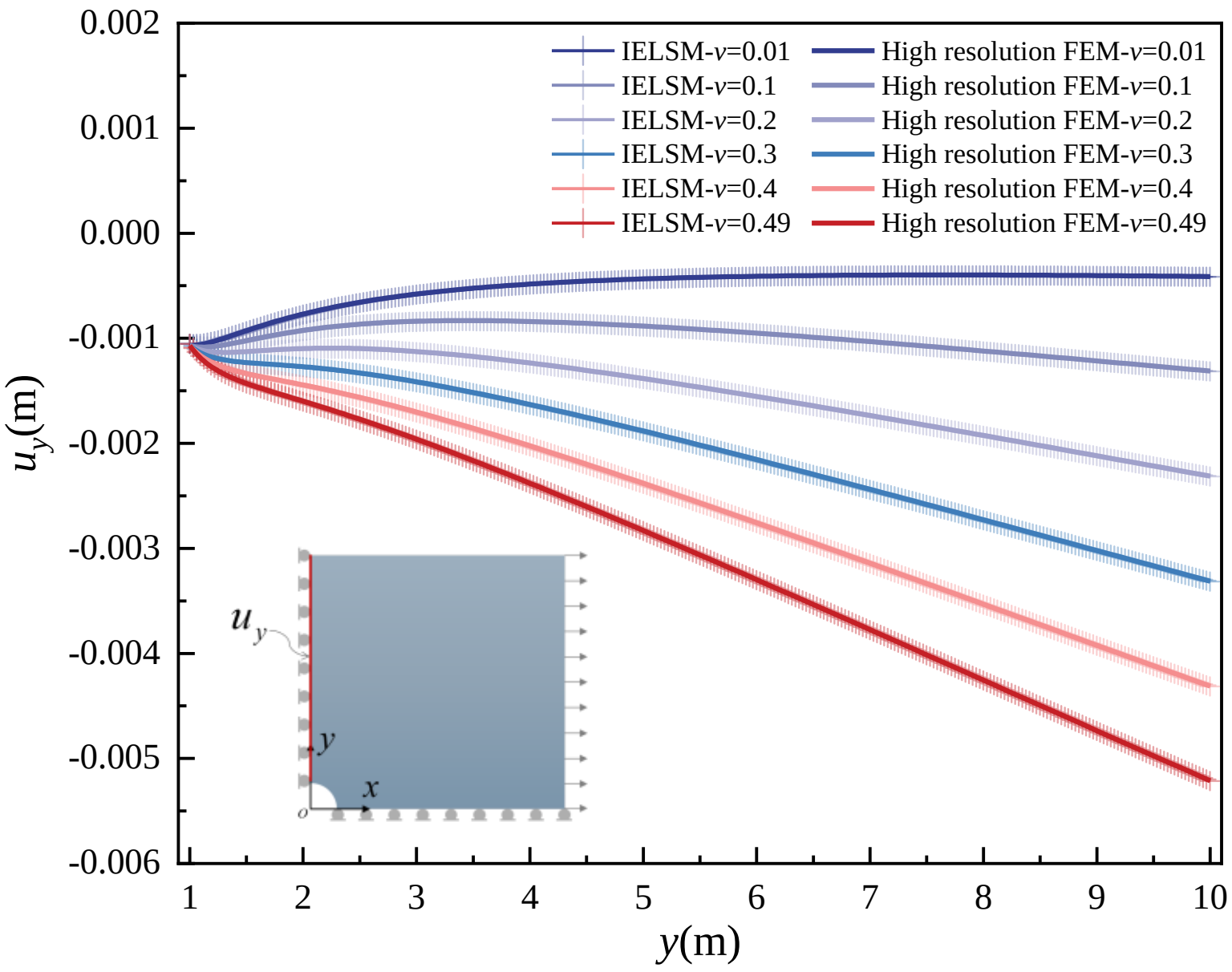

Fig. 21 Circular hole plate: comparison of IELSM and high-resolution FEM results for varying Poisson's ratios.

### 5.4 Centre cracked plate

An infinite center-cracked plate under plane stress conditions is modeled via finite-domain approximation. Quarter-symmetry is employed with domain dimensions $H\times H$ = 10×10m and half-crack length $a$ = 2m, as depicted in Fig. 22. Material properties are $E$ = 1GPa with Poisson's ratio $\nu$ = 0.2. Uniform tensile stress $\sigma$ = 1MPa is applied along top and right boundaries. IELSM mesh densities span $H/L$ = {10, 20, 40, 80, 160}.

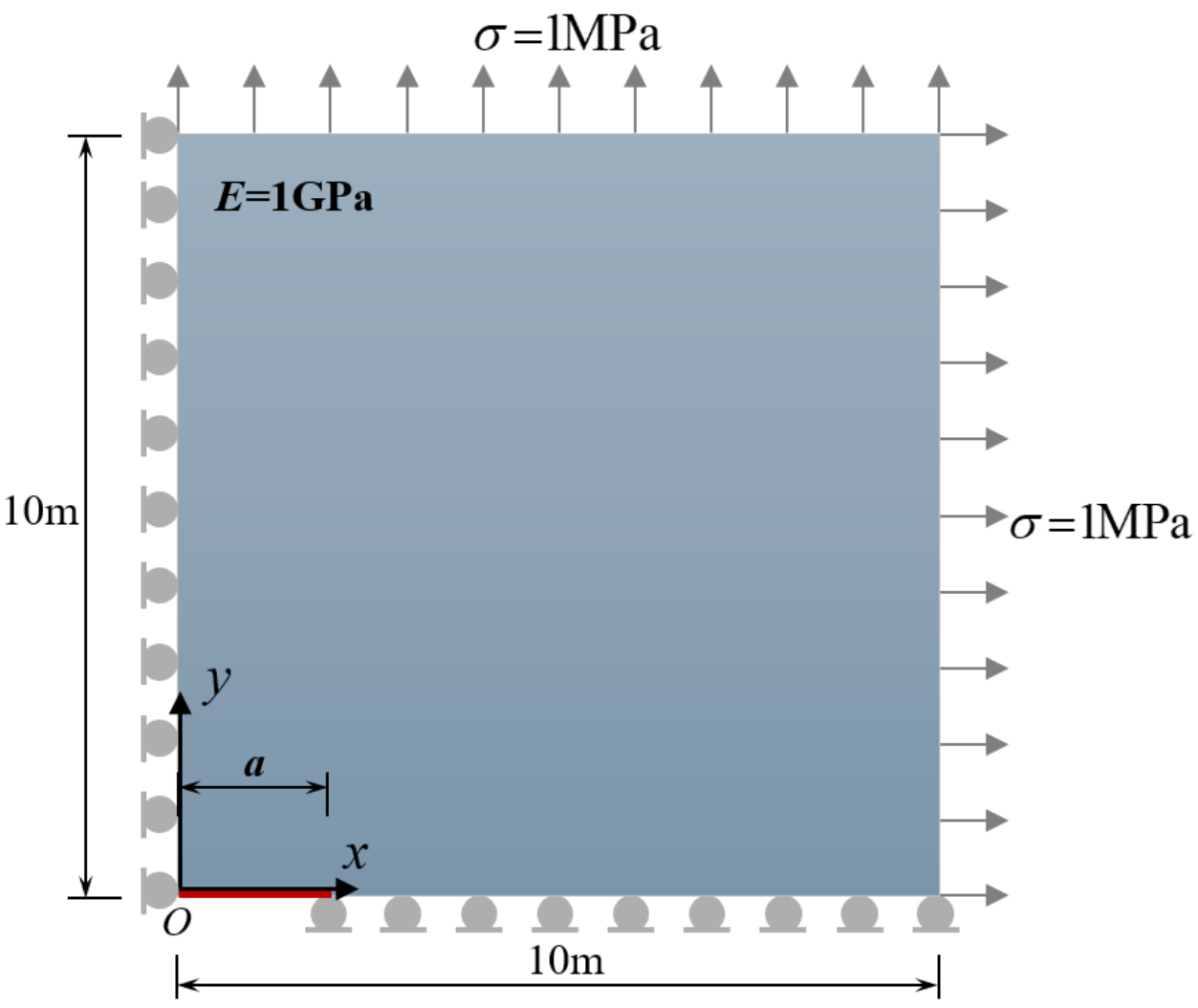

Fig. 22 Centre cracked plate: geometry and boundary conditions.

The analytical crack opening displacement for infinite domains is (Zhao and Zhao, 2012):

$$u_y(x) = \frac{2\sigma}{E}\sqrt{a^2 - x^2} \tag{52}$$

In the convergence analysis, the displacement extrapolation method (Chan et al., 1970) was also employed to evaluate the crack-tip mode-I stress intensity factor (SIF). Based on the asymptotic displacement field near a mode-I crack tip, one obtains:

$$K_{\mathrm{I}}^{*} = \frac{2G}{1+\kappa}\sqrt{\frac{2\pi}{\rho}} \tag{53}$$

Here, $K_{\mathrm{I}}^{*}$ denotes the apparent stress intensity factor evaluated at nodes near the crack tip. By computing $K_{\mathrm{I}}^{*}$ for different values of $\rho$, a set of data points $(\rho_i, K_{\mathrm{I},i}^{*})$ on the crack faces is obtained. A straight line is then fitted to $(\rho_i, K_{\mathrm{I},i}^{*})$ using the least-squares method and extrapolated to $\rho = 0$; the intercept of this line yields the desired $K_{\mathrm{I}}$. Specifically,

$$K_{\mathrm{I}} \approx \frac{\sum\rho_i\sum\rho_i K_{\mathrm{I},i}^{*} - \sum\rho_i^{2}\sum K_{\mathrm{I},i}^{*}}{(\sum\rho_i)^2 - n\sum\rho_i^{2}}, \quad (i = 1, 2, \ldots, n) \tag{54}$$

Where $\rho_i$ and $K_{\mathrm{I},i}^{*}$ are the polar distance and the apparent stress intensity factor at the $i$-th point on the crack faces, respectively, and $n$ is the number of sampling points (nodes) on the crack.

Convergence analysis for $\nu$=0.2, shown in Fig. 23($a$), demonstrates IELSM's mesh convergence by comparing crack-face opening displacement ($u_y$) distributions along the crack surface with analytical solutions. At identical mesh resolution ($H/L$ = 160), IELSM solutions coincide exactly with FEM Q4 element predictions. Fig. 23($b$) displays displacement points used for SIF calculation. Derived SIF values and relative errors are presented in Table 4, showing convergence toward the theoretical value $K_0 = \sigma\sqrt{a\pi}$ .

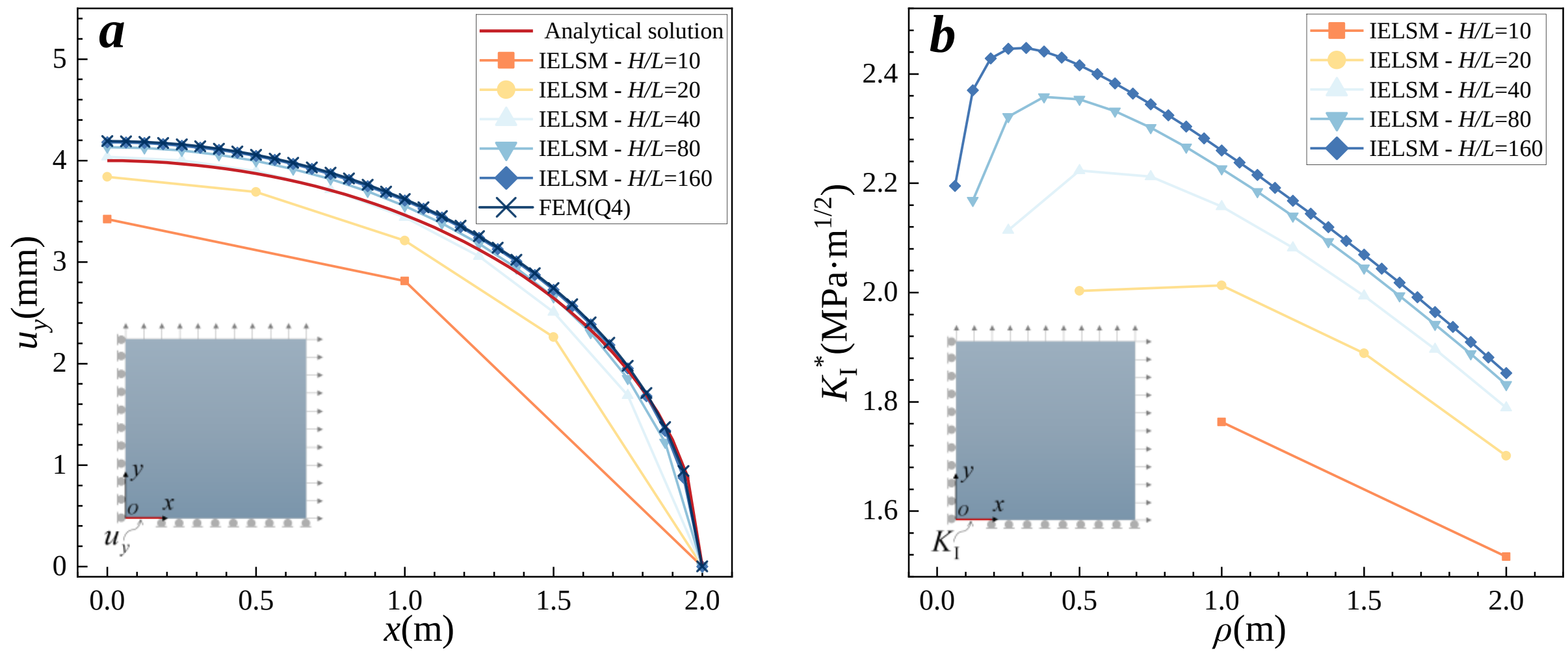

Fig. 23 Centre cracked plate:($a$) convergence analysis ($v$=0.2); ($b$) displacement points for SIF calculation.

Table 4 Centre cracked plate: SIF values computed via IELSM with varying mesh densities.

| $H/L$ | 10 | 20 | 40 | 80 | 160 |
|---|---|---|---|---|---|
| $K_I$ (MPa×m$^{1/2}$) | 2.0092 | 2.1589 | 2.3072 | 2.4278 | 2.5150 |
| $K_0$ (MPa×m$^{1/2}$) | 2.5066 | 2.5066 | 2.5066 | 2.5066 | 2.5066 |
| Rel. Error (%) | 19.85 | 13.87 | 7.85 | 3.15 | 0.33 |

To further compare the displacement fields predicted by IELSM and the Q4 element in FEM, the complete displacement-field contour maps for $v$ = 0.2 are presented (see Fig. 24). The results show that the displacement fields predicted by IELSM and the FEM Q4 element are in full agreement.

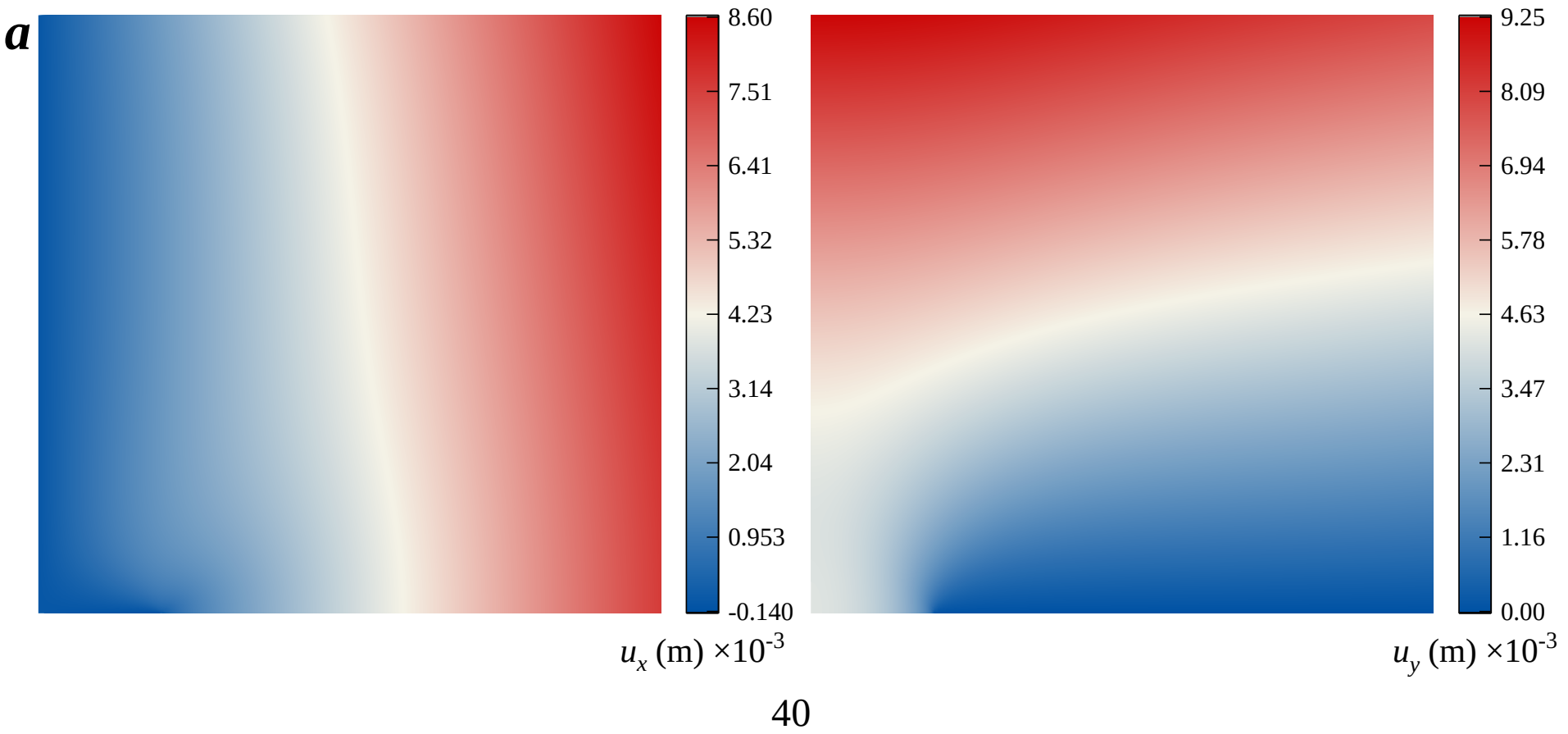

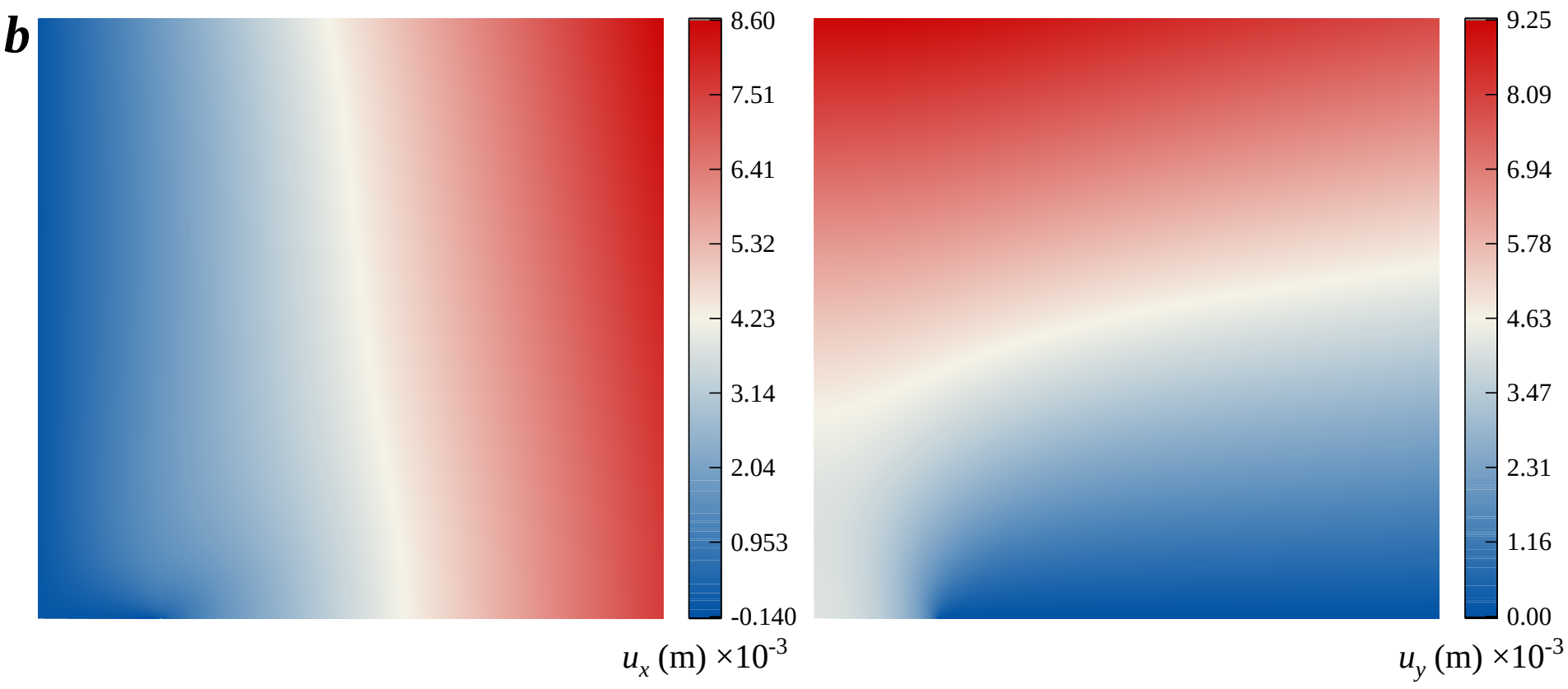

Fig. 24 Centre cracked plate: (*a*) FEM-predicted displacement field; (*b*) IELSM-predicted displacement field ($v = 0.2$).

Stress field predictions within crack tip singularity zones are examined in Fig. 25, showing radial distributions of (*a*) $\sigma_{xx}$, (*b*) $\sigma_{yy}$, and (*c*) $\tau_{xy}$ within $0 \le r \le 0.25a$. Analytical near-tip stresses (Williams, 1957) are:

$$\begin{cases} \sigma_{xx}(\rho,\theta) = \dfrac{K_{\mathrm{I}}}{\sqrt{2\pi\rho}} \cos\dfrac{\theta}{2}(1 - \sin\dfrac{\theta}{2}\sin\dfrac{3\theta}{2}) \\ \sigma_{yy}(\rho,\theta) = \dfrac{K_{\mathrm{I}}}{\sqrt{2\pi\rho}} \cos\dfrac{\theta}{2}(1 + \sin\dfrac{\theta}{2}\sin\dfrac{3\theta}{2}) \\ \tau_{xx}(\rho,\theta) = \dfrac{K_{\mathrm{I}}}{\sqrt{2\pi\rho}} \cos\dfrac{\theta}{2}\sin\dfrac{\theta}{2}\cos\dfrac{3\theta}{2} \end{cases} \tag{55}$$

Under biaxial tension ($\sigma$ = 1MPa), bilinear displacement field IELSM predictions closely match FEM Q4 results, with the maximum stresses near the singularity reaching 5 MPa ($\sigma_{yy}$) and 7 MPa ($\sigma_{xx}$). Linear displacement field IELSM underestimates these peaks (3 MPa and 4.5 MPa, respectively). Both IELSM variants deviate similarly from analytical shear stress predictions, suggesting limitations in accurately capturing the crack-tip shear stress distribution.

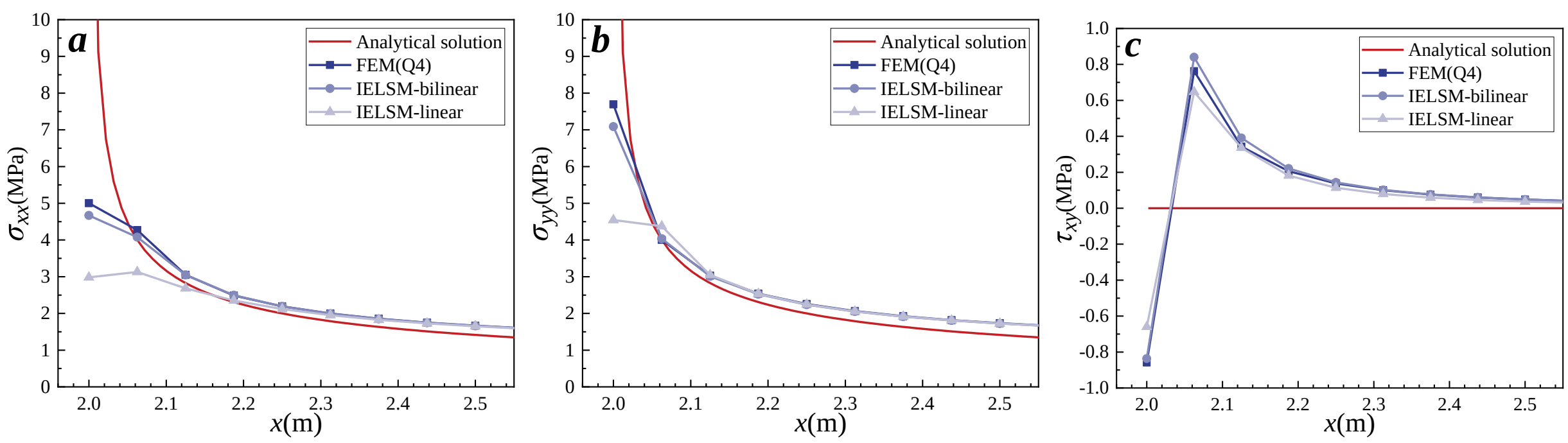

Fig. 25 Centre cracked plate: stress field predictions within crack tip singularity zone ($H/L$ = 160, $v = 0.2$).

## 5.5 Crucifix crack plate

A finite square plate (under plane stress conditions) with central cruciform-shaped cracks subjected to biaxial tension is analyzed via quarter-symmetry (Fig. 26). Domain dimensions are $H \times H = 10 \times 10$m with half-crack length $a$ varying from 2 to 8m. Material properties: $E = 200$GPa, $\nu = 0.286$. Mesh resolutions: $H/L = \{20, 40, 80, 160, 320\}$.

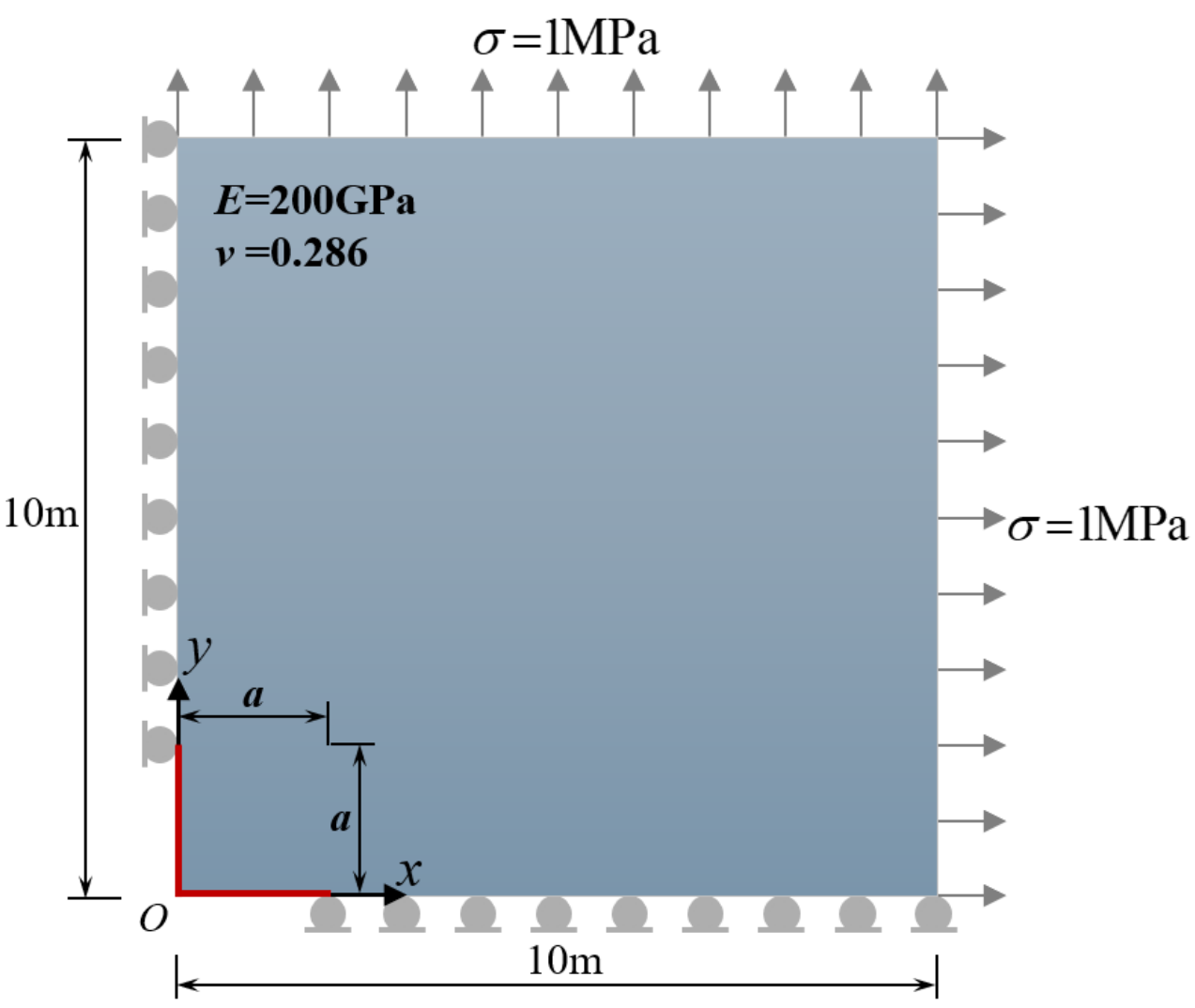

Fig. 26 Crucifix crack plate: geometry and boundary conditions.

Convergence analysis in Fig. 27($a$) demonstrates excellent mesh convergence, with IELSM and FEM Q4 displacement profiles overlapping perfectly at $H/L = 320$. SIF calculation points along the crack face are shown in Fig. 27($b$). Computed SIF values and relative deviations against reference solutions are summarized in Table 5. The reference value, taken from Cheung et al. (1992), represents the normalized stress intensity factor $K_I/K_0 = 0.8800$, where $K_I$ is the stress intensity factor for the crucifix crack tip. The results in Table 5 confirm IELSM's convergence with mesh refinement.

Table 5 Crucifix crack plate: SIF values computed via IELSM (displacement extrapolation method, $a/H = 0.2$).

| $H/L$ | 20 | 40 | 80 | 160 | 320 |
|---|---|---|---|---|---|
| $K_I (\text{MPa} \times \text{m}^{1/2})$ | 1.8564 | 1.9607 | 2.0475 | 2.1112 | 2.1547 |
| $K_I / K_0$ | 0.7406 | 0.7822 | 0.8168 | 0.8423 | 0.8596 |
| Rel. Error (%) | 15.84 | 11.11 | 7.178 | 4.29 | 2.32 |

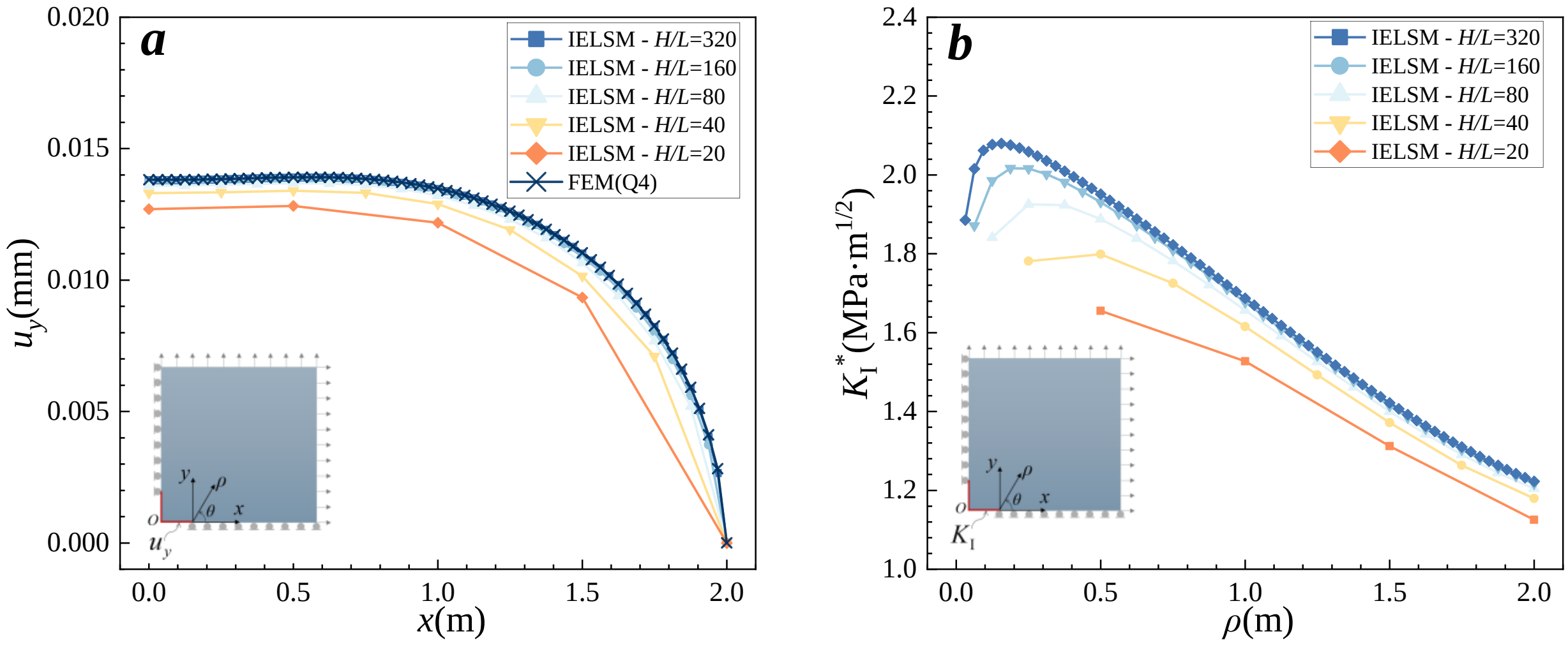

Fig. 27 Crucifix crack plate: (*a*) convergence analysis; (*b*) displacement points for SIF calculation.

Apparent SIF distributions along crack faces for varying crack lengths ($a/H$ = 0.2~0.8) are presented in Fig. 28. Corresponding SIF values and relative deviations against reference solutions are compiled in Table 6, with maximum deviation limited to 2.32%, confirming IELSM's precision in capturing crack opening displacements for SIF extraction.

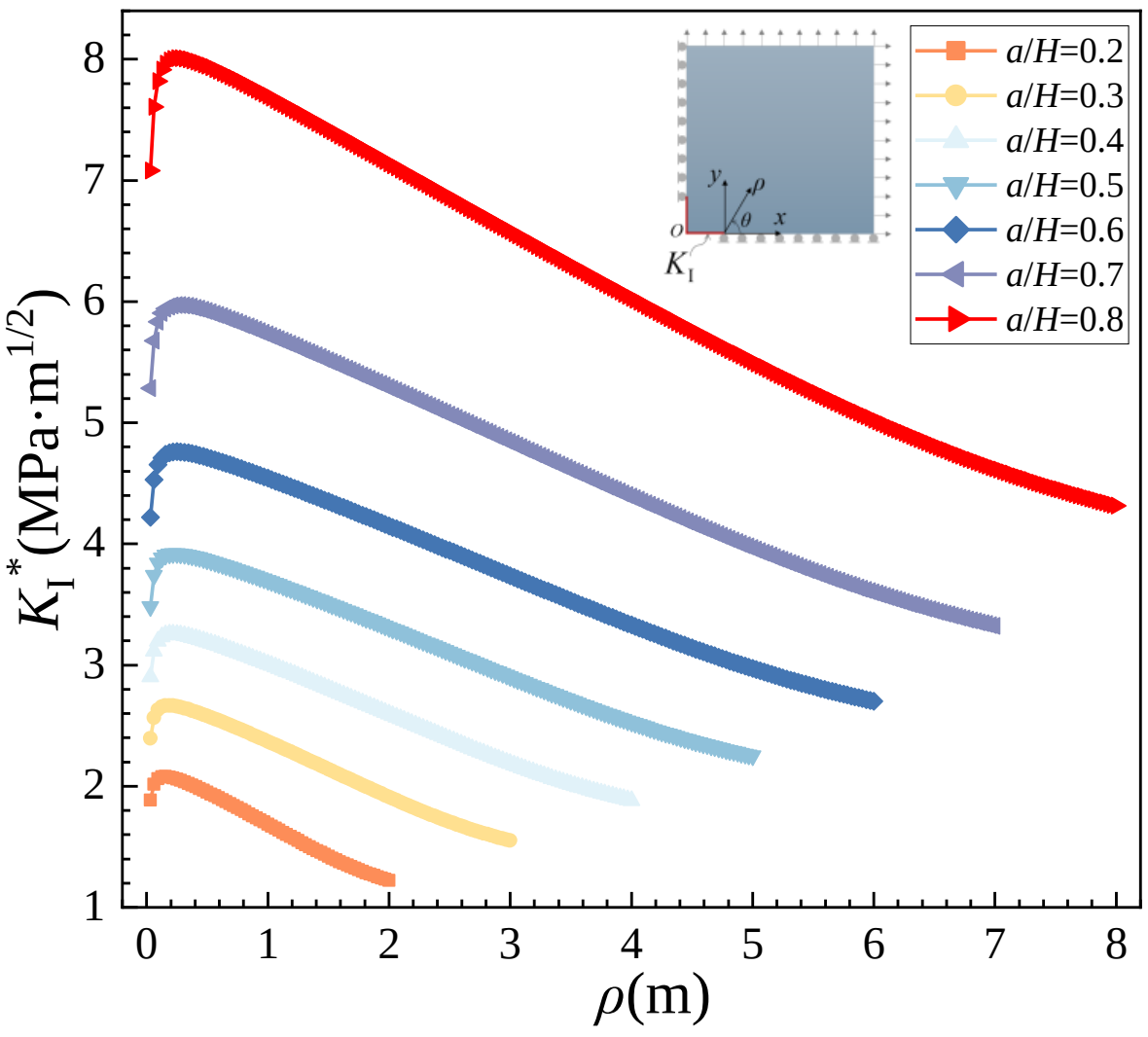

Fig. 28 Crucifix crack plate: apparent $K_I^*$ distributions along crack faces for varying $a/H$.

Table 6 Crucifix crack plate: SIF values for varying crack length ratios.

| $a$/H | 0.2 | 0.3 | 0.4 | 0.5 | 0.6 | 0.7 | 0.8 |
|---|---|---|---|---|---|---|---|
| $K_I$ (MPa×m$^{1/2}$) | 2.1547 | 2.7508 | 3.3494 | 4.0222 | 4.4921 | 6.0794 | 8.0952 |
| $K_0$ (MPa×m$^{1/2}$) | 2.5066 | 3.0700 | 3.5449 | 3.9633 | 4.3416 | 4.6895 | 5.0133 |
| $K_I/K_0$ | 0.8596 | 0.8960 | 0.9449 | 1.0149 | 1.0347 | 1.2964 | 1.6148 |
| $K_I/K_0$ from Ref. (Cheung et al., 1992) | 0.8800 | 0.9092 | 0.9537 | 1.0223 | 1.1300 | 1.2866 | 1.4857 |
| Rel. Error (%) | 2.32 | 1.45 | 0.93 | 0.73 | 0.84 | 0.76 | 0.87 |

Displacement field congruence between IELSM and FEM Q4 elements ($H/L$ = 320) is illustrated in Fig. 29. Strain field comparisons in Fig. 30 reveal that bilinear displacement field IELSM closely matches FEM solutions well, while linear displacement field IELSM underestimates strains near crack tips but agrees with the other solutions in uniform strain regions.

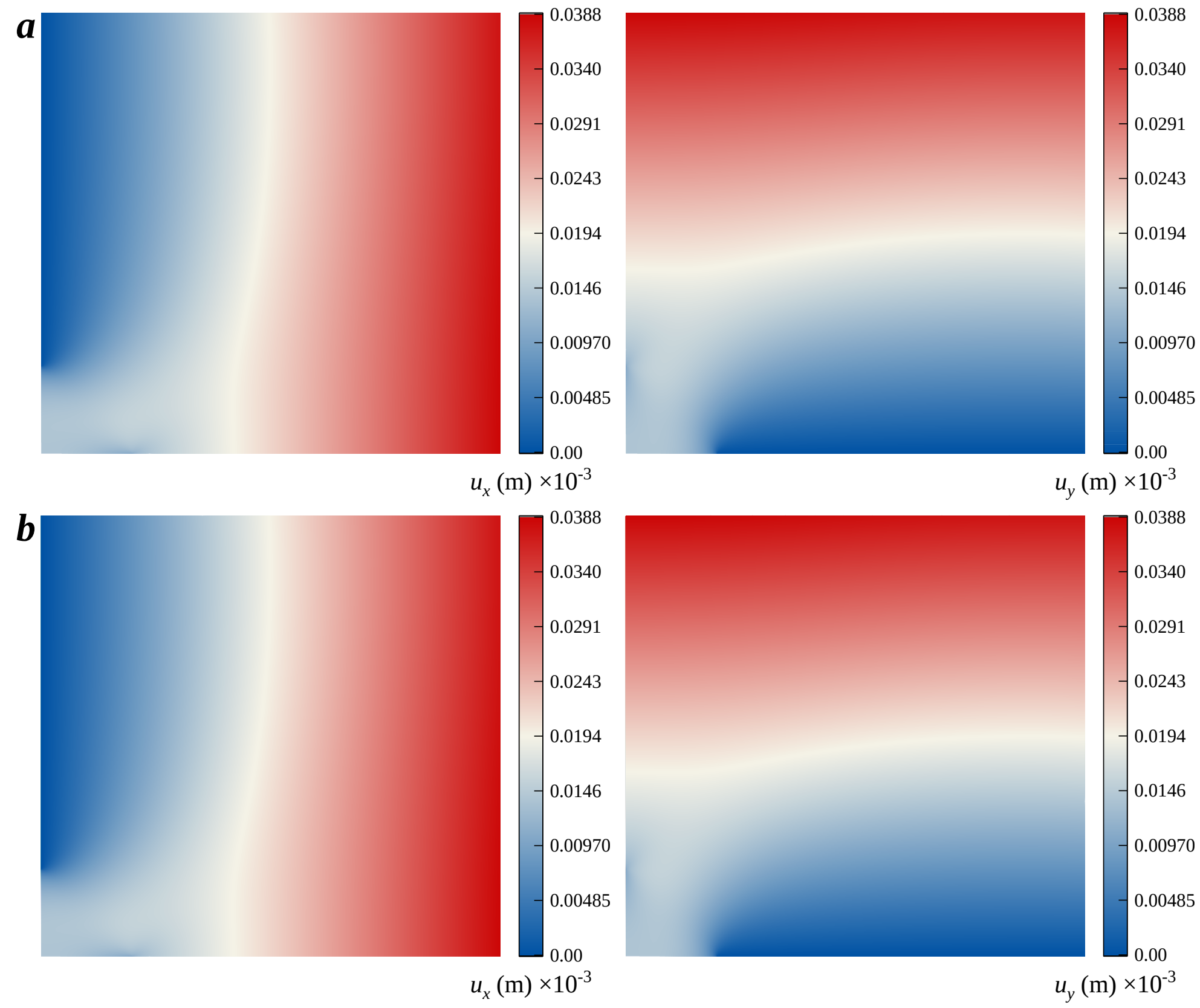

Fig. 29 Crucifix crack plate: (*a*) FEM-predicted displacement field; (*b*) IELSM-predicted displacement field.

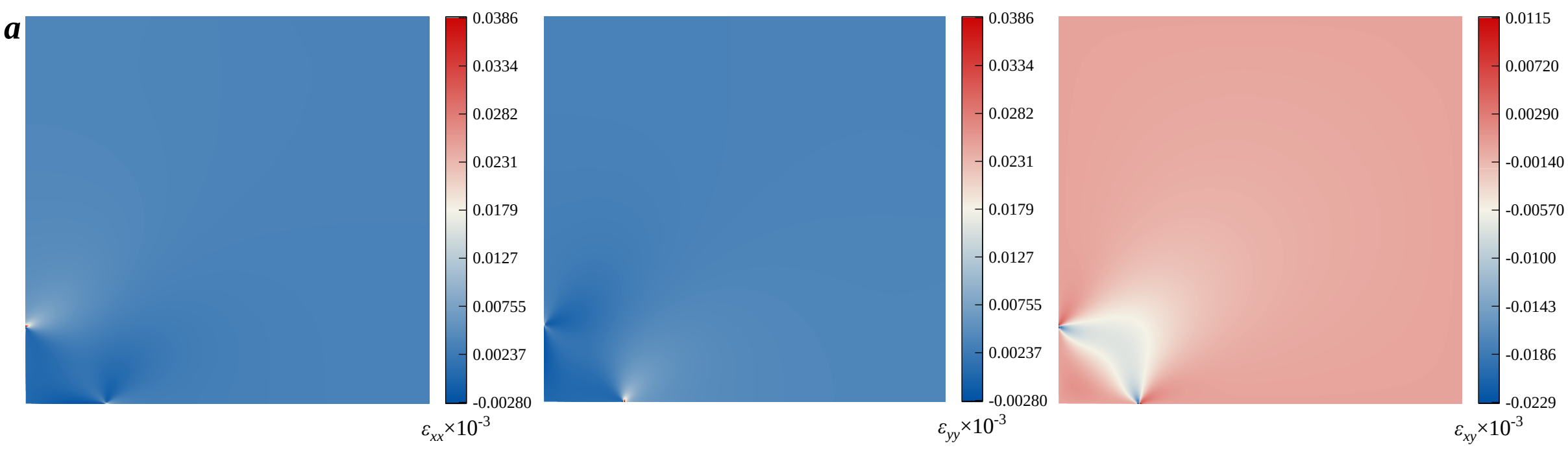

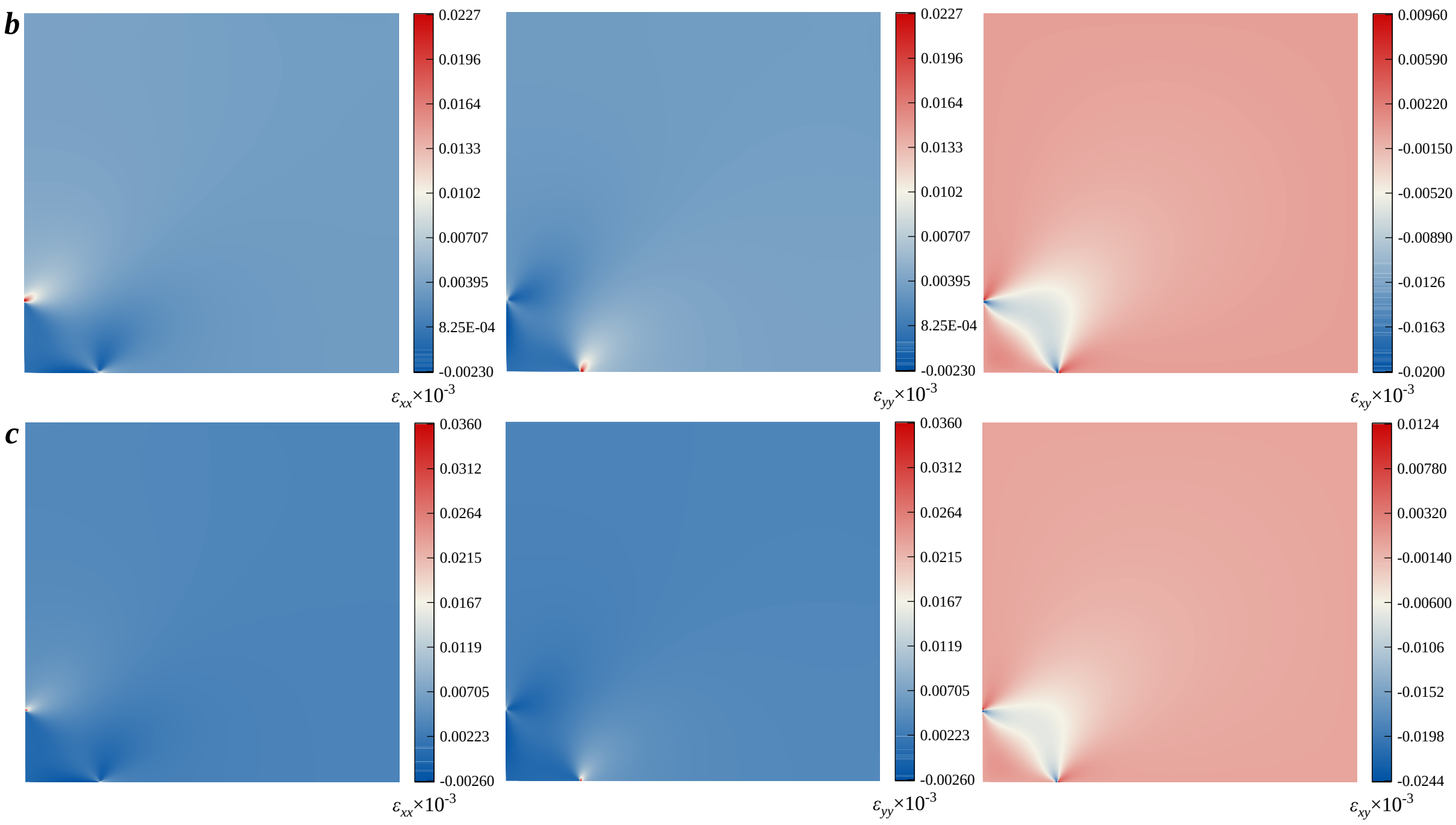

Fig. 30 Crucifix crack plate: (*a*) FEM strain field; (*b*) linear displacement field IELSM strain field; (*c*) bilinear displacement field IELSM strain field.

Stress singularity characteristics are analyzed in Fig. 31, showing radial distributions of stress components within crack tip zones for $a/H$ = 0.2. Under biaxial tension, bilinear displacement field IELSM predictions align with FEM Q4 results, achieving 6MPa ($\sigma_{yy}$) and 9MPa ($\sigma_{xx}$) peak stresses. Linear displacement field IELSM again underestimates these values (3MPa and 5.5MPa). All three methods exhibit comparable shear stress oscillations within stress concentration zones.

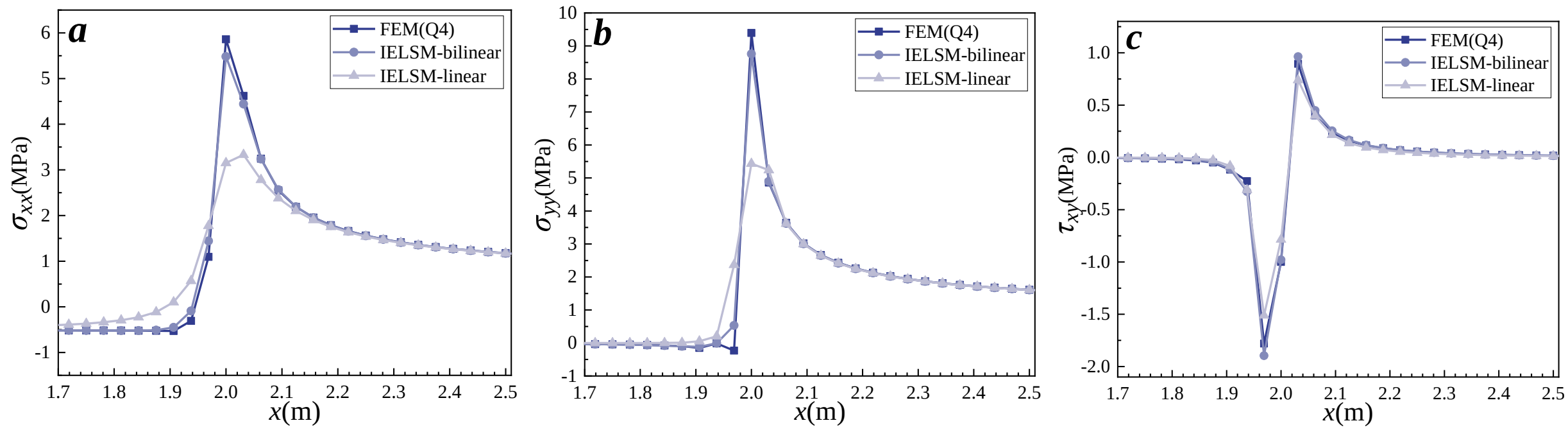

Fig. 31 Crucifix crack plate: stress component distributions within crack tip singularity zone ($H/L$=320, $a/H$=0.2).

## 6.0 Conclusion

This study establishes a novel Isotropic Elastic Lattice Spring Model (IELSM) that fundamentally resolves the long-standing limitation of fixed Poisson's ratio in classical lattice spring methods. Through theoretical derivation, numerical implementation, and comprehensive validation including crack-tip stress field analysis, the following key conclusions emerge:

(1). **Core theoretical framework**: IELSM's core innovation lies in the explicit introduction of additional bulk modulus parameters alongside traditional axial springs. This augmentation establishes an exact, physically meaningful connection between discrete spring networks and continuum elasticity theory. Analytical derivation demonstrates that additional volumetric constraints precisely compensate for the energy deficiency in classical LSM formulations, allowing independent specification of Poisson's ratio while maintaining essential rotational invariance properties.

(2). **Complete parameter mapping**: IELSM provides exact analytical relationships between parameters $\{k_1^n(k_2^n),\ k^v\}$ and continuum elastic constants $\{E,\ \nu\}$. The model spans the full physically admissible Poisson's ratio range, including unusual material behaviors such as auxetic materials (negative Poisson's ratio) and incompressible solids ($\nu\to0.5$), while requiring only two independent parameters consistent with isotropic elasticity theory.

(3). **Numerical implementation framework**: A computationally efficient decomposition scheme translates complex additional volumetric constraints into equivalent combinations of standard mechanical elements. This decomposition not only simplifies implementation but also enables straightforward integration with existing finite element solvers. The discrete formulation of IELSM provides a convenient foundation for later incorporation of fracture criteria and damage models.

(4). **Validation and performance**: The model's efficacy is validated through comprehensive benchmark tests. IELSM demonstrates excellent convergence characteristics and computational accuracy across various loading conditions and material properties. Crucially, the formulation suppresses the hourglass zero-energy modes inherent to standard bilinear quadrilaterals while maintaining a stable eigenvalue spectrum across all higher deformation modes. Particularly noteworthy is its performance in crack-tip stress singularity analysis, where it accurately captures stress intensity factors and matches theoretical solutions with remarkable precision.

**Future Work**: (1) three-dimensional extension; (2) crack initiation and propagation; (3) general quadrilateral element extension; (4) anisotropic material modeling; (5) geometrically nonlinear formulation; (6) hyperelastic constitutive modeling

In summary, IELSM represents a significant advancement in discrete lattice modeling by combining theoretical rigor with practical utility. Although the current formulation is limited to square lattice configurations and thus cannot accommodate arbitrary quadrilateral elements, this work fundamentally overcomes a key limitation that has persisted since Hrennikoff's original framework method. By expanding the applicability of lattice spring models to a broader range of engineering materials and structural problems while maintaining computational advantages essential for large-scale simulations, IELSM effectively bridges lattice spring method with continuum mechanics principles. This achievement provides a versatile framework for elastic boundary value problems and crack analysis, promising to advance both theoretical understanding and engineering applications.

**Acknowledgements**

DM thanks the supports of the National Natural Science Foundation of China (12002247) and the Fundamental Research Fund for the Central Universities of China (WUT: 2021IVB013).

**Appendix A**

***A.1.* Derivation of Poisson's Ratio Constraint for Classical LSM**

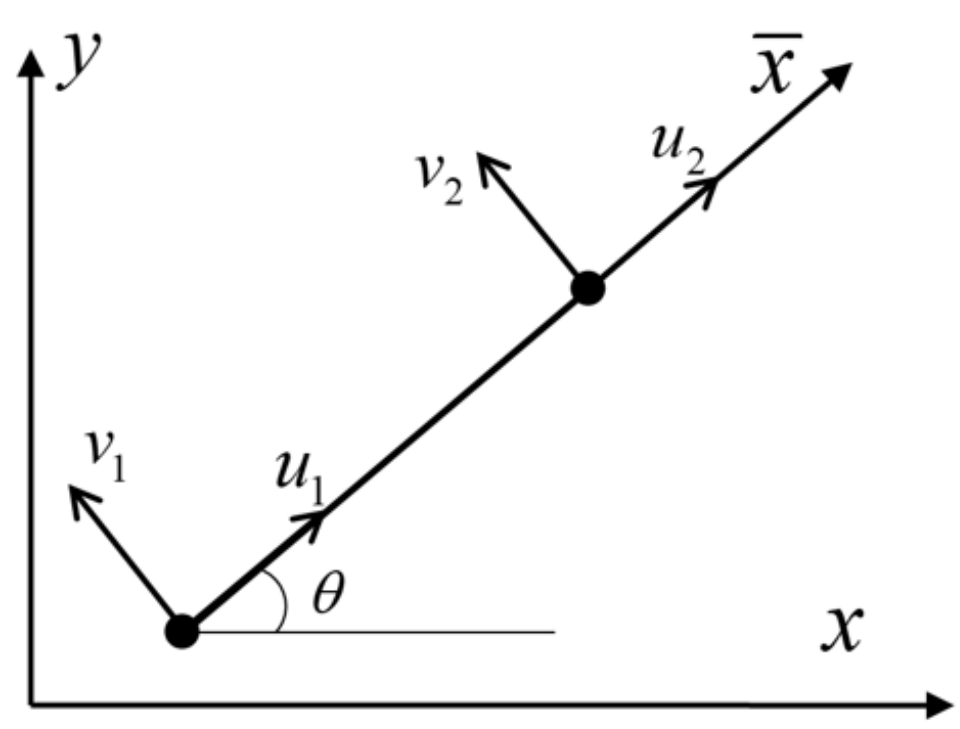

Fig. A1 Illustrates a two-dimensional rod element in both local and global coordinate systems.

Consider a rod element oriented at angle $\theta$ relative to the $x$-axis, with length $L$, cross-sectional area $A$, and Young's modulus $E$, defining its axial stiffness as $k^n = EA/L$. The axial strain of the rod can be obtained through projection of strain components along the rod's axial direction (Griffiths and Mustoe, 2001):

$$\varepsilon_\theta = \cos^2\theta\ \varepsilon_{xx} + \sin\theta\cos\theta(\varepsilon_{xy}+\varepsilon_{yx}) + \sin^2\theta\ \varepsilon_{yy} \quad \text{(A.1)}$$

where $\varepsilon_\theta$ represents the axial strain of the rod, while $\varepsilon_{ij}$ denote the global strain components.

For a rod element containing only an axial spring, the strain energy depends solely on the relative axial displacement of the rod:

$$\begin{aligned} W(\theta) &= \frac{k^n(u_2-u_1)^2}{2} = \frac{k^n(\varepsilon_\theta L)^2}{2} \\ &= \frac{k^n L^2(\cos^2\theta\ \varepsilon_{xx} + \sin\theta\cos\theta(\varepsilon_{xy}+\varepsilon_{yx}) + \sin^2\theta\ \varepsilon_{yy})^2}{2} \end{aligned} \quad \text{(A.2)}$$

where $u_i$ and $v_i$ represent nodal displacements in the local coordinate system of the rod. The strain energies for rods of length $L$ and $\sqrt{2}L$ are expressed as:

$$\begin{cases} W_1 = \dfrac{k_1^n L^2(\cos^2\theta\ \varepsilon_{xx} + \sin\theta\cos\theta(\varepsilon_{xy}+\varepsilon_{yx}) + \sin^2\theta\ \varepsilon_{yy})^2}{2} \\ W_2 = k_2^n L^2(\cos^2\theta\ \varepsilon_{xx} + \sin\theta\cos\theta(\varepsilon_{xy}+\varepsilon_{yx}) + \sin^2\theta\ \varepsilon_{yy})^2 \\ W_{\mathrm{LSM}} = W_1(\theta) + W_1(\theta+\pi/2) + W_1(\theta+\pi) + W_1(\theta-\pi/2) + W_2(\theta+\pi/4) + W_2(\theta-\pi/4) \end{cases} \quad \text{(A.3)}$$

Substituting (A.2) into (A.3) yields Eq. (2). The representative volume of the unit cell corresponds to its area, $L^2$, thus the strain energy density is $V = W_{\mathrm{tot}}/L^2$. Differentiating the strain energy density $V$ with respect to strain components yields stress components:

$$\text{LSM}\begin{cases}\sigma_{xx}=\dfrac{\partial V}{\partial \varepsilon_{xx}}=\dfrac{1}{2}(k_1^n+k_2^n)(3\varepsilon_{xx}+\varepsilon_{yy})\\ \sigma_{yy}=\dfrac{\partial V}{\partial \varepsilon_{yy}}=\dfrac{1}{2}(k_1^n+k_2^n)(\varepsilon_{xx}+3\varepsilon_{yy})\\ \tau_{xy}=\dfrac{1}{2}\left(\dfrac{\partial V}{\partial \varepsilon_{xy}}+\dfrac{\partial V}{\partial \varepsilon_{yx}}\right)=\dfrac{1}{2}(k_1^n+k_2^n)(\varepsilon_{xy}+\varepsilon_{yx})\end{cases} \tag{A.4}$$

Comparison with two-dimensional isotropic elasticity theory:

$$\underbrace{\begin{aligned}\sigma_{xx}&=\frac{E(1-v)}{(1+v)(1-2v)}\varepsilon_{xx}+\frac{Ev}{(1+v)(1-2v)}\varepsilon_{yy}\\ \sigma_{yy}&=\frac{Ev}{(1+v)(1-2v)}\varepsilon_{xx}+\frac{E(1-v)}{(1+v)(1-2v)}\varepsilon_{yy}\\ \tau_{xy}&=\frac{E}{2(1+v)}\gamma_{xy}\end{aligned}}_{Plane\ strain}\quad \underbrace{\begin{aligned}\sigma_{xx}&=\frac{E}{(1-v^2)}\varepsilon_{xx}+\frac{Ev}{(1-v^2)}\varepsilon_{yy}\\ \sigma_{yy}&=\frac{Ev}{(1-v^2)}\varepsilon_{xx}+\frac{E}{(1-v^2)}\varepsilon_{yy}\\ \tau_{xy}&=\frac{E}{2(1+v)}\gamma_{xy}\end{aligned}}_{Plane\ stress} \tag{A.5}$$

yields the following relationships:

$$Plane\ strain\begin{cases}\dfrac{E(1-v)}{(1+v)(1-2v)}\equiv\dfrac{3}{2}(k_1^n+k_2^n)\\ \dfrac{Ev}{(1+v)(1-2v)}\equiv\dfrac{1}{2}(k_1^n+k_2^n)\\ \dfrac{E}{2(1+v)}\equiv\dfrac{1}{2}(k_1^n+k_2^n)\end{cases},\ Plane\ stress\begin{cases}\dfrac{E}{(1-v^2)}\equiv\dfrac{3}{2}(k_1^n+k_2^n)\\ \dfrac{Ev}{(1-v^2)}\equiv\dfrac{1}{2}(k_1^n+k_2^n)\\ \dfrac{E}{2(1+v)}\equiv\dfrac{1}{2}(k_1^n+k_2^n)\end{cases} \tag{A.6}$$

Solving these equations gives Eq. (3).

## Appendix B

### *B.1.* Magnitude Analysis of Linear and Quadratic Displacement Terms in Volumetric Strain

Consider a square representative unit as illustrated in Fig. 4, with displacements ($u_i$, $v_i$) at four nodes. Taking the element centroid $P(L/2, L/2)$ as the reference point with displacement ($u$, $v$), we employ Taylor expansion of the displacement field. Assuming the displacement field is continuously differentiable within the element, for node $P_1(L, L)$ we have:

$$\begin{aligned} u_1 &= u + \frac{L}{2}\frac{\partial u}{\partial x} + \frac{L}{2}\frac{\partial u}{\partial y} + \frac{L^2}{8}\frac{\partial^2 u}{\partial x^2} + \frac{L^2}{8}\frac{\partial^2 u}{\partial y^2} + \ldots \\ v_1 &= v + \frac{L}{2}\frac{\partial v}{\partial x} + \frac{L}{2}\frac{\partial v}{\partial y} + \frac{L^2}{8}\frac{\partial^2 v}{\partial x^2} + \frac{L^2}{8}\frac{\partial^2 v}{\partial y^2} + \ldots \end{aligned} \tag{B.1}$$

Under small deformation assumptions, retaining first-order approximations:

$$u_1 = u + \frac{L}{2}\frac{\partial u}{\partial x} + \frac{L}{2}\frac{\partial u}{\partial y}, \quad v_1 = v + \frac{L}{2}\frac{\partial v}{\partial x} + \frac{L}{2}\frac{\partial v}{\partial y} \tag{B.2}$$

Similar Taylor expansions can be derived for other nodal displacements:

$$u_2 = u - \frac{L}{2}\frac{\partial u}{\partial x} + \frac{L}{2}\frac{\partial u}{\partial y}, \quad v_2 = v - \frac{L}{2}\frac{\partial v}{\partial x} + \frac{L}{2}\frac{\partial v}{\partial y} \tag{B.3}$$

$$u_3 = u - \frac{L}{2}\frac{\partial u}{\partial x} - \frac{L}{2}\frac{\partial u}{\partial y}, \quad v_3 = v - \frac{L}{2}\frac{\partial v}{\partial x} - \frac{L}{2}\frac{\partial v}{\partial y} \tag{B.4}$$

$$u_4 = u + \frac{L}{2}\frac{\partial u}{\partial x} - \frac{L}{2}\frac{\partial u}{\partial y}, \quad v_4 = v + \frac{L}{2}\frac{\partial v}{\partial x} - \frac{L}{2}\frac{\partial v}{\partial y} \tag{B.5}$$

Substituting (B.2)~(B.5) into the linear term of volumetric strain yields:

$$\begin{aligned} \varepsilon_v^l &= \frac{1}{2L}(u_1 - u_2 - u_3 + u_4 + v_1 + v_2 - v_3 - v_4) \\ &= \frac{1}{2L}\left( \begin{aligned} &\left(u + \frac{L}{2}\frac{\partial u}{\partial x} + \frac{L}{2}\frac{\partial u}{\partial y}\right) - \left(u - \frac{L}{2}\frac{\partial u}{\partial x} + \frac{L}{2}\frac{\partial u}{\partial y}\right) - \left(u - \frac{L}{2}\frac{\partial u}{\partial x} - \frac{L}{2}\frac{\partial u}{\partial y}\right) + \left(u + \frac{L}{2}\frac{\partial u}{\partial x} - \frac{L}{2}\frac{\partial u}{\partial y}\right) \\ &+ \left(v + \frac{L}{2}\frac{\partial v}{\partial x} + \frac{L}{2}\frac{\partial v}{\partial y}\right) + \left(v - \frac{L}{2}\frac{\partial v}{\partial x} + \frac{L}{2}\frac{\partial v}{\partial y}\right) - \left(v - \frac{L}{2}\frac{\partial v}{\partial x} - \frac{L}{2}\frac{\partial v}{\partial y}\right) - \left(v + \frac{L}{2}\frac{\partial v}{\partial x} - \frac{L}{2}\frac{\partial v}{\partial y}\right) \end{aligned} \right) \\ &= \frac{\partial u}{\partial x} + \frac{\partial v}{\partial y} \end{aligned} \tag{B.6}$$

For the nonlinear term:

$$
\begin{aligned}
\varepsilon_v^{nl} &= \frac{1}{2L^2}(u_1v_2 - u_1v_4 - v_1u_2 + v_3u_2 - u_3v_2 + u_3v_4 + v_1u_4 - v_3u_4) \\
&= \frac{1}{2L^2}\left((u_1 - u_3)(v_2 - v_4) - (v_1 - v_3)(u_2 - u_4)\right) \\
&= \frac{\left(\left(u + \frac{L}{2}\frac{\partial u}{\partial x} + \frac{L}{2}\frac{\partial u}{\partial y}\right) - \left(u - \frac{L}{2}\frac{\partial u}{\partial x} - \frac{L}{2}\frac{\partial u}{\partial y}\right)\right)\left(\left(v - \frac{L}{2}\frac{\partial v}{\partial x} + \frac{L}{2}\frac{\partial v}{\partial y}\right) - \left(v + \frac{L}{2}\frac{\partial v}{\partial x} - \frac{L}{2}\frac{\partial v}{\partial y}\right)\right)}{2L^2} \\
&\quad - \frac{\left(\left(v + \frac{L}{2}\frac{\partial v}{\partial x} + \frac{L}{2}\frac{\partial v}{\partial y}\right) - \left(v - \frac{L}{2}\frac{\partial v}{\partial x} - \frac{L}{2}\frac{\partial v}{\partial y}\right)\right)\left(\left(u - \frac{L}{2}\frac{\partial u}{\partial x} + \frac{L}{2}\frac{\partial u}{\partial y}\right) - \left(u + \frac{L}{2}\frac{\partial u}{\partial x} - \frac{L}{2}\frac{\partial u}{\partial y}\right)\right)}{2L^2} \\
&= \frac{1}{2L^2}\left(\left(L\frac{\partial u}{\partial x} + L\frac{\partial u}{\partial y}\right)\left(-L\frac{\partial v}{\partial x} + L\frac{\partial v}{\partial y}\right) - \left(L\frac{\partial v}{\partial x} + L\frac{\partial v}{\partial y}\right)\left(-L\frac{\partial u}{\partial x} + L\frac{\partial u}{\partial y}\right)\right) \\
&= \frac{1}{2}\left(-\frac{\partial u}{\partial x}\frac{\partial v}{\partial x} + \frac{\partial u}{\partial x}\frac{\partial v}{\partial y} - \frac{\partial u}{\partial y}\frac{\partial v}{\partial x} + \frac{\partial u}{\partial y}\frac{\partial v}{\partial y} + \frac{\partial v}{\partial x}\frac{\partial u}{\partial x} - \frac{\partial v}{\partial x}\frac{\partial u}{\partial y} + \frac{\partial v}{\partial y}\frac{\partial u}{\partial x} - \frac{\partial v}{\partial y}\frac{\partial u}{\partial y}\right) \\
&= \frac{\partial u}{\partial x}\frac{\partial v}{\partial y} - \frac{\partial u}{\partial y}\frac{\partial v}{\partial x}
\end{aligned} \tag{B.7}
$$

This expression demonstrates that nonlinear terms in volumetric strain correspond to second-order displacement gradients. Under small deformation assumptions, these higher-order terms can be neglected, yielding the simplified volumetric strain:

$$
\varepsilon_v = \varepsilon_v^l = \frac{1}{2L}(u_1 - u_2 - u_3 + u_4 + v_1 + v_2 - v_3 - v_4) \tag{B.8}
$$